# HYDROGEN PRODUCTION FROM POLYMERIC ORGANIC SOLIDS VIA ATMOSPHERIC PRESSURE NONTHERMAL PLASMA

A Dissertation Presented

By

Benard Tabu

Submitted to the College of Engineering,

University of Massachusetts Lowell,

in partial fulfillment of the requirements for the degree of

DOCTOR OF PHILOSOPHY

ENERGY ENGINEERING

May 2023

# HYDROGEN PRODUCTION FROM POLYMERIC ORGANIC SOLIDS VIA ATMOSPHERIC PRESSURE NONTHERMAL PLASMA

BY

BENARD TABU
B.S. EDUCATION (PHYSICAL), GULU UNIVERSITY, UGANDA (2010)
M.S. PHYSICS, MAKERERE UNIVERSITY, UGANDA (2017)

SUBMITTED IN PARTIAL FULFILLMENT OF THE REQUIREMENTS FOR THE DEGREE OF DOCTOR OF PHILOSOPHY
ENERGY ENGINEERING,
UNIVERSITY OF MASSACHUSETTS LOWELL

**Author's Signature:** ________________________ **Date:**_____________________

**Proposal Supervisor**
**Name:** Prof. Juan Pablo Trelles
**Signature:** ________________________________________

**Proposal Committee Members:**
**Name:** Prof. Hsi-Wu Wong
**Signature:** ________________________________________

**Name:** Prof. John Hunter Mack
**Signature:** ________________________________________

**Name:** Prof. Maria Carreon
**Signature:** ________________________________________

# HYDROGEN PRODUCTION FROM POLYMERIC ORGANIC SOLIDS VIA ATMOSPHERIC PRESSURE NONTHERMAL PLASMA

BY

BENARD TABU

ABSTRACT OF A DISSERTATION SUBMITTED TO THE FACULTY OF THE
DEPARTMENT OF MECHANICAL ENGINEERING
IN PARTIAL FULFILLMENT OF THE REQUIREMENTS

FOR THE DEGREE OF
DOCTOR OF PHILOSOPHY
ENERGY ENGINEERING
UNIVERSITY OF MASSACHUSETTS LOWELL
2023

Dissertation Advisor: Juan Pablo Trelles, Ph.D.
Associate Professor, Department of Mechanical Engineering

# ABSTRACT

The potential of using hydrogen as a sustainable energy carrier is attributed to its high energy density and its utilization without $CO_2$ emissions. Existing technologies mainly produce hydrogen thermochemically via natural gas reforming or electrochemically through water splitting. Organic solid feedstocks rich in hydrogen, such as biomass and plastic waste, are under-utilized for this purpose. Approaches based on low-temperature atmospheric pressure plasma powered by renewable electricity could lead to the production of green hydrogen more viably than current approaches, leading to sustainable alternatives for upcycling plastic and biomass waste. This doctoral research dissertation focuses on the production of hydrogen from solids via atmospheric nonthermal plasma. First, two low-temperature atmospheric pressure plasma reactors, based on transferred arc (transarc) and gliding arc (glidarc) discharges and depicting complementary operational characteristics, are designed, built, and characterized to produce hydrogen from low-density polyethylene (LDPE) as a model plastic waste. Experimental results show that hydrogen production rate and efficiency increase monotonically with increasing voltage level in both reactors. Despite the markedly different modes of operation of the reactors, their hydrogen production performance metrics are comparable. The maximum hydrogen production efficiency and minimum energy cost are 0.16 mol/kWh and 3100 kWh/kg $H_2$, respectively, for the transarc reactor and 0.15 mol/kWh and 3300 kWh/kg $H_2$, respectively, for the glidarc reactor. Based on these findings, a Streamer Dielectric-Barrier Discharge (SDBD) reactor is devised to produce hydrogen and carbon co-products from LDPE and cellulose, the latter as a model of biomass waste feedstock. Spectroscopic and electrical diagnostics and modeling are used to estimate representative properties of the plasma, including electron and excitation temperatures, number density, and power consumption. Cellulose and LDPE are plasma-treated for different treatment times to characterize the evolution of the hydrogen production process. The maximum hydrogen production efficiency and minimum energy cost for cellulose treated by the SDBD reactor are 0.8 mol/kWh and 600 kWh/kg of $H_2$, respectively, representing approximately twice the efficiency and half the energy cost attained during the SDBD treatment of LDPE. Solid products are characterized via scanning electron microscopy, revealing the distinct morphological structure of the two feedstocks treated, as well as by elemental analysis. The results demonstrate that SDBD plasma is effective at producing hydrogen from cellulose and LDPE at atmospheric pressure conditions in relatively low temperatures, rapid response, and compact processes.

# DEDICATION

I dedicate this work to my lovely children Rwotomiya Xavier, Oyella Marion Harmony, Aber Jacinta, Adyero Julia, and Trinity, and my wife, Acan Nancy, for their tireless support, love, and prayers during this doctoral program. To my parents Akech Julia and Oloya Sarafino; my auntie Adyero Grace; and my siblings Nickson, Francis, Richard, Monica, Alfred, Paul, Gladys, and Gloria.

# ACKNOWLEDGMENT

I want to take this opportunity to thank my advisor, Prof. Juan Pablo Trelles, for his continuous fatherly mentorship and support. He created time for me whenever needed and challenged and motivated me to improve my academic and professional abilities. I am grateful to him for improving my academic ability and responsibility as my graduate experience progressed.

I also would like to thank Fulbright Foreign Student Program for partial sponsorship. The US Embassy officials in Kampala and Gulu University, thanks for all your assistance.

I would like to thank UMass Lowell for providing me with the opportunity and the facilities to pursue my research. Furthermore, my sincere thanks go to members of my dissertation committee, Dr. Maria Carreon, Dr. John Hunter Mack, and Dr. Hsi-Wu Wong, for their support. Furthermore, mentors such as Prof. Carter Keough, Prof. Thomas Walter, David, and Holly, thank you so much; you empowered me to thrive at UMass Lowell.

I also like to thank all my friends in Re-Engineered Energy Laboratory: Dassou, Valentin, Tlegen, Benard, Ephraim, Visal, Andrew, and Kevin for helping me to succeed with my research.

Finally, I would like to thank all my friends, colleagues, and families in the US, Uganda, and worldwide for your support. Henry, Rich & Maureen, Paul, Monica, Moreen, Heba, and George, long live all of you!

# LIST OF FIGURES

## CHAPTER 1:

## CHAPTER 2:

# CHAPTER 1: INTRODUCTION

## 1.1 Motivation: Green hydrogen for storage and organic waste valorization

Over the last decade, global energy consumption increased steadily by 20% to 630 quadrillion Btu in 2020, as illustrated by the US Energy Information Administration (EIA)[1] in Fig. 1a, and it is projected to increase by 30% by 2040 [2][3]. The increasing energy consumption is probably driven by the rising global population and industrialization. US EIA [1] estimates that in 2021, over 80% of current energy systems rely on fossil-based fuels, as depicted in Fig. 1b. This dependency on fossil fuels releases over 39.5 Gigatons of carbon dioxide ($CO_2$) [4], an amount that is expected to increase unless the production and utilization of alternative energy sources, such as green hydrogen, are scaled-up.

Hydrogen ($H_2$) is an energy carrier that releases energy and water when reacted with oxygen. It is naturally embedded in chemical compounds such as natural gas, biomass, plastics, and water [5]. The conventional production of $H_2$ via steam-methane reforming generates a significant amount of $CO_2$ [6]. To achieve carbon-free production of $H_2$, electrolysis is a promising option, mainly when powered by renewable electricity. However, electrolytic approaches account for only 4% of the current $H_2$ production due to their relatively high cost. Other environmentally benign methods in development to produce hydrogen include photocatalysis, photobiological water splitting, and photochemical water splitting [7]. However, the production of $H_2$ from solids, particularly biomass and plastic waste, remains largely unexplored. Geyer *et al.* [8] projected an accumulation of over 25 billion metric tons of global plastic waste by 2050. Similarly, an estimated 100 billion metric tons of biomass is generated annually globally [9]. The predominant methods of incineration and landfill for managing solid waste are deleterious to human health and the environment. For instance, plastic waste interferes with terrestrial and aquatic ecosystems leading to animal mortality.

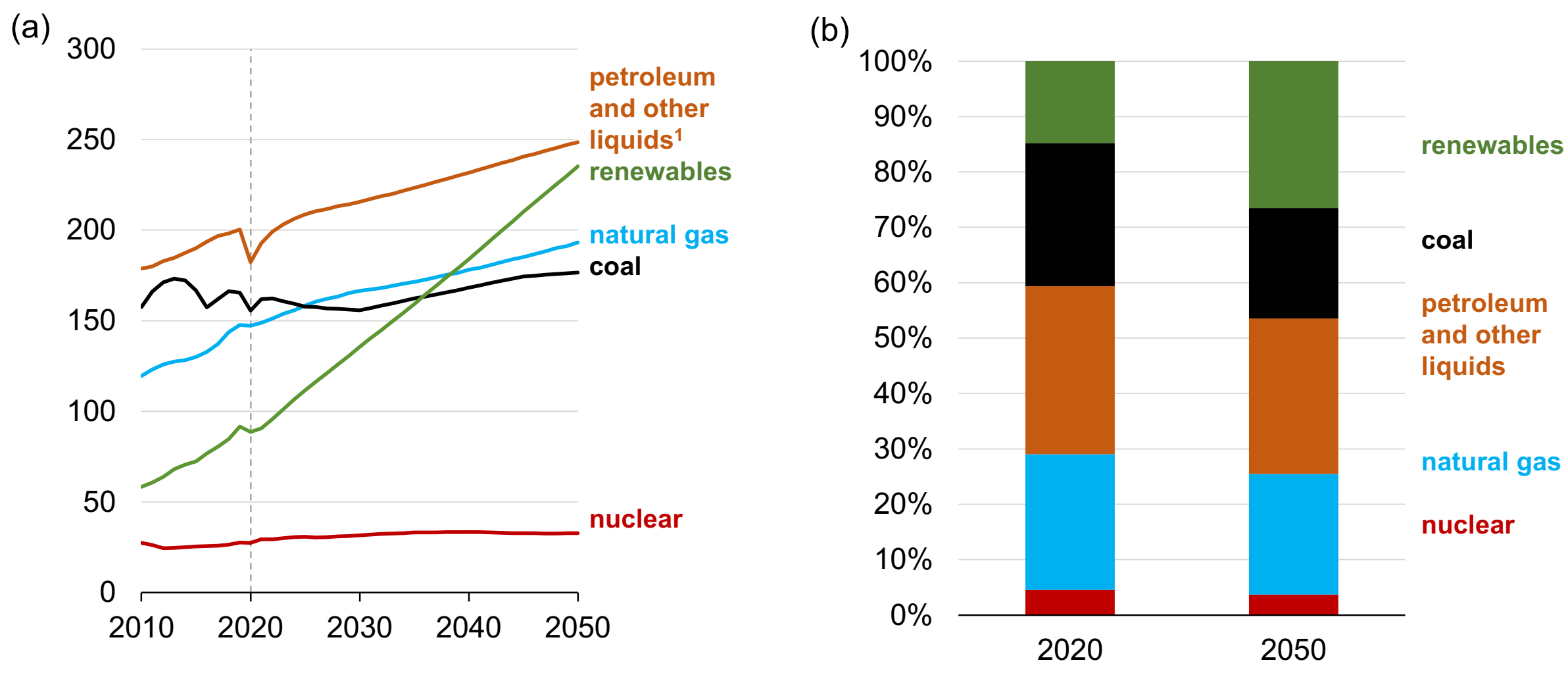

**Fig. 1. Global primary energy consumption by energy source.** (a) energy consumption in quadrillions of British thermal units (b) share of primary energy consumption by energy source [1].

Biomass and plastic waste are potential sources of green hydrogen if harnessed in environmentally benign manners. Traditional thermochemical approaches to produce $H_2$ from biomass and plastic waste, mainly pyrolysis or gasification, present relatively low energy efficiency and limited selectivity [10] and are prone to produce volatile organic compounds, $NO_x$, $CO_2$, and secondary waste. Additionally, electrochemical methods to produce hydrogen from biomass and plastic waste require replenishing electrolytes and electrodes [11], making them more expensive than thermochemical approaches [12]. Approaches that rely on low temperature and atmospheric operation, such as nonthermal plasma, can be more viable than current methods [13], [14] by depicting greater efficiency, and selectivity and being simpler or cost-effective. Furthermore, plasma-based approaches would lead to green $H_2$ production if powered by renewable electricity. This research focuses on the production of hydrogen using low-temperature atmospheric nonthermal plasma as an initial step towards $H_2$ production from both plastic and biomass via the direct use of renewable electricity.

Plasma is a partially ionized gas consisting of electrons, ions, and neutrals. Some of the atmospheric pressure plasma sources are presented in Fig. 2. They are broadly classified as thermal and nonthermal plasma sources. Thermal plasma is characterized by high electron number density and high temperature (< 10000 K) of both the electron and heavy species. For instance, arc [15] , inductively coupled plasma (ICP) [16], and microwave [17] are some examples of thermal plasma.

However, in nonthermal plasma, electrons are highly energetic and have much higher temperatures than heavy species. These electrons initiate and drive the chemical reactions leading to increased energy efficiency, compact footprint, and high selectivity. Examples include glidarc [18], transarc [18], and glow discharge [19] , as depicted in Fig 2. It should be noted that some sources, such as gliding arc and microwave, may have dual characteristics of thermal and nonthermal, depending on the regime of power operations.

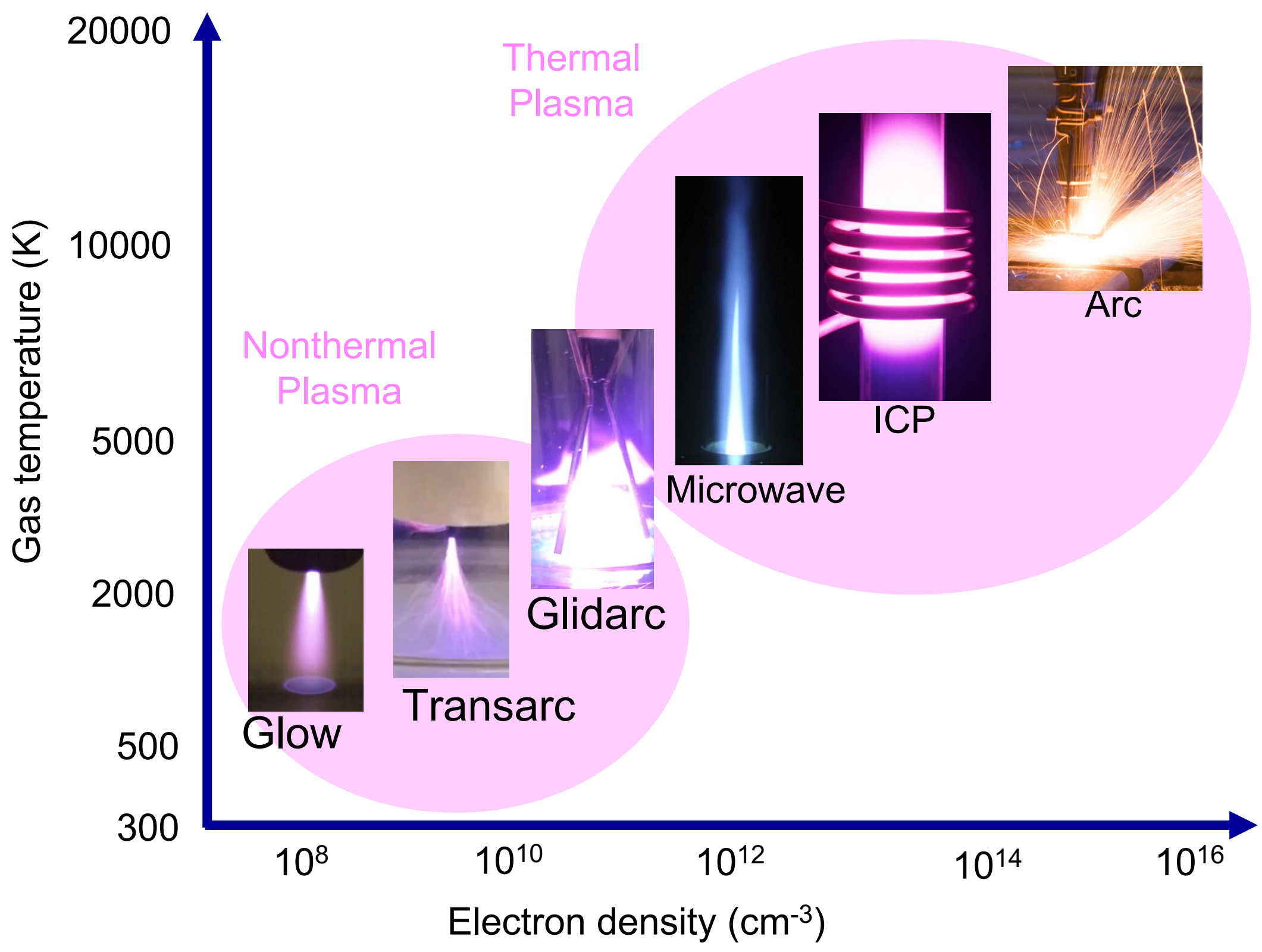

**Fig. 2. Sources of atmospheric pressure plasma.** Classification of thermal and nonthermal plasma based on electron number density and gas temperature.

## 1.2 Plasma-based hydrogen production from solids

Plasma-based approaches for hydrogen production from solids do not require oxidizing agents and have compact footprints attributed to the high reactivity of plasma fluxes [20]–[22]. The use of thermal plasma approaches for solid waste valorization has been studied extensively, leading to the construction of several plants worldwide. For example, Li *et al.* [23] cited over 27 thermal plasma plants constructed for solid waste valorization, with the majority in the United States of America and Japan. The high energy density and temperatures (6000 to 20000 K or higher) of thermal plasma processes [24] are desirable for numerous applications such as welding, plasma cutting, thermal sprays, and solid waste treatment. However, the low energy efficiency and selectivity limit the application of thermal plasma to processes such as hydrogen production from polymeric solids. In contrast to thermal plasma, nonthermal plasma is suitable for low-temperature and high-selectivity processes, as indicated by the growing interest in nonthermal atmospheric plasma for solid waste valorization [25].

Atmospheric pressure nonthermal plasma processes generally depict greater selectivity and energy efficiency than thermal plasma methods [26]–[28]. Moreover, nonthermal plasma processes operating at atmospheric pressure are highly desirable due to potentially lower capital cost and operating expenses (e.g., no need for vacuum systems) and compatibility with other unit operations (e.g., separation, condensation) [27]. Furthermore, the highly energetic electrons (typically depicting temperatures of 1 eV = 11600 K or higher) of nonthermal plasma ionize, dissociate, and excite a significant fraction of the molecules, leading to high reactivity when interacting with solid feedstock leading to the production of gaseous products. Yao and collaborators [28] studied the hydrogenolysis of polyethylene to light hydrocarbons using an atmospheric pressure nonthermal plasma reactor based on dielectric barrier discharge (DBD) over solid catalysts and using hydrogen and argon as the working gases. They obtained over 95% selectivity of lower alkanes ($C_1$-$C_3$) and lower fractions (< 5%) of unsaturated hydrocarbons. Furthermore, their results showed that introducing a catalyst (Pt/C or SAPO-34) significantly improved the energy efficiency but had minimal influence on the product formation rate. Aminu *et al*. [26] used two-stage pyrolysis/low-temperature plasma catalytic processes based on DBD to produce hydrogen and syngas (a mixture of mainly hydrogen and carbon monoxide) from plastic waste. They concluded that low-temperature plasma enhanced total gas production and hydrogen yield compared to the catalysis-only process. Moreover, syngas selectivity was greatest at 1 minute of operation, after which it declined due to the predominance of pyrolysis reactions. Diaz-Silvarrey *et al*. [29] pyrolyzed high-density polyethylene using a nitrogen DBD. They observed a significant increase in syngas production at moderate temperatures, i.e., from 15 wt% to 44 wt% at 600 $^{o}$C. Ahmed *et al*. [30] critically reviewed plasma-based approaches for the decomposition of hydrocarbons and suggested using nonthermal plasma as an alternative to conventional catalytic decomposition methods. Although promising results have been reported in the literature, the potential of nonthermal plasma for plastic waste and biomass valorization is largely unexplored.

This research focuses on hydrogen production from low-density polyethylene (LDPE) and cellulose using nonthermal atmospheric pressure plasma. Cellulose and LDPE are used as models of plastic and biomass feedstock, respectively. LDPE comprises long hydrocarbon chains with short branches, usually between 0.5 and 1 million carbon units [35]. It is widely used for packaging, thin-film coatings, pipes, and cables production and is the main contributor to plastic waste [35], [36]. Cellulose is a linear chain of repeated anhydroglucose rings $(C_6H_{10}O_5)_n$, usually between 10000 to 15000 long, depending on the cellulose source material. The anhydroglucose units are bonded covalently by 1,4' glycosidic links, which provide mechanical stiffness [37]. Cellulose is considered the most common organic compound on earth [38], naturally embedded in wood, hemp, cotton, crop residues, and linen [36], [39]. This research is envisioned as an initial step towards producing hydrogen from plastic and biomass via the direct use of renewable electricity.

## 1.3 Goal and Objectives

The goal of this doctoral research is to *experimentally study the use of atmospheric pressure nonthermal plasma to produce hydrogen from polymeric organic solids, particularly low-density polyethylene as a plastic waste model and cellulose as a representative biomass feedstock.*

To achieve this research goal, the following objectives are performed:

1. Design and characterize nonthermal plasma reactors to produce hydrogen from organic polymeric solids. First, two nonthermal plasma reactors, namely transferred arc (transarc) and gliding arc (glidarc) are designed, built, and characterized to produce hydrogen from LDPE cellulose. The thermal model of SolidWorks flow simulation is used to assess the performance of the reactor designs. The results show that the reactors can be operated at near room temperature despite the high temperature in the reactor chamber. Second, a streamer dielectric barrier discharge (SDBD) based on pin-to-plate is designed and built to produce hydrogen and carbon co-products from LDPE and Cellulose.
2. Assess the performance of the reactors to produce hydrogen from polyethylene and cellulose. The performance of transarc and glidarc in the production of hydrogen from LDPE via low-temperature atmospheric pressure plasma is evaluated in terms of hydrogen production and hydrogen production efficiency. The results show that the maximum hydrogen production efficiency and minimum energy cost are 0.16 mol/kWh and 3100 kWh/kg $H_2$, respectively, for the transarc reactor and 0.15 mol/kWh and 3300 kWh/kg $H_2$, respectively, for the glidarc reactor. Furthermore, the maximum hydrogen production efficiency and minimum energy cost for cellulose treated by the SDBD reactor are 0.8 mol/kWh and 600 kWh/kg of $H_2$, respectively, representing approximately twice the efficiency and half the energy cost attained during the SDBD treatment of LDPE.

3. Evaluate the performance of the reactors to produce carbon co-products. The solid samples before and after plasma treatment are evaluated using optical imaging, field emission scanning electron microscopy, and CHN analysis. The pristine solid sample consists of entanglement of cellulose fibers that are long and well-intact with empty spaces leading to high porosity and weak dielectric strength. However, the plasma-treated cellulose has fragmented fibers which consist of protruded fibrils of diameter 50 nm, which are loose and visible leading to weak structural strength. On the other hand, pristine LDPE is highly dense and nonporous, contributing to its strong dielectric strength and, subsequently, greater power consumption. In contrast, the plasma-treated LDPE has shallow dimples with micro-grains well embedded in the sample.

## 1.4 Summary of the thesis contents.

The thesis follows the multi-monograph format, with, Chapter 1 explaining the background and motivation of the research. The other chapters, except Chapter 4, are presented as published journal publications or manuscripts already submitted for peer review. Each publication is part of the work that forms the dissertation.

Chapter 2 presents the hydrogen production from LDPE via atmospheric pressure nonthermal plasma. Two novel reactors, namely, transarc and glidarc, are designed, built, and characterized to produce hydrogen from LDPE. SolidWorks flow simulation based on a thermal model is used to assess the performance of reactors design. The results show that the hydrogen production from LDPE is comparable despite the markedly different modes of operation between the two reactors. This work is summarized in the following published article:

Tabu, B., Akers, K., Yu, P., Baghirzade, M., Brack, E., Drew, C., Mack, J.H., Wong, H.W. and Trelles, J.P., 2022. Nonthermal atmospheric plasma reactors for hydrogen production from low-density polyethylene. *International Journal of Hydrogen Energy*, *47*(94), pp.39743-39757.

In Chapter 3, hydrogen production from cellulose and low-density polyethylene via atmospheric nonthermal plasma is experimentally evaluated. A streamer dielectric barrier discharge (SDBD) is designed, built, and experimentally used for extracting hydrogen. The electrical model is developed to determine the actual plasma power consumed. Spectroscopic diagnostics are used to determine excitation temperature, electron temperature, and electron number density. The results show that the hydrogen production of cellulose doubled that of LDPE despite comparable power consumed. The effort is summarized in the submitted manuscript:

**Tabu, B.**, Veng, V., Morgan, H., Das, S.K., Brack, E., Alexander, T., Mack, J.H., Wong, H.W., and Trelles, J.P., 2023. Hydrogen from cellulose and low-density polyethylene via atmospheric pressure nonthermal plasma. *International Journal of Hydrogen Energy*.

Chapter 4 presents the summary, conclusions, and recommendations for further work.

# CHAPTER 2: NONTHERMAL ATMOSPHERIC PLASMA REACTORS FOR HYDROGEN PRODUCTION FROM LOW-DENSITY POLYETHYLENE

## Abstract

Hydrogen is largely produced via natural gas reforming or electrochemical water-splitting, leaving organic solid feedstocks under-utilized. Plasma technology powered by renewable electricity can lead to the sustainable upcycling of plastic waste and production of green hydrogen. In this work, low-temperature atmospheric pressure plasma reactors based on transferred arc (transarc) and gliding arc (glidarc) discharges are designed, built, and characterized to produce hydrogen from low-density polyethylene (LDPE) as a model plastic waste. Experimental results show that hydrogen production rate and efficiency increase monotonically with increasing voltage level in both reactors, with the maximum hydrogen production of 0.33 and 0.42 mmol/g LDPE for transarc and glidarc reactors, respectively. For the transarc reactor, smaller electrode-feedstock spacing favors greater hydrogen production, whereas, for the glidarc reactor, greater hydrogen production is obtained at intermediate flow rates. The hydrogen production from LDPE is comparable despite the markedly different modes of operation between the two reactors.

## 2.1 Introduction

The global energy demand is projected to increase by 56% by 2040, driven by population growth and industrialization, particularly in developing countries [1, 2]. Since fossil fuels are responsible for 80% of global energy demands [3], greenhouse gas emissions and their adverse impacts are also expected to increase unless the production and use of alternative energy sources, such as green hydrogen, are scaled-up [4]. Hydrogen not only has the highest energy density (120 MJ/kg) of all fuels [5], but it also does not produce $CO_2$, the main greenhouse gas, when reacted with oxygen [5–7]. Furthermore, hydrogen is one of the most abundant elements in the earth's crust [9]. However, hydrogen does not occur naturally; instead, it is embedded in water, hydrocarbons, and solid organic compounds such as biomass and plastics [7–9].

The increasing global production of plastics, which surpassed 360 million tons in 2018 [8, 9], has led to a dramatic increase in plastic waste, polluting the environment and interfering with ecosystems [11, 12]. Incineration, the dominant approach to deal with plastic waste, leads to $CO_2$ emissions and is prone to emit volatile organic compounds deleterious to human health [17]. Strategies to valorize plastic waste, such as recycling and particularly its utilization as a source of hydrogen, could have a primary role in dealing with the disposal of plastic waste.

Traditional routes to produce hydrogen from plastic waste are mainly divided between thermochemical and electrochemical methods. In thermochemical approaches, heat is supplied to plastic waste to attain high temperatures (typically -3000 $^0$C) that promote desired chemical conversion reactions [18]. This can either be done in the absence of oxygen via pyrolysis [16–18] or in the presence of a controlled amount of oxygen through gasification [13, 17]. In electrochemical methods, plastic waste is converted directly or indirectly by reduction-oxidation reactions within electrochemical cells [18, 19]. Since both the electrolytes and electrodes require replenishing [23], electrochemical methods are generally more expensive than thermochemical approaches [24]. Even though thermochemical processes are widely used in plastic waste treatment, these processes typically depict low rates of hydrogen production, limited selectivity [24, 25], and low energy efficiency due to energy spent in auxiliary functions, such as cooling of gas products. Methods for plastic waste treatment based on low temperature and atmospheric pressure operation, such as nonthermal (low-temperature) plasma processes, have the potential to be more viable than current approaches [26–28]. Moreover, if powered by renewable electricity (e.g., wind or solar photovoltaic power), plasma-based techniques would mitigate $CO_2$ emissions associated with plastic waste treatment.

Plasma, i.e., partially ionized gas constituted of free electrons and heavy species (ions, atoms, and molecules), generated at (near) atmospheric pressure conditions is broadly classified as either thermal or nonthermal [30]. In thermal plasma, electrons and heavy species are in thermal equilibrium and therefore depict the same temperature, usually ranging from 6 000 to over 20 000 K [31]. In contrast, in nonthermal plasma, the temperature of free electrons is high (1 eV ~ 11600 K or higher) compared to the heavy species temperature (e.g., a few hundred Celsius), resulting in a state of nonthermal equilibrium [25, 26].

Plasma-based approaches for plastic waste treatment generally do not require oxidizing agents, given the high reactivity promoted by plasma species. Moreover, atmospheric pressure plasma processes often have compact footprints thanks to the high fluxes of reactive species [27, 28]. The application of thermal plasma to plastic waste treatment has been studied to a significant extent, even leading to the construction of pilot plants [23, 36]. The high energy density and high temperature of thermal plasma processes are desirable for applications such as thermal sprays, welding, plasma cutting, and solid waste treatment. However, for processes that require selective treatment of reactants with relatively low melting points, such as hydrogen production from plastics, high-temperature operations may be undesirable as they may lead to limited energy efficiency or complex installations [37].

Approaches based on nonthermal plasma potentially have greater energy efficiency and selectivity than thermal plasma processes [30–32]. Furthermore, nonthermal plasma processes operating at atmospheric pressure are highly desirable due to potentially lower capital and operating expenses (e.g., no need for vacuum systems) and compatibility with other unit operations [40]. Yao and collaborators [41] studied the hydrogenolysis of polyethylene to light hydrocarbons using an atmospheric pressure nonthermal plasma reactor based on dielectric barrier discharge (DBD) over solid catalysts and using hydrogen and argon as the working gases. They obtained over 95% selectivity of lower alkanes ($C_1$-$C_3$) and low fractions (< 5%) of unsaturated hydrocarbons. Furthermore, their results showed that introducing a catalyst (Pt/C or SAPO-34) significantly improved the energy efficiency but had minimal influence on the product formation rate. Aminu *et al*. [39] used a two-stage pyrolysis/low-temperature plasma catalytic process, also based on DBD, to produce hydrogen and syngas (a mixture of mainly hydrogen and carbon monoxide) from plastic waste. They concluded that low-temperature plasma enhanced the total gas production and hydrogen yield compared to the catalysis-only process. Also, syngas selectivity was greatest at 1 minute of operation, after which it declined due to the predominance of pyrolysis reactions. Diaz-Silvarrey *et al*. [15] pyrolyzed high-density polyethylene using a nitrogen DBD. They observed a significant increase in syngas production at moderate temperatures, i.e., from 15 wt% to 44 wt% at 600 °C. Xiao and collaborators [26] recovered hydrogen and aromatics from polypropylene waste via plasma-catalytic pyrolysis and noted an increment in the gas products of 18 wt% with 4.19 mmol/g $H_2$ formed. Ahmed *et al*. [42] critically reviewed plasma-based approaches for decomposing hydrocarbons and suggested using nonthermal plasma as an alternative to conventional catalytic decomposition methods. Although promising results have been reported in the literature, the potential of nonthermal plasma for plastic waste valorization is largely unexplored.

This article focuses on hydrogen production from low-density polyethylene (LDPE) using nonthermal atmospheric plasma. In addition to hydrogen, other co-products such as methane, ethylene, ethyne, propane, and larger molecular hydrocarbons have been reported from the processing of similar organic polymeric feedstock, such as high-density polyethylene (HDPE) [15], polyethylene [28], and polypropylene [43]. The present study focuses on the design and characterization of the plasma reactors to produce hydrogen from LDPE, and therefore hydrogen is treated as the main product. LDPE comprises long hydrocarbon chains with short branches, usually between 0.5 and 1 million carbon units [44]. It is widely used for packaging, thin-film coatings, pipes, and cable production and is a primary component of global plastic waste [44]. This research is envisioned as an initial step toward valorizing plastic waste via the direct use of renewable electricity at atmospheric pressure and low-temperature conditions and with minimal auxiliary reactants. Section 2 presents the design of two nonthermal atmospheric pressure plasma reactors to produce hydrogen from LDPE. The experimental characterization of the reactors, encompassing electrical, fluid flow, and chemical diagnostics, is shown in section 3. Section 4 discusses the performance of the two reactors in terms of hydrogen production rate, production efficiency, and their correlation with operational parameters. Concluding remarks are presented in section 5.

## 2.2 Nonthermal plasma reactors

### 2.2.1 Reactors design

Two reactors are designed, built, and characterized for hydrogen production from atmospheric nonthermal plasma. The reactors are based on transferred arc (transarc) and gliding arc (glidarc) electrical discharges, and present complementary operational characteristics. Schematics of the reactors are presented in Fig. 1.

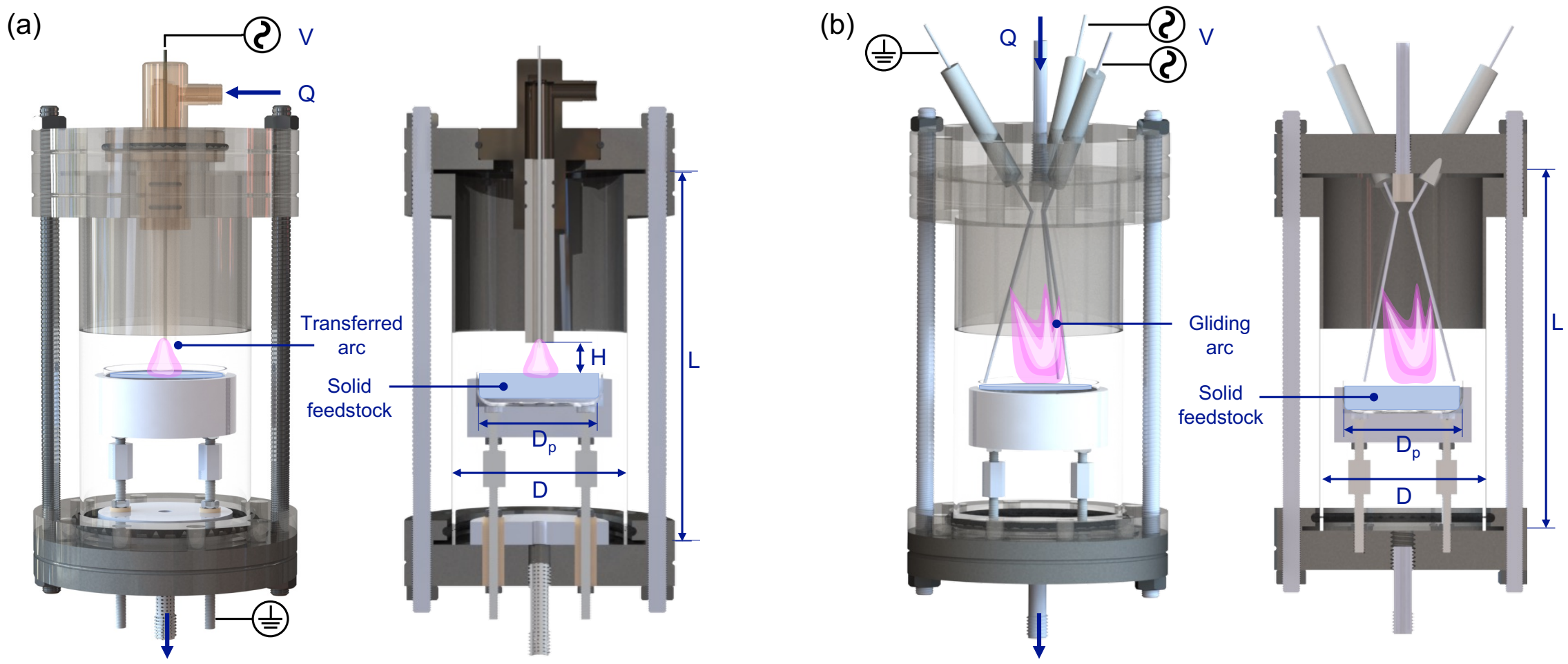

**Fig. 1. Plasma reactors for hydrogen production from polyethylene.** Assembled design and cross-section view of the (a) transferred arc (transarc) reactor and the (b) gliding arc (glidarc) reactor. The reactors' main operating parameters are the electrode-feedstock spacing *H*, flow rate, *Q*, and voltage level *V*.

The transarc reactor (Fig. 1a) has a pin-to-plate configuration with a powered tungsten electrode placed perpendicularly above an aluminum disc, which acts as the ground electrode and support of the crucible holding the solid feedstock (LDPE sample). The name *transferred arc* stems from the electric current being *transferred* from the powered electrode to the feedstock. Thus, the feedstock is electrically coupled to the plasma. The distance between the tip of the powered electrode and the upper surface of the feedstock, denoted as $H$, is used as a control parameter. The gas nozzle is made of high-temperature resin fitted with a ceramic (alumina) bushing.

The glidarc reactor (Fig. 1b) consists of tri-prong equally spaced tungsten electrodes diverging 135$^{o}$ with a gliding length of 30 mm. This electrode configuration generates a Y-shaped arc at the minimum inter-electrode separation distance. Two electrodes are powered by a separate power supply and the third electrode is set as ground. The name *gliding arc* stems from the fact that the generated arc *glides* along the electrodes due to the combined effects of advection of the gas inflow and the buoyancy of the low-density plasma. The minimum inter-electrode separation is 6 mm, as used in the glidarc reactor by Dassou *et al*. [45]. The glidarc plasma is electrically decoupled from the feedstock, making it suitable for treating a continuous stream of feedstock and surfaces.

Both reactors are powered by high voltage alternating current (AC) power supplies, delivering up to 300 W of output power with an independent frequency control from 20 to 70 kHz. The power supplies are voltage-controlled by setting the voltage level ($V$) from 0 to 100%, leading to a maximum voltage output (for zero load) from 1 to 40 kV. Nitrogen is used as a processing gas, injected with a flow rate ($Q$). The power supply voltage level $V$ and flow rate $Q$ are control parameters for both reactors. The reactor chambers have a diameter $D$ = 76.2 mm and height $L$ = 150 mm and a quartz section to allow optical access. The residence time $t_{res}$ of the gas is therefore given by:

$$t_{res} = \frac{\pi D^2 L}{4Q}. \quad (1)$$

The solid LDPE samples have a fixed mass of 10 g and are placed inside a quartz plate 55 mm in diameter and 15 mm in height. The quartz plate with the feedstock is fitted in a cylindrical aluminum holder. The holder acts as the ground electrode for the transarc reactor, but it is electrically de-coupled in the glidarc reactor.

Computational Fluid Dynamics (CFD) thermal-fluid models created in SolidWorks Flow Simulation [46] are used to evaluate the effect of control parameters on the operation of the reactors. The models describe the plasma as a volumetric heat source approximated as a solid with 100% porosity (i.e., no inertial resistance to fluid transport) in chemical equilibrium (i.e., species composition and material properties are a function of the local temperature only). For the transarc reactor, the plasma is approximated as a rectangular cylinder of 1.6 mm diameter connecting the tip of the powered electrode to the feedstock. Whereas for the glidarc reactor, the plasma volume is approximated as a truncated pyramid with a triangular cross-section 50 mm long, approximately filling the inter-electrode space. Convective heat transfer boundary conditions, specified with an outside temperature of 300 K and a convective heat transfer coefficient of 25 $W/m^2K$, are imposed over all the outer surfaces of the reactors. Given the chemical equilibrium assumption, no chemical kinetics associated with the plasma or the interaction between the plasma and the feedstock are explicitly included in the models. Instead, the thermal-fluid models describe fluid flow and thermal characteristics throughout the reactors (reactor chamber, solid feedstock, and auxiliary components). Given the nonthermal nature of the generated plasma in the reactors, only a portion of the consumed power is dissipated as heat. The amount of thermal power (dissipated heat) is an input to the models. Therefore, the models describe the operation of the reactors as a function of the control parameters inflow rate $Q$ and thermal power dissipated by the plasma (assumed correlated with $V$). Representative results of the thermal-fluid models are presented in Fig. 2.

Fig. 2 shows the geometry of the computational thermal-fluid models, as well as velocity and temperature distributions for representative operating conditions, namely dissipated thermal power of 1.75 W (i.e., 5% of 35 W, a representative value of power consumed by the plasma), nitrogen flow rate $Q$ = 0.1 slpm, and electrode-feedstock spacing $H$ = 5 mm for the transarc; and input power of 1.4 W (i.e., 5% of 28 W), nitrogen flow rate $Q$ = 2 slpm, and electrode-feedstock spacing $H$ = 5 mm for the transarc and glidarc. The transarc temperature distribution (Fig. 2b) is highest at the center of the plasma volume and decreases uniformly with increasing radial distance. The highest temperature on the feedstock surface is ~ 750 K. The relatively high temperatures in the transarc simulations are attributed to the relatively small plasma volume, which leads to increased thermal power per unit volume. The temperature distribution for the glidarc reactor presents a three-fold symmetry, which suggests non-uniform heating of the feedstock (Fig. 2e), with the highest temperature over the feedstock close to 300 K. The simulation predicts a significantly greater area of plasma interaction with the feedstock's surface as compared to the transarc reactor. This observation is complemented by the isosurface temperature distributions shown in Fig. 2a and 2d.

.

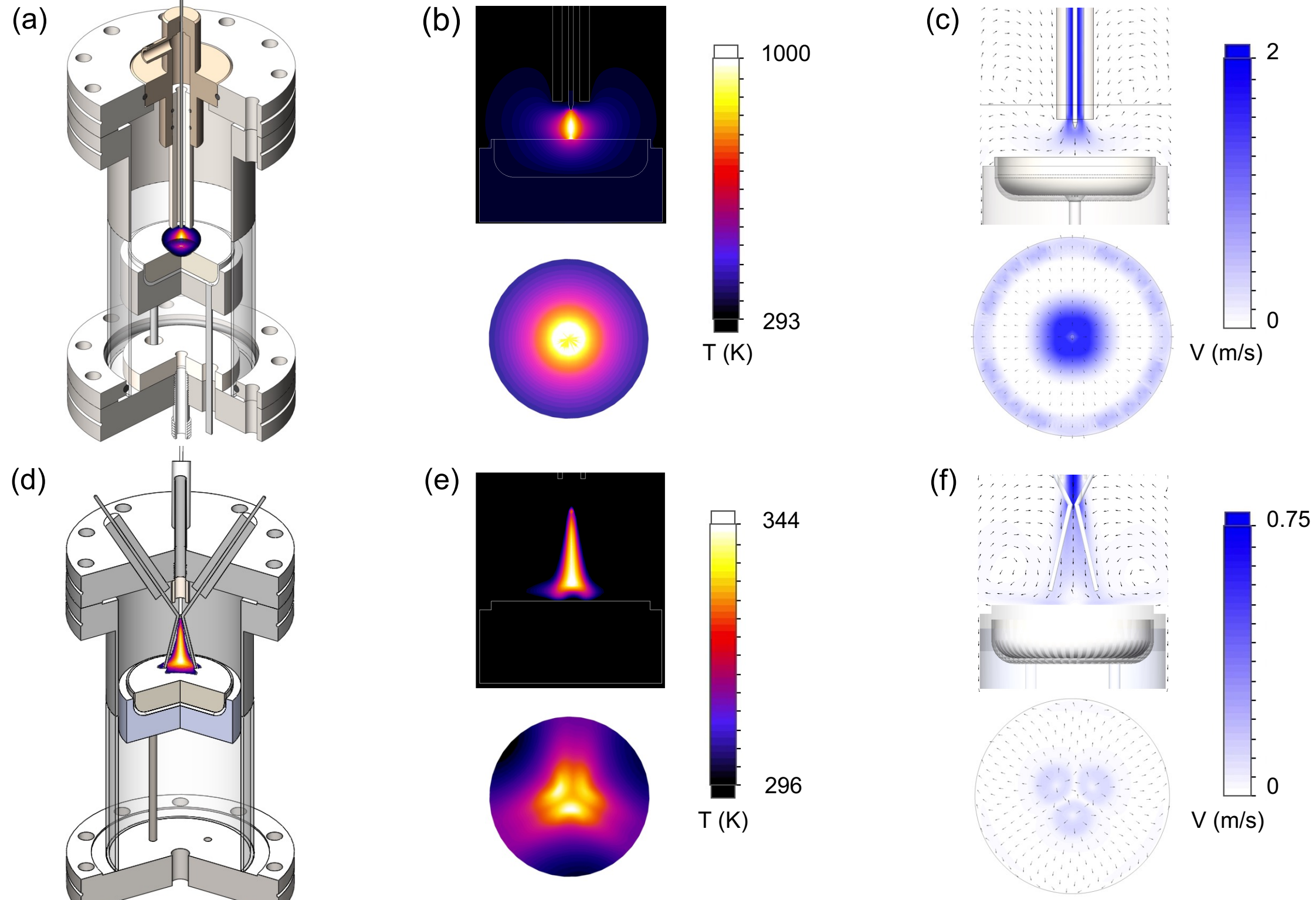

**Fig. 2. Computational thermal-fluid reactor models**. Transarc reactor: (a) design schematic, (b) temperature distribution, and (c) velocity distribution for 1.75 W of heat dissipation, $Q$ = 0.1 slpm, and $H$ = 5 mm. Glidarc reactor: (d) design schematic, (e) temperature distribution, and (f) velocity for 1.4 W of heat dissipation, $Q$ = 2 slpm, and $H$ = 5 mm.

Despite the high temperatures in the plasma volume, particularly for the transarc, the temperature near the reactors' walls is close to the ambient temperature of 300 K, irrespective of the amount of imposed thermal power. This suggests that the reactors can operate at or near room temperature without forced cooling. The flow fields in Fig. 2c and 2f show that the axial gas inflow leads to the formation of vortex rings near the sample's surface in both reactors. These vorticial structures are characterized by relatively long residence times and may lead to the recombination of gas products emanating from the feedstock. Moreover, the higher velocity at the center of the transarc reactor indicates the potential formation of a crater-like pattern at the center of the feedstock. In contrast, the low-velocity magnitude of the three-fold way over the feedstock surface observed in the glidarc simulations suggests a more uniform treatment. These model predictions are contrasted against experimental observations in section 3.4 and section 4.1, respectively.

### 2.2.2 Operational characteristics

The expected operational characteristics of the reactors obtained with the thermal-fluid models as a function of dissipated thermal power and flow rate are shown in Fig. 3. For the transarc reactor, the average surface temperature increases linearly with dissipated thermal power per unit volume (Fig. 3a) and has minimal dependence on flow rate. The slight difference in the average surface temperature for the flow rate of 2 and 4 slpm at $0.8\times10^3$ W/cm$^3$ is ascribed to the computational error of the simulation. The glidarc reactor's average surface temperature varies directly with dissipated thermal power, but inversely with flow rate, as shown in Fig. 3b. A higher flow rate leads to enhanced convective cooling, which reduces the amount of heat deposited on the substrate. The average heat flux over the feedstock for both reactors (Fig. 3c and Fig. 3d) follows the same trends as the average surface temperature. These simulation results suggest that the transarc reactor can operate with small flow rates compared to those needed for the glidarc reactor, whose operation is very sensitive to flow rate. Based on these results, the experimental characterization of the reactors uses voltage level $V$ (assumed proportional to thermal power dissipation) for both reactors, electrode-feedstock spacing $H$ for the transarc reactor, and flow rate $Q$ for the glidarc reactor, as main operational parameters.

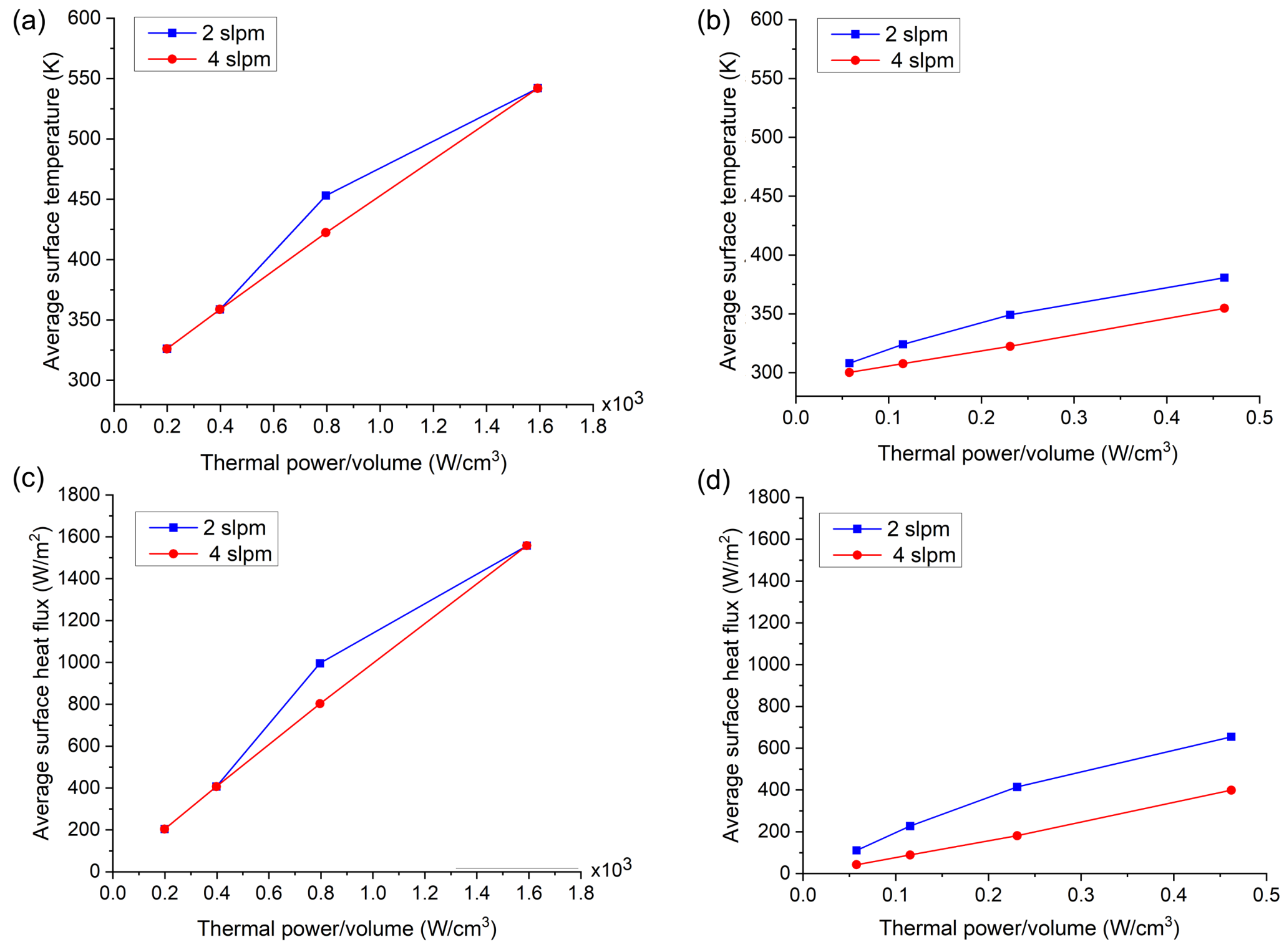

**Fig. 3. Operational characteristics predicted by thermal-fluid models.** Average surface temperature for the (a) transarc and (b) glidarc reactors and average surface heat flux for the (c) transarc, and (d) glidarc reactors for varying thermal power density (proportional to $V$) and flow rate $Q$.

## 2.3 Characterization of reactors

### 2.3.1 Experimental set-up

The experimental set-ups are depicted in Fig. 4, one for characterizing the operation of the reactors (Fig. 4a) and the other for Schlieren imaging (Fig. 4b). The reactors are powered by high voltage AC power supplies (PVM500-2500 Plasma Power Generator) with peak-to-peak voltage from 1 to 40 kV, and 25 mA peak current. As indicated in section 2, the transarc reactor operates with a single power supply, whereas the glidarc reactor utilizes two power supplies. Two Alicat mass flow controllers regulate the gas flow rate through the reactor; one is used for the nitrogen inflow and the other for the gas products outflow. A Tektronix Oscilloscope (TBS 2104) equipped with a current probe (P6021A) and high voltage probe (P6015A) are used to measure the electrical characteristics of the reactors' operation. The gas products are analyzed by a Shimadzu GC-2014 Gas Chromatography (GC) system.

The Schlieren imaging set-up allows the visualization of refractive index variations, which depict density gradients in the test medium. The set-up consists of collimator and de-collimator lenses (with focal lengths of 30 and 50 cm, respectively) aligned with a light fiber-optic and halogen source (250 W), the test medium (center of the plasma region within the reactor chamber), a knife-edge, and a high-speed camera (Edgertronic SC2+). The knife-edge adjusts the system's sensitivity while the high-speed camera captures the density gradient variation of the test medium. The test medium comprises the plasma interacting with either feedstock (LDPE) or an inert (quartz disc) sample within cross-shaped reactor chambers with flat quartz windows (to prevent optical distortions by the curvature of cylindrical quartz chambers). The high-speed camera is configured with shutter speed and frame rate of 1/8500 s and 8000 fps, respectively, to visualize the transarc plasma. To visualize the glidarc plasma, due to its dynamic nature with a gliding period in the order of milliseconds, a lower frame rate of 500 fps is used. The optical visualization of the operation of both reactors uses a camera adjusted to 1080 p resolution and 30 fps.

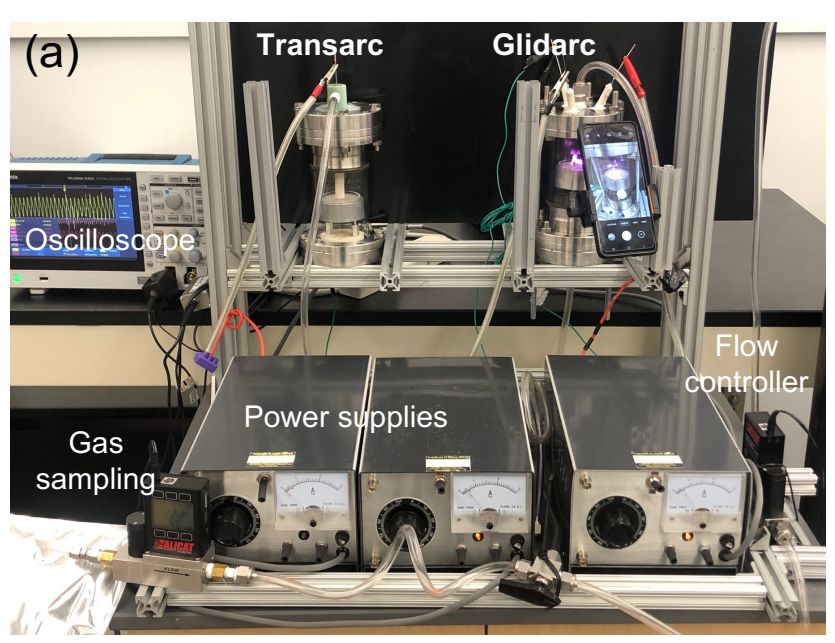

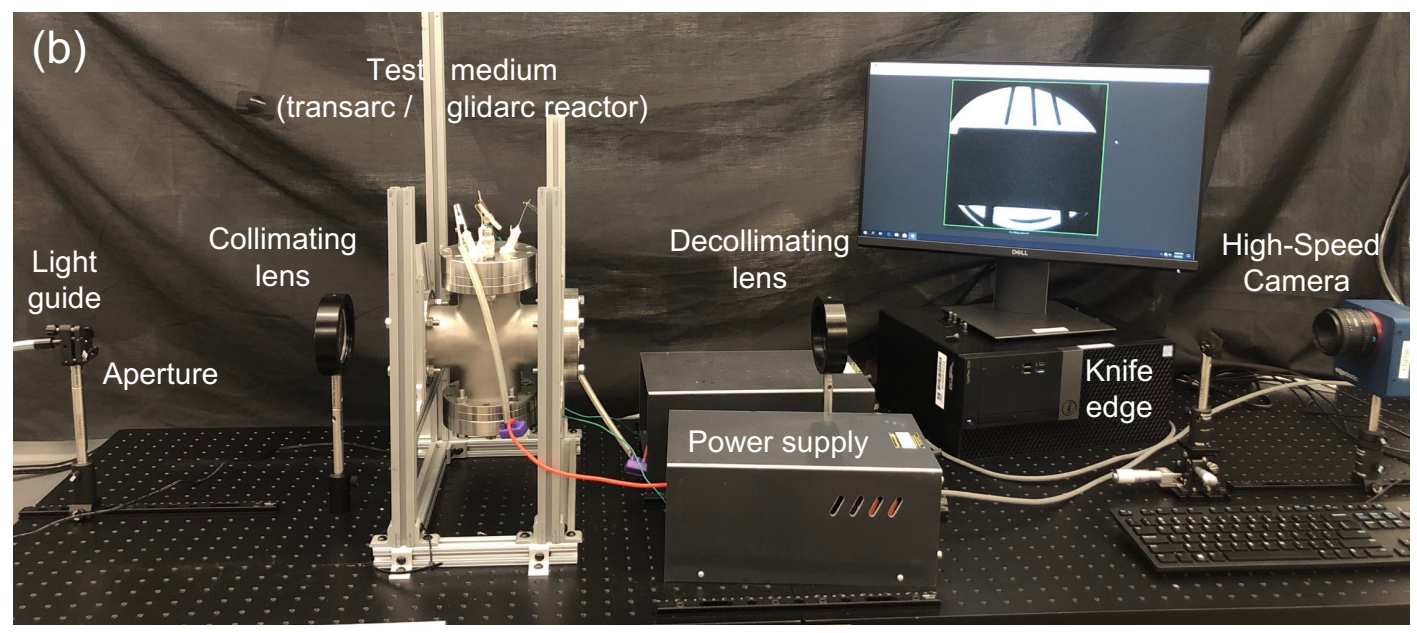

**Fig. 4. Experimental set-ups.** (a) characterization of the operation of the reactors during hydrogen production and for (b) optical and Schlieren visualization.

### 2.3.2 Optical imaging

Optical characterization of the operation of the transarc reactor is conducted for varying voltage level $V$ and flow rate $Q$, as depicted in Fig. 5. The electrode-feedstock spacing $H$ is kept fixed at 10 mm. The minimum and the maximum $V$ are first determined for each $Q$. Given that the computational characterization of the transarc reactor showed a limited effect on the flow rate (Fig. 3), three relatively small flow rates, i.e., 0.1, 0.5, and 1.0 slpm, are used. These flow rates correspond to residence times $t_{res}$ of approximately 410, 82, and 41 s, respectively. The minimum and maximum voltage levels are set equal to 6 and 30%, respectively, for all flow rates. The minimum voltage level at a given $Q$ leads to faintly visible plasma (i.e., corona discharge). The intensity of the discharge increases with voltage level leading to the transition from corona discharge ($V$ = 6% to 10%) to glow ($V$ = 10% to 20%), and then to arc/streamer discharge ($V$ > 20%). Discernably, a flow rate of 1 slpm produces a less-intense discharge with a slightly larger divergence of the plasma column than the discharges at 0.1 and 0.5 slpm. The intensity and divergence of the discharge are identified as critical parameters for hydrogen production. Hence, based on the optical characterization results, a fixed value of $Q$ = 0.1 slpm and $V$ = 20%, and 30% are chosen for the hydrogen production experiments. To expand the range of characterization of the transarc reactor operation, given its minor sensitivity to $Q$, the electrode-feedstock spacing $H$ is set to either 5 or 10 mm in the hydrogen production experiments.

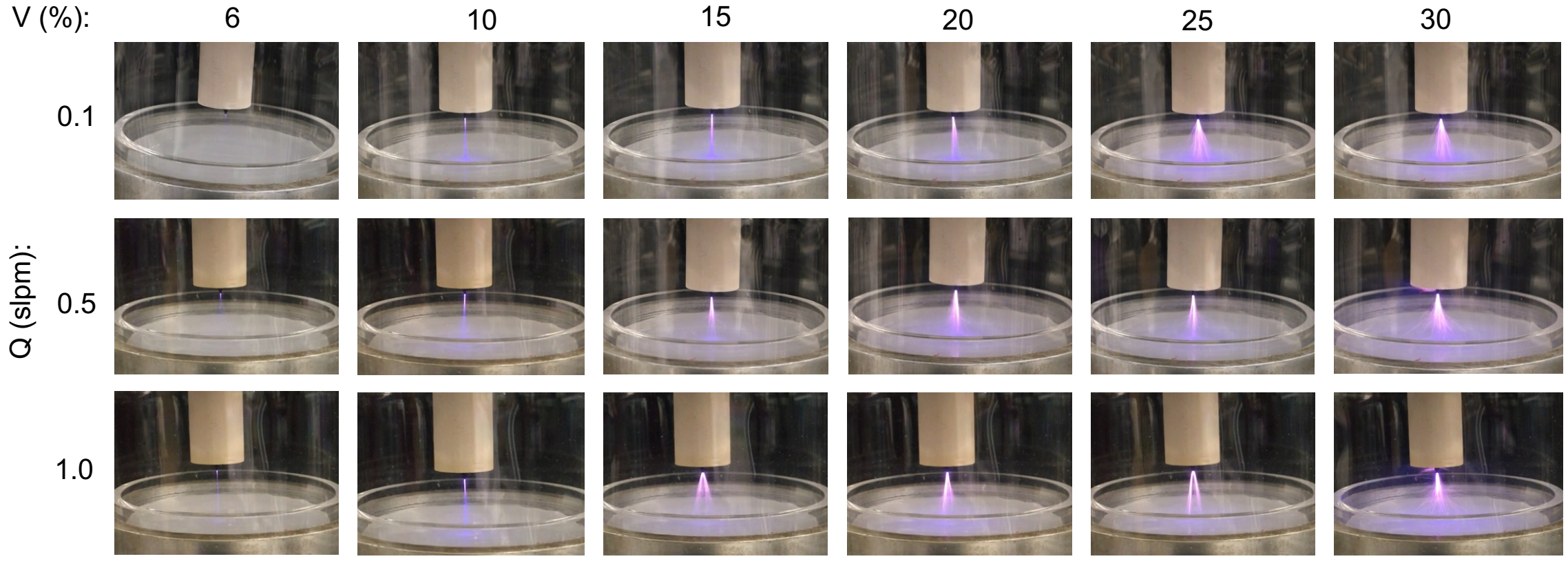

**Fig. 5. Characterization of the transarc reactor.** Optical imaging for varying flow rate Q and voltage level $V$ for $H$ = 10 mm, depicting a streamer discharge in all the operating conditions.

In contrast to the transarc reactor, the glidarc reactor requires higher flow rates to establish appropriate interaction between the plasma and feedstock. Therefore, larger $Q$ values, i.e., 2, 3, 4, and 6 slpm, are used in the experimental characterization. These flow rates correspond to residence times $t_{res}$ of 20.5, 13.7, 10.3, and 6.8 s, respectively. The lowest and highest $V$ is 40% and 75% across all flow rates, respectively. The results of the characterization of the operation of the glidarc reactor by optical imaging are shown in Fig. 6. The intensity of the tri-prong arc proportionally increases with voltage level for all investigated flow rates. The plasma does not interact with the feedstock for the highest flow rate of $Q$ = 6 slpm and $V \geq$ 60% or flow rates $Q$ < 2 slpm at any $V$. This behavior is a characteristic of gliding arc discharges, whose dynamics depend on the balance between buoyancy and advective forces (due to the low density of the plasma and due to the drag by the gas flow, respectively). For $Q$ < 6 slpm, the impingement of the plasma on the feedstock is more pronounced for $V \geq 60\%$, suggesting that greater $V$ would favor greater hydrogen production. Therefore, for the hydrogen production tests, the glidarc reactor is operated at higher voltage levels of $V$ = 65% and 75%, and flow rates $Q$ = 2 and 4 slpm.

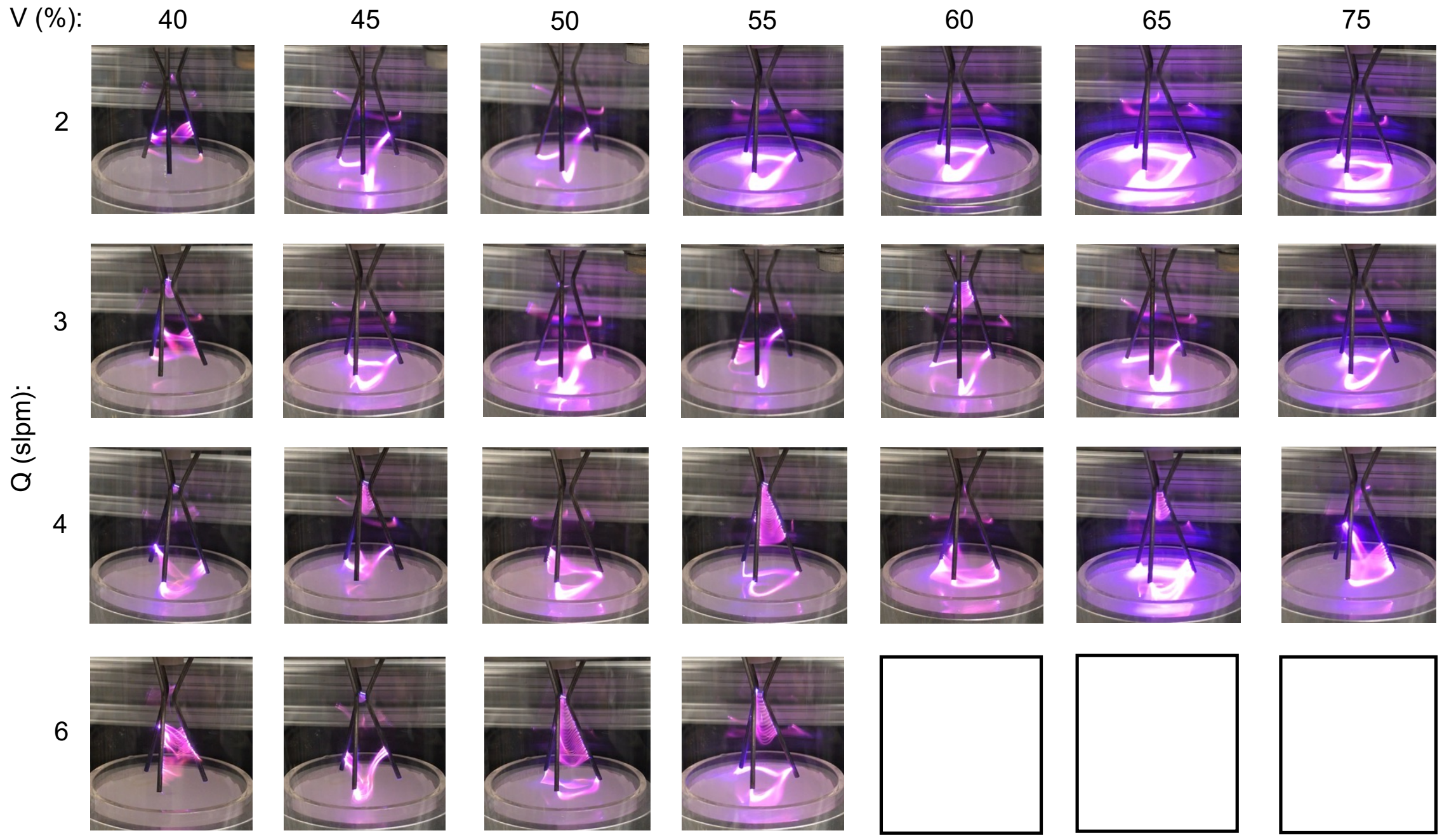

**Fig. 6. Characterization of the glidarc reactor.** Optical visualization for varying flow rate $Q$ and voltage level $V$. (No plasma is generated for $Q$ = 6 slpm and $V$ = 60% and higher).

### 2.3.3 Plasma-feedstock interaction

The solid feedstock used in the hydrogen production experiments consists of commercial LDPE pellets (average diameter of 3 mm) from Millipore Sigma (Sigma Aldrich, 428043). LDPE pellets totaling 10 g are melted at 180 $^0$C in the quartz plate ($D_p$

= 55 mm diameter, see Fig. 2) using an electrical heater (Fisherbrand, 100-120 V) and then allowed to solidify to yield a solid LDPE sample of approximately 6 mm thickness.

The hydrogen production experiments consist of treating the solid LDPE sample with nitrogen plasma in either the transarc or glidarc reactor for 30 minutes. Gas product samples are extracted at 5-minute intervals throughout the experiments. The experiment for each set of operating conditions (i.e., $V$ and $H$ for the transarc and $V$ and $Q$ for the glidarc) is repeated three times. The variation in results is quantified by the error bars (i.e., standard error of the mean). In the hydrogen production experiments, the transarc reactor is operated under a low flow rate of 0.1 slpm while varying voltage level ($V$ = 20% or 30%) and electrode-feedstock spacing ($H$ = 5 or 10 mm). For the glidarc reactor, a fixed electrode-feedstock spacing of 5 mm is used with varying flow rate ($Q$ = 2 or 4 slpm) and voltage level ($V$ = 65% or 75%). These conditions are selected based on results in section 2 and section 3.3.

Representative images of the operation of the reactors at the beginning (0.5 min) and the end (30 min) of the hydrogen production experiments are shown in Fig. 7. Optical images of the transarc at 0.5 and 30 minutes of operation are depicted in Fig. 7a and 7b. The results show that for $H$ = 5 mm, the plasma presents a stable and intense glow, whereas for $H$ = 10 mm, the plasma appears filamentary, representative of arc/streamer conditions. This filamentary arc covers a broader area of the sample's surface, potentially leading to greater hydrogen production (section 4). The LDPE sample melted after ~ 5 minutes of operation. The yellow glow by the end of the experiment for $H$ = 5 mm and $V$ = 30% (Fig. 7b) can be attributed to the emission from carbon particles. For the larger spacing of 10 mm, a filamentary discharge weakly impinges the surface of the LDPE, and the plasma characteristics (size, emission, and dynamics) minimally change during the duration of the experiments.

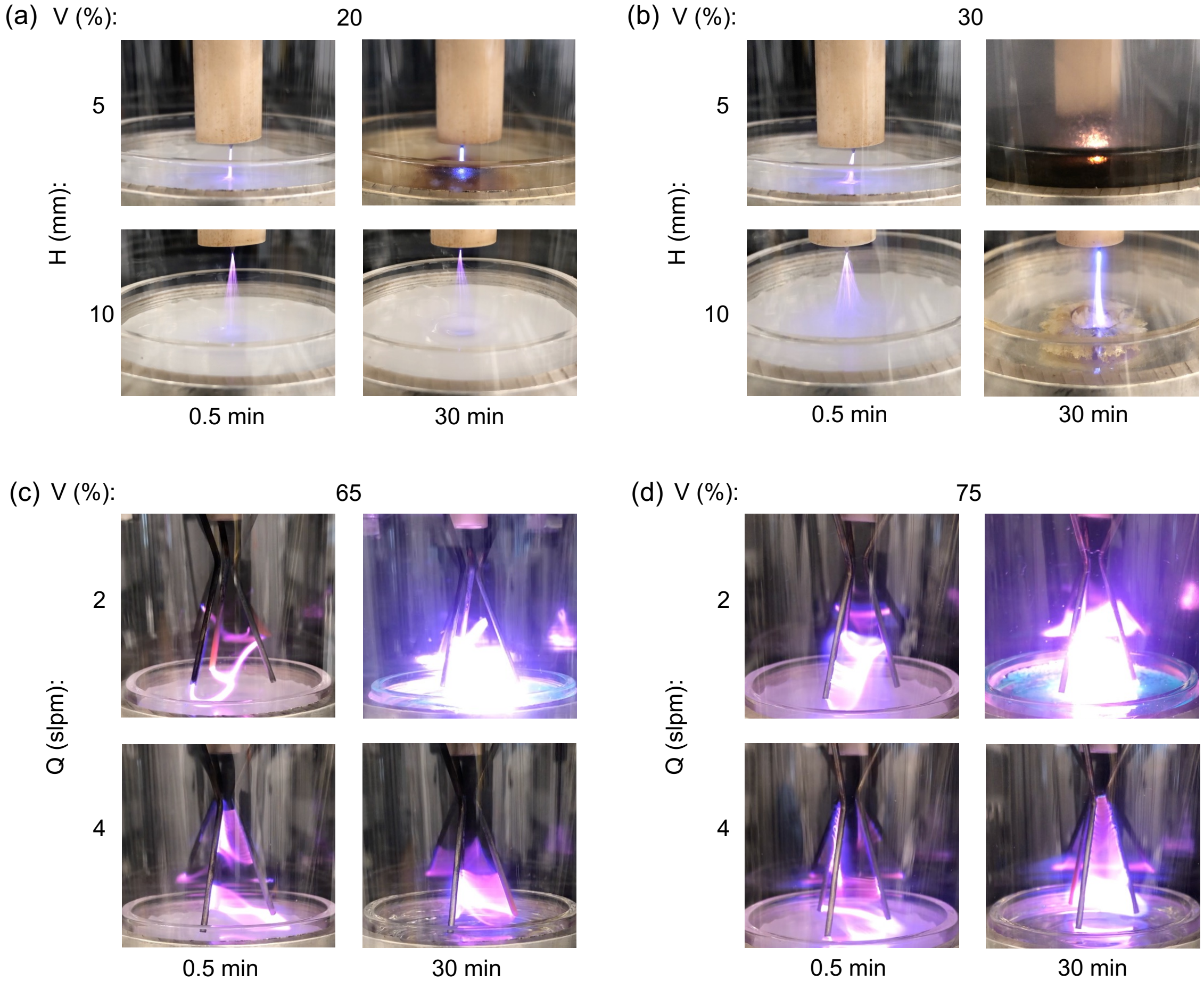

**Fig. 7. Reactors operation during hydrogen production from LDPE**. Transarc reactor at (a) the beginning and (b) the end of the experiment. Glidarc reactor at (c) the beginning and (d) the end of the experiment.

The glidarc reactor generates a tri-prong arc that impinges on the feedstock at the end of each gliding period, as shown in Fig. 7c and 7d. The intensity of the glidarc plasma is higher for $Q$ = 2 slpm, which is credited to the longer residence time, leading to pronounced interaction with the LDPE sample and, consequently, higher hydrogen production (section 4). The experiments show significant differences between the plasma near the initial and final portions of the LDPE treatment. This is attributed to the formation of gaseous products, heating of the feedstock, and heating of electrodes. The blue-green glow over the sample's surface for $Q$ = 2 slpm and $V$ = 75%, which is not observed under any other operational condition, suggests the formation of hydrocarbons.

### 2.3.4 Schlieren imaging

Schlieren imaging allows resolving the flow dynamics inside the reactor and unveils potential relationships between plasma dynamics and hydrogen production. Schlieren imaging results of the transarc reactor interacting with the inert (quartz) and LDPE samples are presented in Fig. 8a and Fig. 8b, respectively. The transarc plasma interacting with the inert and LDPE samples, as indicated by the horizontal (green) arrow, is faintly visible for every experimental condition. The interaction of the transarc plasma with the inert sample generates mild turbulence, which is weakly visible in Fig. 8a. However, the transarc plasma interaction with the LDPE sample (Fig. 8b) produces significant turbulence over the surface of the feedstock, as indicated by the vertical (purple) arrow. Given that Schlieren imaging resolves mass density gradients within the flow and that hydrogen is significantly lighter than nitrogen (the working gas), the observed turbulence is probably due to hydrogen emanating from the surface of the feedstock. The more significant turbulence observed for the condition of $H$ = 5 mm and $V$ = 30% is consistent with greater hydrogen production (discussed in section 4.2).

Schlieren imaging results of the glidarc reactor reveal the gliding of the tri-prongs arc (indicated by the horizontal green arrow) and its eventual impingement onto the sample, as shown in Fig. 8c and Fig. 8d for the inert and LDPE samples, respectively. The occurrence of turbulence is not captured by Schlieren imaging due to the significantly slower dynamics of the glidarc than those for the transarc (i.e., a 16 times lower frame rate is used to capture the dynamics of the glidarc than that used for the transarc). The interaction of the glidarc plasma with the inert sample (Fig. 8c) does not generate any glow, and the arc extinguishes on reaching the sample's surface. This suggests that the inert sample does not produce a significant amount of hydrogen when interacting with plasma.

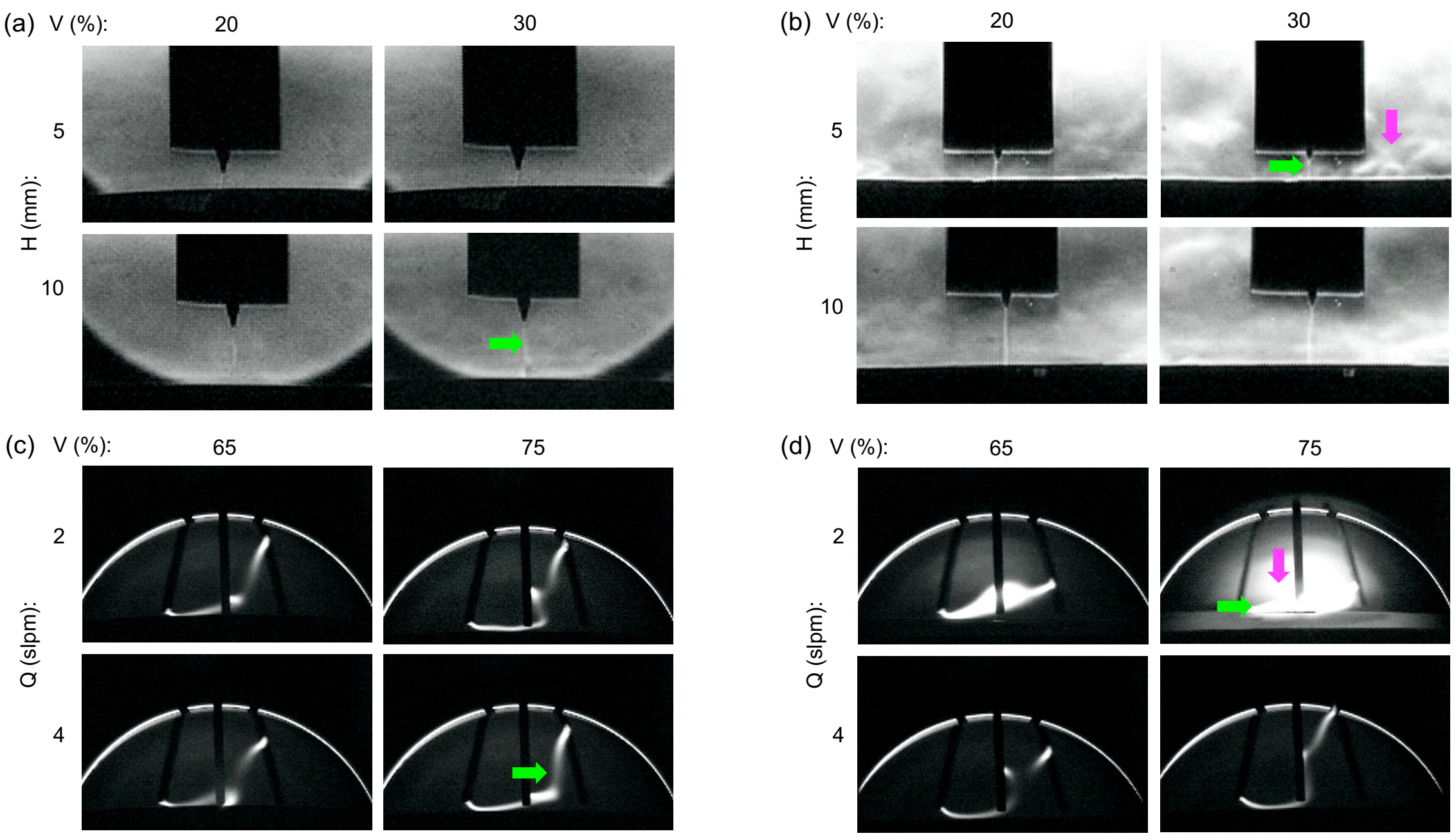

**Fig. 8. Schlieren imaging of the operation of the reactors.** The transarc plasma interacting with (a) inert and (b) LDPE samples. Glidarc plasma interacting with (c) inert and (d) LDPE samples. The horizontal arrows indicate the location of the plasma column, and the vertical arrows indicate the formation of turbulent flow originating from the surface of the sample, likely due to the production of hydrogen.

In contrast, when the glidarc plasma interacts with the LDPE sample, a pronounced glow is observed, indicated by the vertical (purple) arrow shown in Fig. 8d. The pronounced glow observed for $Q$ = 2 slpm is probably attributed to hydrogen emanating from the feedstock, which correlates with greater hydrogen production (section 4.2). The horizontal arrows indicate the location of the plasma column, and the vertical arrows show the formation of turbulent flow from the substrate, likely due to hydrogen production.

## 2.4 Hydrogen production from LDPE

### 2.4.1 Sample characterization

The treated samples depict the extent of interaction between the plasma and the LDPE feedstock. Fig. 9 shows the LDPE samples before (Fig. 9a) and after 30 minutes of treatment (Fig. 9b for the transarc and Fig. 9c for the glidarc) for the selected values of operational parameters. For the transarc, the white surface of the pristine LDPE sample develops a dark-brown color after treatment, especially for $H$ = 5 mm (Fig. 9b). The significantly darker and more extensive region of the sample treated using $H$ = 5 mm and $V$ = 30% implies a more significant plasma-LDPE interaction, consistent with the observed greater turbulence (Fig. 8b). The dark color suggests the formation of carbon compounds over the treated feedstock surface, consistent with the emission of carbon particles implied by the results in Fig. 7b. For $H$ = 10 mm, the weak streamer discharge generated leads to the formation of a crater at the center of the sample. This crater formation is suggested by the computational simulation results in section 2.2, which show concentrated temperature, heat flux, and velocity at the center of the feedstock (e.g., Fig. 2b).

Figure 9c shows the samples after 30 minutes of treatment in the glidarc reactor. The samples melt within five minutes of the experiment under all the selected operational conditions. The darkening of the sample's surface is more significant for the lower flow rate ($Q$ = 2 slpm) and higher voltage level ($V$ = 75%), consistent with the enhanced interactions between the plasma and the feedstock revealed by Schlieren imaging (section 3.4). Higher voltage levels lead to greater plasma power deposited on the feedstock, as suggested by the higher heat fluxes in the simulation results in Fig. 3d. The lower flow rate of 2 slpm leads to a longer interaction time between the reactive plasma species and the feedstock (which can be assumed proportional to $t_{res}$). In contrast, the higher flow rate of 4 slpm, which leads to a shorter $t_{res}$ of 10.3 s and greater convective cooling of the plasma, results in minor darkening of the feedstock, as observed in Fig. 9c. This is also supported by the simulation results shown in section 2.2, in which the heat flux and temperature in the glidarc reactor are higher for $Q$ = 2 slpm as compared to $Q$ = 4 slpm due to lower convective cooling. As discussed in section 4.3, the lower flow rate and higher voltage level used in the glidarc reactor led to more significant plasma-feedstock interaction, which favors greater hydrogen production.

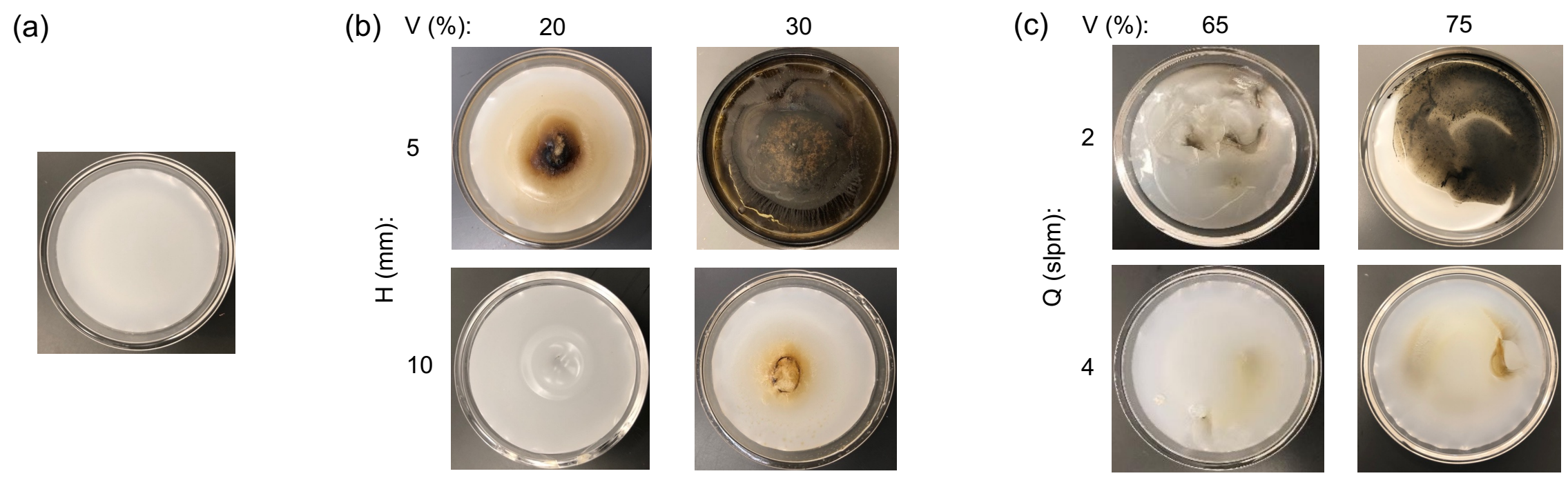

**Fig. 9. LDPE sample treatment.** (a) Pristine sample before plasma treatment and samples after 30 minutes of plasma treatment under representative operating conditions in the (b) transarc and (c) glidarc reactors.

### 2.4.2 Electrical characterization

The electrical characterization of the reactors helps assess the dynamics of the plasma and determine their role in hydrogen production. The transarc plasma produces a smooth sinusoidal voltage signal of up to 22.5 kV peak-to-peak (pp) with a sharply varying current of frequency ~ 25 kHz. The sharply varying current is characteristic of filamentary (streamer) discharges. The glidarc plasma generates smooth sinusoidal signals of frequency ~ 22.5 kHz and an instantaneous voltage of up to 6 kV pp, which is in phase with the current.

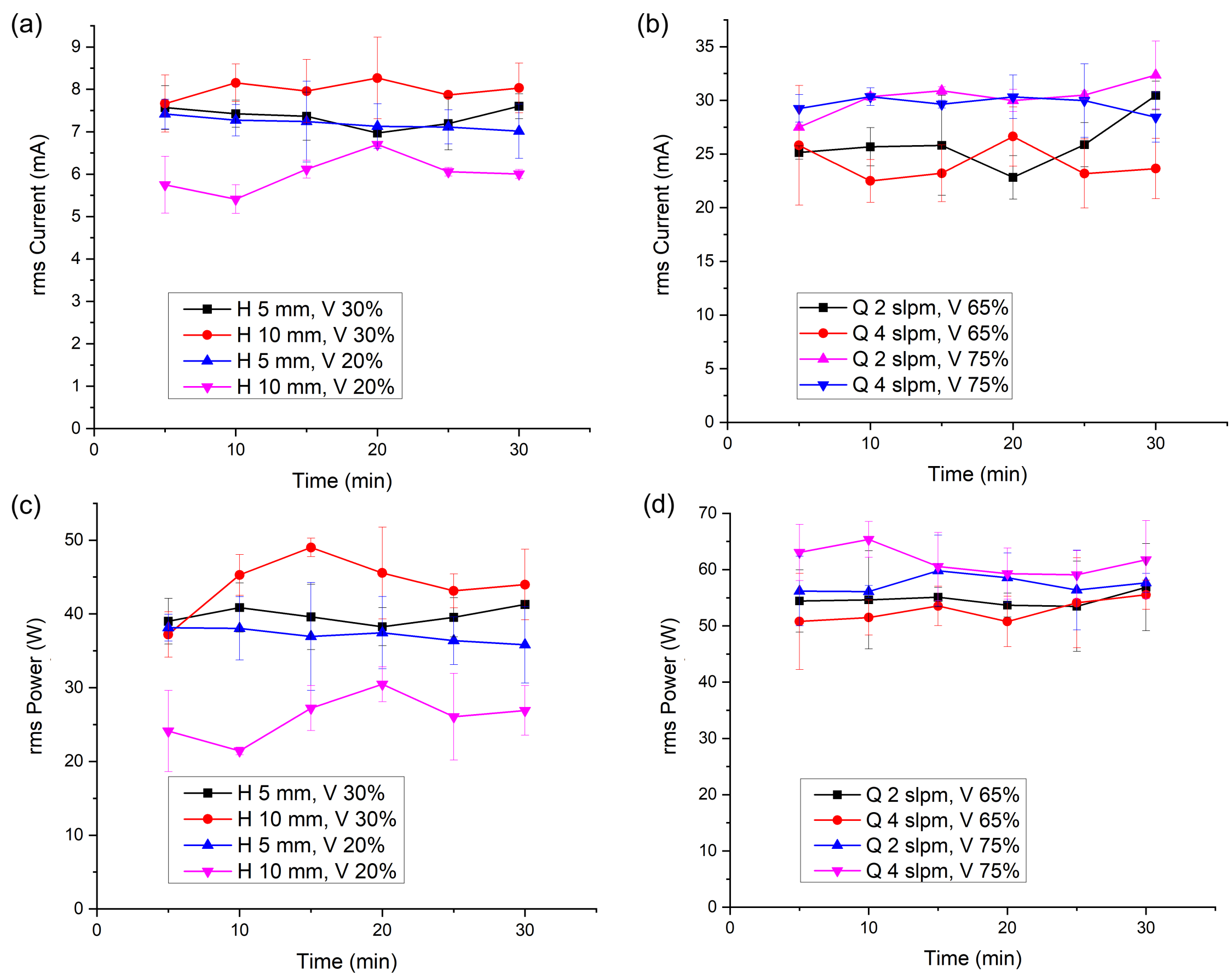

**Fig. 10. Electrical characteristics of reactors operation**. Root-mean-square (rms) current as a function of time for the (a) transarc and (b) glidarc reactors and rms power for the (c) transarc and (d) glidarc reactors.

The overall electrical characteristics as a function of operating conditions are presented in Fig. 10. The root-mean-square (rms) current increases with increasing voltage level for both reactors (Fig. 10a and Fig. 10b), resulting in increased power deposited over the feedstock. Also, for the transarc, a higher rms current implies higher electron flux onto the feedstock, which can likely increase the probability of cleavage of carbon-hydrogen (C-H) bonds and consequently increase hydrogen production. In comparing the results in Fig. 10a with those in Fig. 10b, it is to be noted that the glidarc reactor utilizes two power supplies and hence uses significantly greater current than the transarc reactor. The rms power as a function of time for both reactors and the different experimental conditions are presented in Fig. 10c and Fig. 10d. In both reactors, voltage level $V$ is the main parameter determining the consumed power, which increases with $V$ irrespective of other conditions (i.e., electrode-feedstock spacing $H$ and flow rate $Q$). The transarc plasma rms power (Fig. 10b) increases slightly with voltage level and fluctuates minimally during the experiments for all the operational conditions tested. Similarly, as for the glidarc reactor, the rms power for the glidarc reactor increases slightly with voltage level. The maximum rms power is obtained for $Q$ = 4 slpm and $V$ = 75%, as shown in Fig. 10d. Additionally, the rms power of the glidarc reactor shows negligible variation across the different conditions tested, and it is approximately 20 W higher than that of the transarc. Despite the glidarc reactor's sensitivity to flow rate, the effect of flow rate on rms power is negligible. However, the residence time $t_{res}$ is shorter at higher flow rates, limiting the interaction time between plasma species and the feedstock and potentially lowering hydrogen production.

### 2.4.3 Process performance

The main performance metrics of the process are the hydrogen production rate and the hydrogen production efficiency (i.e., hydrogen production rate per unit power). The hydrogen production rate ($P_r$) is defined as:

$$P_r = C_{out}Q, \tag{2}$$

where $C_{out}$ is the molar concentration of hydrogen in the outflow stream. The hydrogen production efficiency ($\eta_e$) is given by:

$$\eta_e = \frac{P_r}{P_{rms}}, \tag{3}$$

where $P_{rms}$ is the rms power consumed by the reactor.

Results for hydrogen production rate $P_r$ as a function of process time and operational parameters for both reactors are shown in Fig. 11. The results show that $P_r$ increases with increasing voltage level in both reactors. The transarc reactor (Fig. 11a) attains the mean maximum production rate of 6.6 mmol/h (0.33 mmol/g LDPE) at $H$ = 5 mm, $Q$ = 0.1 slpm, and $V$ = 30%; while the maximum average production rate for the glidarc reactor (Fig. 11b) is 8.4 mmol/h (0.42 mmol/g LDPE) at $Q$ = 2 slpm, $H$ = 5 mm, and $V$ = 75%. These conditions for maximum hydrogen production correspond to those observed by Schlieren visualization (section 3.4), namely, the greatest turbulence for the transarc and the greatest plasma-substrate interaction for the glidarc, respectively. In general, greater hydrogen production is attributed to the larger amount of power deposited over the LDPE sample at higher voltage levels. Furthermore, $P_r$ increases with time under higher voltage levels during the first ~15 minutes and then stabilizes. The slight decline in hydrogen production rate after 25 minutes likely suggests the formation of a layer of carbon/char that hinders hydrogen production. The lower hydrogen production rates at the beginning of the experiments (< 10 minutes) under all operational conditions and in both reactors suggest that a portion of the energy is consumed in melting the samples. After melting, the energy deposited by the plasma may lead to a more effective incision of C-H bonds in LDPE, leading to greater $P_r$.

Despite the lower voltage level, the hydrogen production rate for $H$ = 5 mm, $Q$ = 0.1 slpm, and $V$ = 20% is higher than that of $H$ = 10 mm, $Q$ = 0.1 slpm, and $V$ = 30% (lines with triangle and circle marks, respectively, in Fig. 11a). This suggests that shorter electrode-feedstock spacing $H$ leads to greater interaction between the reactive plasma species and the LPDE feedstock, resulting in higher hydrogen production rates. On the contrary, when $H$ is larger, the highly energetic electrons and reactive plasma species lose significant energy due to collisions resulting in quenching or recombination reactions before having the opportunity to interact with the feedstock, leading to lower hydrogen production rates.

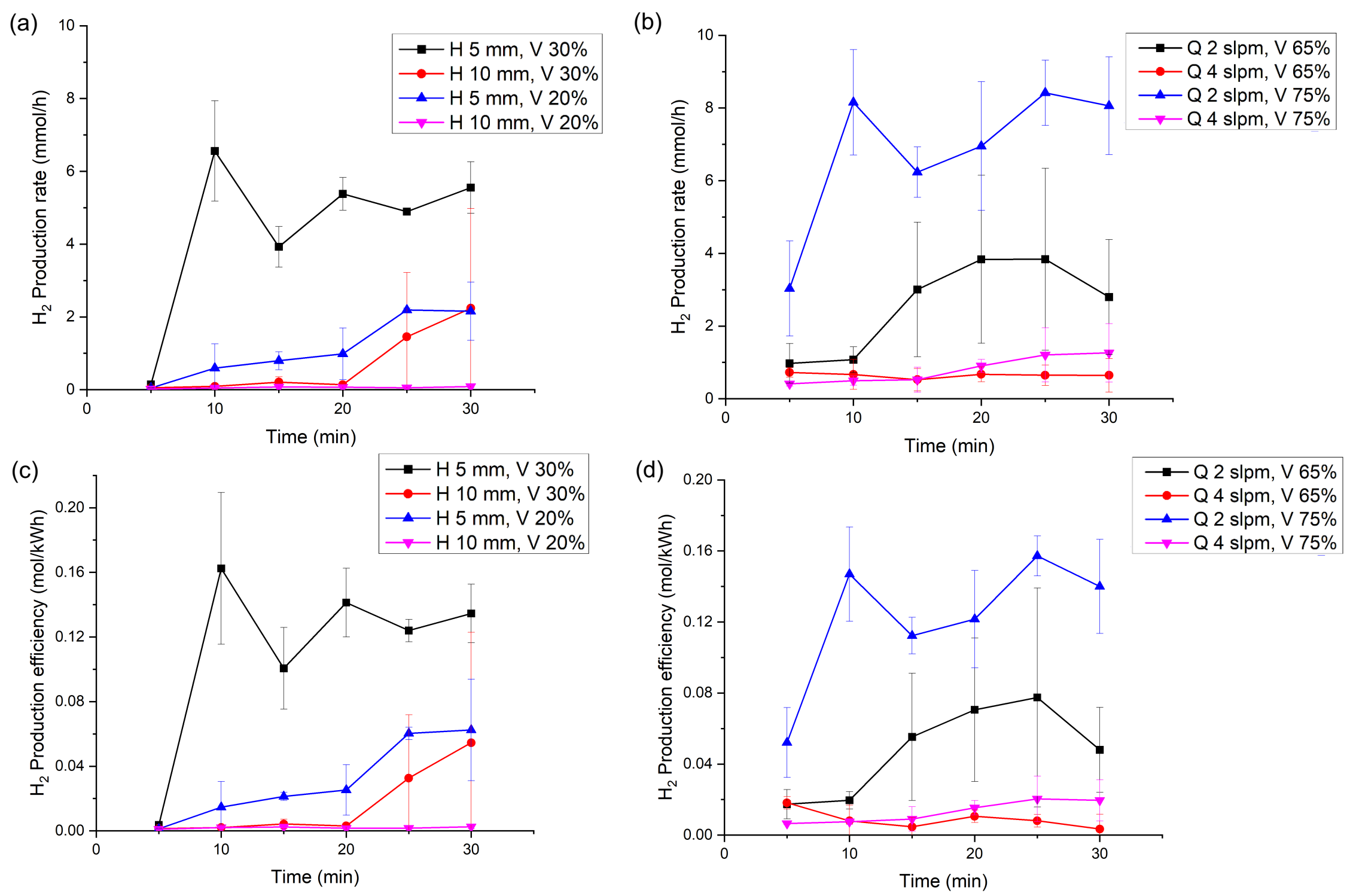

**Fig. 11.Hydrogen production rate and production efficiency.** Hydrogen production rate versus time for the (a) transarc reactor and the (b) glidarc reactor. Hydrogen production efficiency as a function of time for the (c) transarc reactor and (d) the glidarc reactor.

The results in Fig. 11b show that, for the glidarc reactor, greater flow rates lead to lower hydrogen production rates. This is likely due to two effects. First, the shorter residence time $t_{res}$ reduces the probabilities of electrons and excited species reacting with the feedstock. This has been observed by Indarto *et al.* [47] in investigating the effect of working gas flow rate on carbon dioxide conversion using glidarc plasma. Second, intense convective cooling of the glidarc plasma leads to lower heat flux over the feedstock, as depicted in the simulation results in section 2.2. The higher hydrogen production rate for $Q$ = 2 slpm, $V$ = 75%, and $H$ = 5 mm is attributed to longer $t_{res}$, less cooling, and higher deposited power. The slightly higher hydrogen production rate by the glidarc reactor is due to its utilization of two power sources, which effectively increases the electrical power deposited on the feedstock.

To determine how effectively electrical energy is utilized to produce hydrogen from LDPE, hydrogen production efficiency $\eta_e$ is shown in Fig. 11c and Fig. 11d for the transarc and glidarc reactors, respectively. Hydrogen production efficiency increases proportionately with voltage levels for both reactors. High voltage levels lead to greater power deposited onto the feedstock, which increases hydrogen production rate irrespective of the other operating conditions. This observation is consistent with the computational simulation results of greater heat flux onto the feedstock (section 2.2). Additionally, as observed in the hydrogen production results in Fig. 11a and Fig. 11b, electrode-feedstock spacing and flow rate significantly affect the hydrogen production efficiency in the transarc and the glidarc reactor, respectively. The transarc reactor attains a maximum $\eta_e$ of 0.16 mol/kWh at $H$ = 5 mm, $V$ = 30%, and $Q$ = 0.1 slpm. This maximum $\eta_e$ is comparable to that of the glidarc reactor, which is 0.15 mol/kWh at $Q$ = 2 slpm, $V$ = 75%, and $H$ = 5 mm. Overall, despite markedly different modes of operations, hydrogen production rate and hydrogen production efficiency are similar in both reactors for all operational conditions. This is an important aspect to factor in for the scaling-up of the systems.

The energy cost of hydrogen production for the transarc and the glidarc reactors are 3100 and 3300 kWh/kg $H_2$, respectively. These values are significantly higher than those for steam methane reforming (21.9 kWh/kg $H_2$) and electrolysis (47.6 kWh/kg $H_2$), currently the most energy-efficient approaches to produce hydrogen. The large difference in performance is in part ascribed to the nature of the feedstock. Methane and water vapor are both gaseous feedstock that require less energy in overcoming the weak intermolecular forces as well as cleaving atomic bond energies as compared to solid LDPE. Although hydrogen production from LDPE via low-temperature atmospheric plasma has a higher energy cost of hydrogen production, the process potentially has greater environmental benefits especially if LDPE is sourced from plastic waste and the reactors are powered by renewable energy sources such as wind and solar energy.

### 2.4.4 Correlations between operational parameters and hydrogen production

The expected performance of the reactors, necessary for scaling analyses, can be assessed through correlations between hydrogen production ($P_r$ and $\eta_e$) and operational parameters ($V$, $Q$, $H$) and/or operational characteristics (e.g., rms voltage $V_{rms}$, rms power $P_{rms}$). Correlations of the form $V_{rms}^{a}P_{rms}^{b}H^{c}$ and $V_{rms}^{a}P_{rms}^{b}Q^{c}$ are sought for the transarc and the glidarc reactors, respectively. For dimensional and practical reasons, the exponents were set as $a, b, c \in \{-3, -2, -1, 0, 1, 2, 3\}$. This set of exponents leads to 343 ($7^3$) different parametric combinations. Among these, the conditions with the strongest correlation are identified as those with the greatest correlation coefficient ($R^2$), which are depicted in the results in Fig. 12.

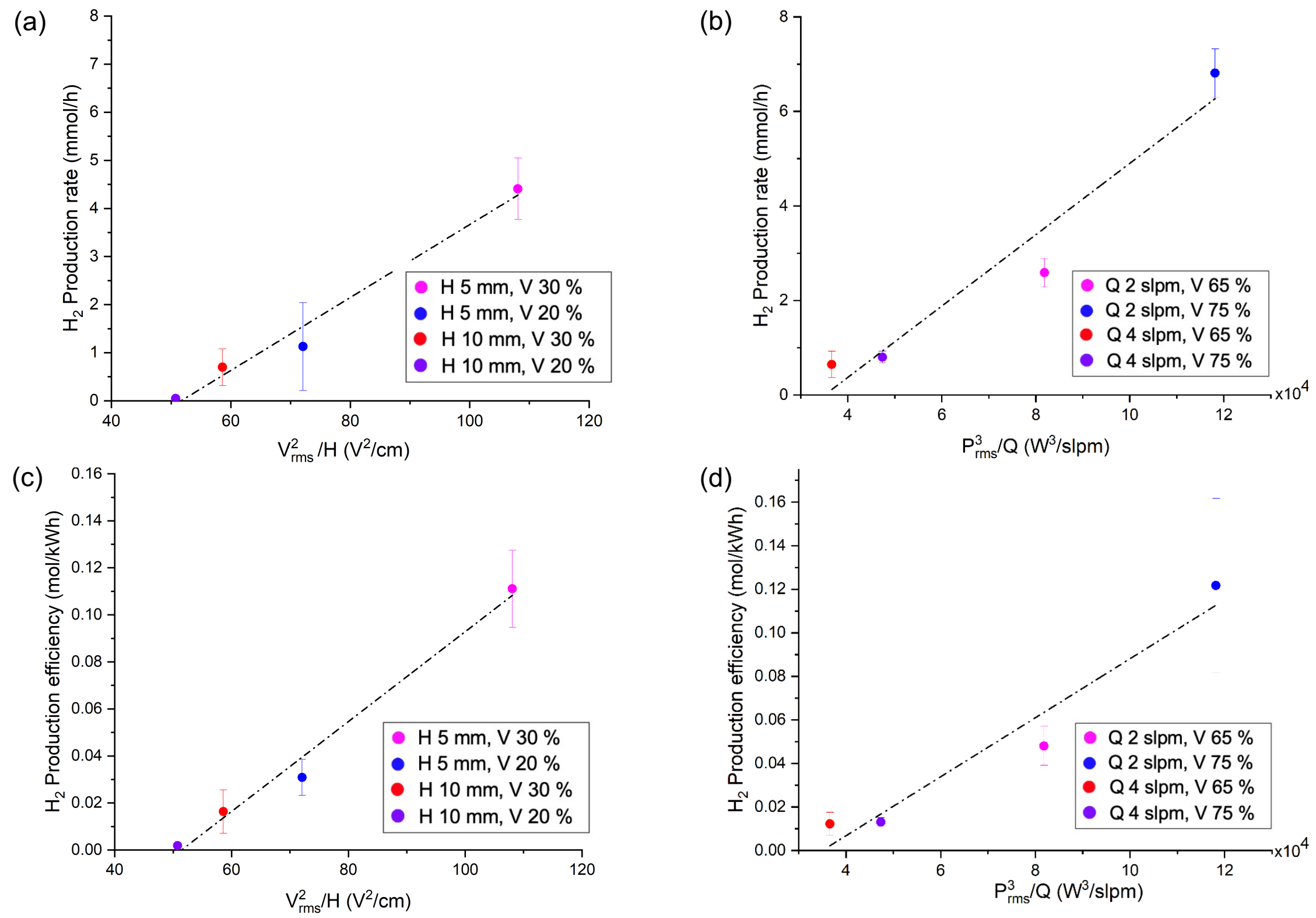

**Fig. 12. Correlations between hydrogen production and operating parameters**. Hydrogen production as a function of (a) equivalent power per unit length for the transarc reactor and (b) power per unit flow rate for the glidarc reactor. Hydrogen production efficiency as a function of (c) equivalent power per unit length for the transarc reactor and (d) power per unit flow rate for the glidarc reactor.

The hydrogen production rate $P_r$ by the transarc reactor strongly depends on the rms voltage and electrode-feedstock spacing $H$ following the relation: $P_r = \alpha_r V_{rms}^2/H - \beta_r$ ($\alpha_r$=7.582×$10^{-4}$, $\beta_r$ = 3.918, $R^2$ = 0.979), which is shown in Fig. 12a. Given the columnar structure of the transarc plasma, the term $V_{rms}^2/H$ can be considered equivalent to plasma power per unit length. Hence, the obtained correlation implies that $P_r$ for the transarc reactor is correlated with plasma power unit length. This result is consistent with the simulation results in section 2.2 indicating a direct dependency of surface heat flux and of average temperature with thermal power per unit volume. The inverse dependency of $P_r$ with electrode-feedstock spacing $H$ is also consistent with the findings in section 4.3. Similarly, hydrogen production efficiency strongly correlates with plasma power per unit length, leading to a linear relationship given by $\eta_e = \alpha_e V_{rms}^2/H - \beta_e$ ($\alpha_e$=1.911×10$^{-5}$, $\beta_e$ = 9.828×10$^{-2}$, $R^2$ = 0.986). For a given plasma power (proportional to $V_{rms}^2$), $\eta_e$ increases monotonically with decreasing inter-electrode feedstock spacing $H$.

Both, hydrogen production rate $P_r$ and production efficiency $\eta_e$ for the glidarc reactor depend on the cube of the rms power $P_{rms}$ and inversely to the flow rate $Q$, as shown in Fig. 12b and Fig. 12d. The inverse relationship between $P_r$ and $Q$ is consistent with the simulation results in section 2.2 in which higher temperature and surface heat flux are observed for the lower flow rate of 2 slpm. A low flow rate produces less convective cooling and higher residence time $t_{res}$, which lead to longer characteristic times for plasma species to interact with the feedstock. The correlation of $P_r$ with $P_{rms}$ and $Q$ is given by $P_r = \alpha_r P_{rms}^3/Q - \beta_r$ ($\alpha_r = 7.546 \times 10^{-5}, \beta_r$ = 2.646, $R^2$ = 0.940). It is to be noted that in the absence of inflow gas ($Q$ = 0 slpm), the generated plasma does not glide down along the electrodes, and hence does not interact with the feedstock (leading to no hydrogen production). Contrastingly, for the larger flow rates, the hydrogen production rate reduces significantly due to rapid cooling of the gas and limited time for plasma species to interact with the feedstock. Therefore, optimal hydrogen production is attained at intermediate values of $Q$, as depicted in the results in Fig. 6. The dependence of the production rate with the cube of rms power suggests a trend that compensates for the significant amount of energy consumed at the beginning of the sample treatment (i.e., slow melting of the top of the sample compared to what is achieved by the transarc), and then the production increases (Fig. 11b). The production efficiency $\eta_e$ of the glidarc reactor depicts a comparable trend as $P_r$ given by the relation $\eta_e = \alpha_e P^3_{rms}/Q - \beta_e$ ($\alpha_e$=1.355×10$^{-6}$, $\beta_e$ = 4.74×10$^{-2}$, $R^2$ = 0.945). The greater residence for lower flow rates implies greater plasma interaction with the feedstock leading to greater production efficiency. The hydrogen production efficiency's dependency on the flow rate is limited to a specified range as no plasma-feedstock interaction is achieved at too low ($Q$ < 2 slpm) or high ($Q$ > 6 slpm) flow rates (see section 3.2).

## 2.5 Conclusions

Two nonthermal plasma reactors with complementary characteristics, based on transarc and glidarc discharges, are designed, developed, and characterized to produce hydrogen from LDPE as a model plastic waste. CFD thermal-fluid models are used to attain expected operational characteristics as functions of design and operation parameters, namely electrode-feedstock spacing, flow rate, and dissipated thermal power – the latter assumed proportional to the voltage level of the power supply. Simulation results identify electrode-feedstock spacing, flow rate, and voltage level as the main process parameters of the reactors. The built reactors are experimentally evaluated using electrical diagnostics, optical and Schlieren imaging, and gas chromatography to quantify hydrogen production. The Schlieren visualization results qualitatively show that hydrogen production correlates with the amount of turbulence over the LDPE feedstock for the transarc reactor and the residence time of the plasma over the feedstock for the glidarc reactor. The experimental evaluation of hydrogen production from LDPE shows that the power consumed by the plasma remains approximately constant throughout the 30 min treatment time. Moreover, the results show that hydrogen production increases proportionally with voltage level in both reactors and that electrode-feedstock spacing and flow rate are the dominant operational parameters in the transarc and the glidarc reactor, respectively. The energy cost of hydrogen production for both reactors is significantly higher than the conventional and most efficient hydrogen production technologies of steam methane reforming and water electrolysis. Hydrogen production and production efficiency correlate linearly with rms voltage squared divided by inter-electrode spacing for the transarc reactor and with the rms power cubed divided by flow rate for the glidarc reactor. The two reactors depict comparable performance in terms of hydrogen production rate and efficiency, despite distinct differences in their operational principle. Overall, the results show that atmospheric pressure nonthermal plasma is effective at producing hydrogen from LDPE.

## Acknowledgments

This work has been supported by the US Army Combat Capabilities Development Command (CCDC) Soldier Center Contracting Division through Contract # W911QY-20-2-0005.

# CHAPTER 3: HYDROGEN FROM CELLULOSE AND LOW-DENSITY POLYETHYLENE VIA ATMOSPHERIC PRESSURE NONTHERMAL PLASMA

## Abstract

The valorization of waste, by creating economic value while limiting environmental impact, can have an essential role in sustainable development. Particularly, polymeric waste such as biomass and plastics can be used for the production of green hydrogen as a carbon-free energy carrier through the use of nonthermal plasma powered by renewable, potentially surplus, electricity. In this study, a Streamer Dielectric-Barrier Discharge (SDBD) reactor is designed and built to extract hydrogen and carbon co-products from cellulose and low-density polyethylene (LDPE) as model feedstocks of biomass and plastic waste, respectively. Spectroscopic and electrical diagnostics, together with modeling, are used to estimate representative plasma properties, namely electron and excitation temperatures, number density, and power consumption. Cellulose and LDPE are plasma-treated for different treatment times to characterize the evolution of the hydrogen production process. Gas products are analyzed using gas chromatography to determine hydrogen production rate, production efficiency, hydrogen yield, selectivity, and energy cost. The results show that the maximum hydrogen production efficiency for cellulose is 0.8 mol/kWh, which is approximately double that for LDPE. Furthermore, the energy cost of hydrogen production from cellulose is 600 kWh/kg of $H_2$, half that of LDPE. Solid products are examined via scanning electron microscopy, revealing the distinct morphological structure of the two feedstocks treated, as well as by elemental composition analysis. The results demonstrate that SDBD plasma is effective at producing hydrogen from cellulose and LDPE at near atmospheric pressure and relatively low-temperature conditions in rapid-response and compact processes.

## 3.1 Introduction

The utilization of fossil-based resources is the major contributor to greenhouse gas emissions leading to environmental pollution and climate change. According to Li [1], 31.5 Gt of $CO_2$ was generated in 2022 despite the low economic activities attributed to the COVID-19 pandemic. Such substantive $CO_2$ emissions are mainly due to the use of fossil fuels, accounting for over 80% of global energy consumption [2, 3]. The use of alternative fuels, particularly green hydrogen (i.e., hydrogen generated using renewable energy [4]) derived from non-fossil feedstock such as plastic and biomass waste, can support the creation of economic value while limiting environmental impacts. The increasing amount of organic polymeric waste, estimated to reach 25 billion metric tons of plastic waste globally by 2050 [5] and 146 billion metric tons of biomass annually [6, 7], can be considered an enormous resource, particularly for the production of hydrogen via processes powered by renewable, potentially surplus, electricity.

Methane-steam reforming and water electrolysis are currently the dominant methods for the production of hydrogen. Water electrolysis is the primary approach for carbon-free hydrogen production (when powered by renewable electricity). Nevertheless, due to its relatively high energy cost compared to methane-steam reforming [8], electrolytic approaches account for only 4% of total hydrogen production [9]. Other environmentally-benign hydrogen production methods in development include photocatalytic, photobiological, and photochemical water splitting [10].

In contrast to the use of methane or water as feedstock, the use of solids, particularly polymeric waste, for hydrogen production remains largely untapped. Aziz *et al.* [11] noted that converting biomass and other organic solid materials to hydrogen is a promising approach due to feedstock availability and could lead to positive economic, social, and environmental impacts. The main methods for the production of hydrogen from solids are pyrolysis and gasification [12, 14]. These methods can be thermo-chemical (using temperature and pressure as the main process parameters), thermo-catalytic (incorporating catalysts), or thermal plasma-based, and generally focus on the production of syngas, a mixture of mainly hydrogen and carbon monoxide. Thermal plasma methods are particularly appealing because they make direct use of electricity, which given the increasing capacity of renewable electricity generation and limited electricity storage, is sometimes available as surplus. Moreover, thermal plasma methods are robust for the treatment of heterogeneous and hard-to-decompose waste streams, require minimal or no consumables, and do not rely on catalysts. Nevertheless, pyrolysis and gasification processes – including those based on thermal plasma – generally operate with low energy efficiency (typically defined as the caloric content of syngas produced per unit energy consumed) and/or low selectivity towards hydrogen production.

Hydrogen production processes based on nonthermal plasma have the potential to provide significantly greater energy efficiency and/or selectivity than those based on thermal plasma. Plasma in industrial applications is typically generated by electrical discharges to bring a working gas (usually inert gases, nitrogen, or air) to a partially-ionized state. In thermal plasma, the constitutive species, namely free electrons and so-called heavy-species (i.e., ions, excited and ground-state atoms, and molecules), are in thermal equilibrium at a relatively high temperature, e.g., near 20000 K for arc discharge plasmas. In contrast, in nonthermal plasma, the free electrons are at significantly higher temperatures (typically between 1 and 10 eV, where 1 eV ~ 11600 K), than the heavy-species (from a few hundred to $< 2000$ K). The thermal nonequilibrium in nonthermal plasmas can translate into processes with higher energy efficiency and/or selectivity [15], by directing the energy of electrons towards desired chemical reactions while limiting the energy carried by the gas species (which, although also driving chemical reactions, manifests as undesired heating). Additionally, nonthermal plasma processes are generally more amenable to compact, modular implementations, which may be favored for distributed (de-centralized) and small-scale installations.

The present study focuses on the use of a nonthermal plasma approach for the production of hydrogen from biomass and plastic waste. The model feedstocks for biomass and plastic waste are cellulose and low-density polyethylene (LDPE), respectively. Cellulose is a linear chain of repeated anhydroglucose rings $(C_6H_{10}O_5)_n$, usually between 10000 to 15000 units long [16], depending on the source material. The anhydroglucose units are bonded covalently by 1,4' glycosidic links, which provide mechanical stiffness [17]. Cellulose is considered the most common organic compound on earth [18], naturally embedded in hemp, cotton, wood, crop residues, linen, etc. [16]. Investigations on the production of hydrogen from cellulose via electricity (rather than heat) have focused on electrochemical routes. Wei *et al*. [19] investigated electrochemically assisted molten carbonate pyrolysis of cellulose in the absence of a catalyst and obtained the maximum hydrogen yield of 8.3 mmol/g of cellulose at the relatively low temperature of 600 $^oC$. Similarly, Zeng *et al*. [20] studied molten salt pyrolysis of cellulose using a mixture of $Na_2CO_3$, $K_2CO_3$, and $Li_2CO_3$ as the electrolyte and achieved a peak hydrogen yield of 3.1 mmol/g of cellulose at 650 $^oC$. Furthermore, they also observed that hydrogen content increases rapidly from 18.05 to 26.19 vol.% as the result of increasing temperature from 650 to 850 $^oC$. In an experimental study of the transient behavior of devolatilization and char reactions during the steam gasification of biomass, Moon *et al*. [21] obtained a hydrogen production rate of about 800 mmol/h in ~ 2 minutes at an operating temperature of 700 $^oC$. Hoang *et al*. [22] characterized hydrogen production from steam gasification of plant-originated lignocellulosic biomass and obtained the maximum hydrogen yield of 55.6 mmol/g of cellulose at the operating temperature of 900 $^oC$.

Regarding the use of plastic waste for hydrogen production, Aminu *et al*. [23] reported a hydrogen production yield of 4.1 mmol/g from polyethylene during a two-stage low-temperature plasma catalytic treatment of plastic waste. Chai *et al*. [24] catalytically pyrolyzed a composite mixture of LDPE and pinewood dust using Ni-CaO-C as a catalyst and obtained an optimal hydrogen yield of 115.3 mmol/g of feedstock (LDPE to Pinewood dust ratio of 1 to 1) and hydrogen selectivity of 86.7% when 5 ml of water was injected, and the reactor was operating at 700 $^oC$. Nguyen and Carreon [25] investigated the catalytic deconstruction of high-density polyethylene (HDPE) via nonthermal plasma, reporting hydrogen yield of 8 mmol/g HDPE and selectivity of about 50%. In prior work by the authors [26], two nonthermal plasma reactors, based on transferred arc (transarc) and gliding arc (glidarc) discharges, were devised and used to produce hydrogen from LDPE. The maximum hydrogen yields were 0.33 and 0.42 mmol/g LDPE, with the corresponding minimum energy cost of 3100 and 3300 kWh/kg of $H_2$, for transarc and glidarc reactors, respectively. The study revealed that, despite the comparable yield and energy cost of these two largely different plasma sources, the transfer of electric current through the feedstock (as in the transarc reactor) leads to more compact and rapid-response processes.

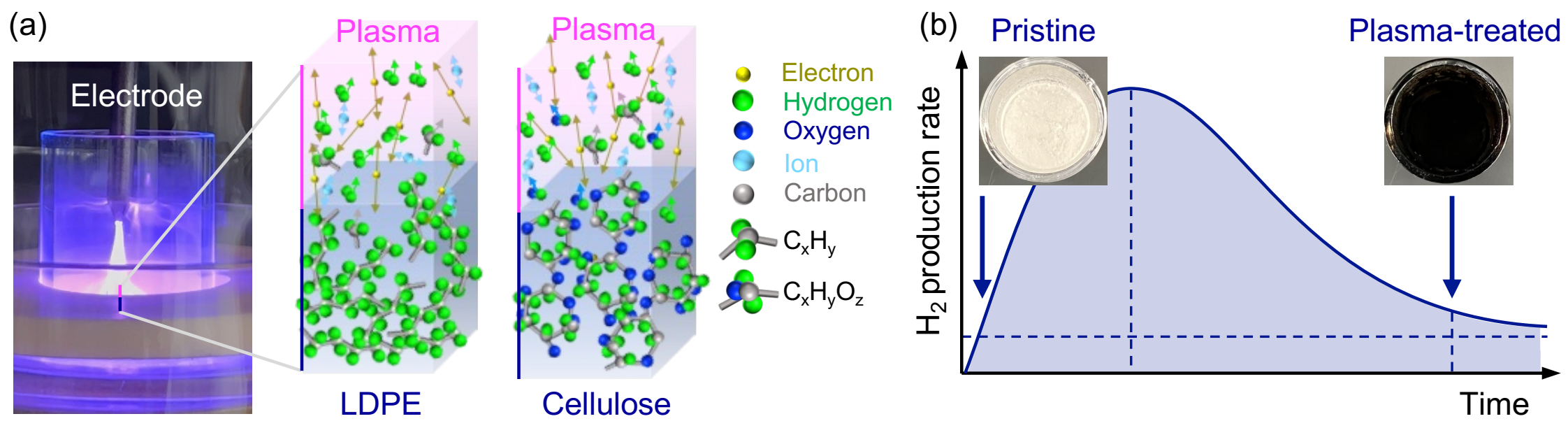

**Fig. 1. An overview of plasma dehydrogenation.** (a) Streamer Dielectric-Barrier Discharge (SDBD) plasma treatment of low-density polyethylene (LDPE) and cellulose, and (b) expected evolution of the hydrogen production rate as a function of treatment time.

In this study, the use of Streamer Dielectric-Barrier Discharge (SDBD) plasma, using argon as the working gas and operating at (near) atmospheric pressure, is investigated to produce hydrogen from cellulose and LDPE as organic polymeric waste models. The approach is schematically summarized in Fig. 1. A high-voltage alternating-current (AC) power supply is used to generate a plasma between a metal electrode and the feedstock placed over a dielectric barrier. SDBD plasma is a highly reactive medium composed of highly energetic electrons, ions, and metastable species. The charged species oscillate along the direction parallel to the electrode due to the imposed AC electric field. These reactive species interact with the cellulose and LDPE molecules, causing chain scissions and the release of hydrogen and low-hydrocarbons as gas products, and the de-hydrogenation or carbonization of the remaining feedstock. Due to the nonthermal nature of SDBD plasma, the reactor operates at a relatively low temperature (< 200 ºC average inside the reactor chamber). Given the decreasing availability of hydrogen to interact with plasma species as the process progresses, it is expected that the hydrogen production rate will initially increase until achieving a maximum and then monotonically decrease as the feedstock gets de-hydrogenated.

The article is organized as follows. Section 2 describes the experimental setup and procedures, as well as the electrical model and spectroscopic diagnostics of the SDBD reactor. In section 3, the results for hydrogen production from cellulose are presented, while section 4 consists of the results of hydrogen production from LDPE. Finally, the concluding remarks are presented in section 5.

## 3.2 Reactor design and characterization

### 3.2.1 Streamer Dielectric Barrier Discharge (SDBD) Reactor

The designed Streamer Dielectric Barrier Discharge (SDBD) reactor to produce hydrogen from polymeric solids is schematically depicted in Fig. 2. The reactor is designed with a pin-to-plate dielectric configuration and aimed to operate with plasma in a streamer (filamentary) discharge mode. The SDBD name is derived from the discharge mode of operation as well as the essential role of the dielectric barrier on the performance of the hydrogen production process. The reactor is powered by an alternating-current (AC) high-voltage power supply with a tungsten pin as the powered electrode (Fig. 2a). The voltage level of the power supply $V$ (from 0 to 100%) and the working gas flow rate $Q$ are the main operating parameters. The reactor's chamber has a diameter $D$ = 76 mm and height of 160 mm (Fig. 2b). The pin electrode is electrically isolated by a ceramic bushing, and plasma is generated between the electrode's tip and the solid feedstock, which is placed in a crucible assembly on top of an aluminum plate acting as the ground electrode. The crucible consists of two quartz dishes - the larger dish has a diameter of 56 mm, while the smaller one, which contains the feedstock, has a diameter $d$ = 20 mm and includes an annular dielectric ring made of alumina. This ensures sufficient electrical insulation to mitigate undesired arcing (i.e., the formation of an electrical discharge circumventing the feedstock). The geometrical dimensions of the crucible assembly are of primary importance in determining the characteristics of the plasma and the performance of the process. Particularly, in the absence of the dielectric barrier, the discharge is weaker, and so is the rate of hydrogen production. The presence of the dielectric increases the amount of electrical power deposited on the feedstock. The main dimensions of the plasma-feedstock-dielectric barrier assembly are shown in Fig. 2c, namely electrode-feedstock spacing $h_p$ = 5 mm, feedstock height $h_f$ (different size of cellulose and LDPE, see section 2.2), and dielectric height $h_d$ = 4.5 mm.

### 3.2.2 Experimental setup and procedures

***Components.*** The SDBD reactor generates a streamer plasma in contact with the solid feedstock, leading to the production of hydrogen and carbon co-products. The experimental setup is depicted in Fig. 3. The reactor is operated by a high-voltage AC power supply (PVM 500-2500) characterized by an adjustable peak-to-peak voltage of 1-40 kV and a maximum current of 25 mA. Argon is used as the processing gas, and two mass flow controllers are used to measure and control the inlet and outlet flow rates. The electrical characteristics of the reactor's electrical circuit are measured using a Tektronix oscilloscope (TBS 2104) connected to a current probe (P6021A) and to a high-voltage probe (P6015A).

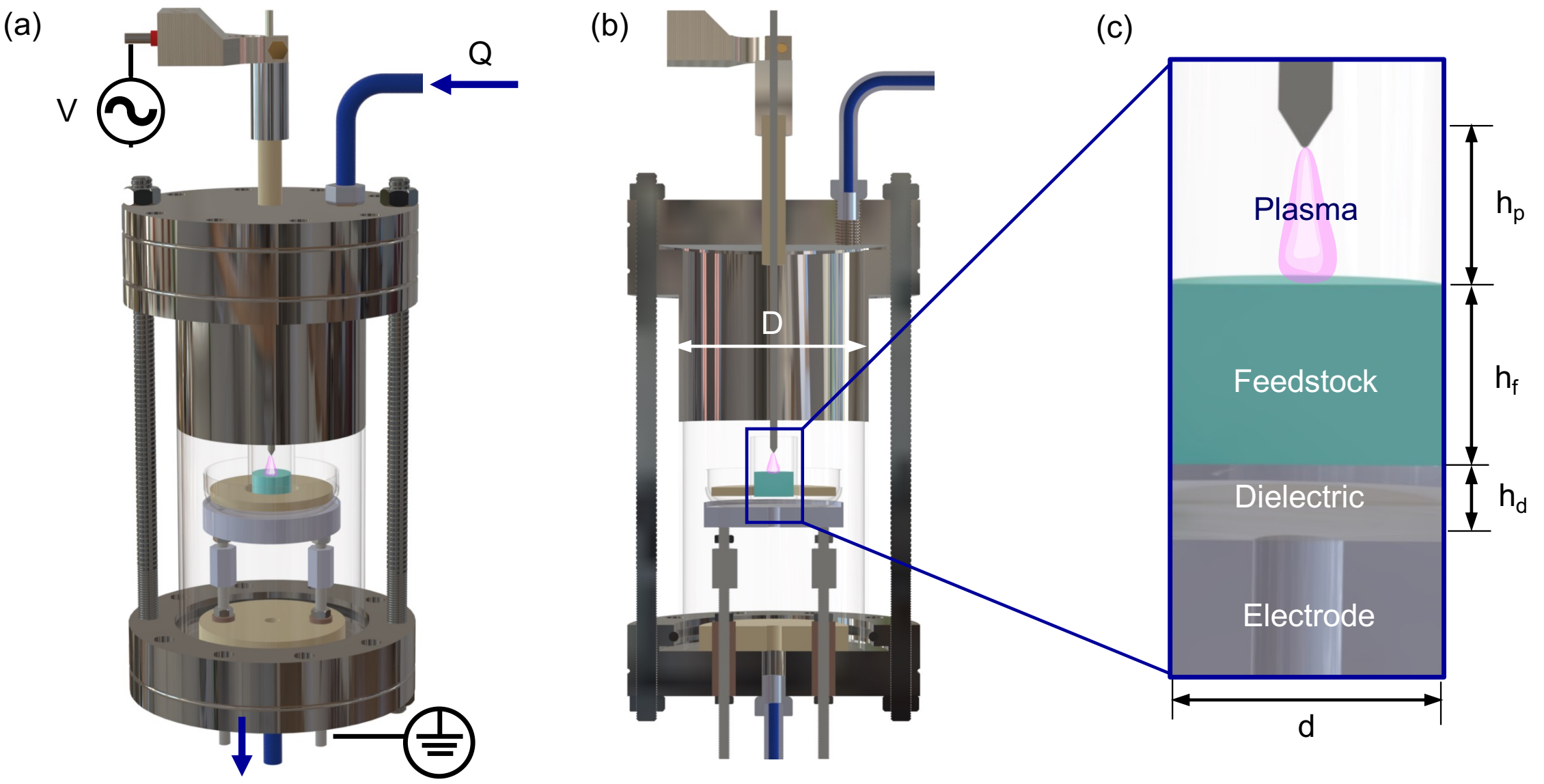

**Fig. 2. SDBD plasma reactor design based on pin-to-plate streamer discharge – dielectric barrier configuration.** (a) Reactor assembly with voltage level from the power supply V and processing gas flow rate $Q$ as the main process parameters. (b) Cross-sectional view of the reactor depicting the reactor chamber's diameter $D$. (c) Zoom-in view of the plasma-feedstock-dielectric region, depicting the electrode-feedstock spacing $h_p$, feedstock heigh $h_f$, dielectric height $h_d$, and diameter of the feedstock holder $d$.

To measure the spectral characteristics (optical emission) of the plasma, an Avantes spectrometer (ULS-2048-USB2) equipped with a set of optical lenses and a fiber optic probe is used. The gas products are collected in sampling bags and subsequently analyzed by a Shimadzu-2014 gas chromatograph. The morphological structure of the solid samples before and after plasma treatment is characterized using a digital camera and a field emission scanning electron microscope (FESEM) JSM 7401 coupled with an energy dispersive x-ray spectroscopy (EDS) detector. The elemental composition of the solids sample is determined by the CHN-elemental analysis.

***Feedstock sample preparation.*** The cellulose feedstock is a powder made of cotton linters commercially available from Millipore Sigma (Supelco, V001141). The sample is prepared by mechanically pressing 1 g of cellulose in the inner crucible plate (diameter $d$ = 20 mm and height $h_f \sim$ 10 mm, see Fig. 2c) using a manual hydraulic pellet press. The LDPE feedstock consists of 1 g of pellets with an average diameter of 3 mm from Millipore Sigma (Sigma Aldrich, 428 043), which are pre-melted at 180 $^{o}$C for 15 min and then re-solidified in the inner crucible (diameter $d$ = 20 mm and height $h_f \sim$ 6 mm, see Fig. 2c).

***Experimental procedure.*** In each experiment, cellulose and LDPE are treated with SDBD plasma for varying treatment times with the same voltage level $V$ = 60% and inlet flow rate $Q$ = 0.01 slpm of argon. The 60% voltage level was chosen as representative of the process leading to root-mean-square input power, $P_{t,rms}$ = 50 W and 53 W for cellulose and LDPE, respectively. Given the faster carbonization of cellulose, the treatment times of the experiments are set as $t_{treatment}$ = { 0.5, 1, 2, 3, 4, 5, 6, and 15 min }. For LDPE, the treatment time is more uniformly spread compared to cellulose and set as $t_{treatment}$ = { 0.5, 1, 2, 4, 6, 8, 10, and 15 min }. The electrical characteristics are measured at different time instants $t_{elec}$ depending on the treatment time. Specifically, for cellulose, for $t_{treatment}$ = 0.5 min, $t_{elec}$ = { 0.2 min }; for $t_{treatment}$ = 1 min, $t_{elec}$ = { 0.2, 0.4 min }; and for $t_{treatment}$ = { 2, 3, 4, 5, and 15 min }, $t_{elec}$ = { 0.5, 1 min }. For LDPE, for $t_{treatment}$ = 0.5 min, $t_{elec}$ = { 0.2 min }; for $t_{treatment}$ = 1 min, $t_{elec}$ = { 0.2, 0.4 min }; and for $t_{treatment}$ = { 2, 4, 6, 10, and 15 min }, $t_{elec}$ = { 0.5, 1 min }. All the generated gas products are collected throughout the plasma treatment (time $t$ from 0 to $t_{treatment}$) with the constant flow rate $Q$ = 0.01 slpm, and after treatment ($t > t_{treatment}$) with a higher flow rate of $Q$ = 0.1 slpm for 15 min for purging. The purging with argon drives all the generated gaseous products into the sampling bag for analysis by gas chromatography (GC). The spectroscopic measurements are carried out 15 seconds from the beginning of each experiment before the emanation of the opaque gaseous products that obscure the optical access of the plasma (see section 3 and section 4).

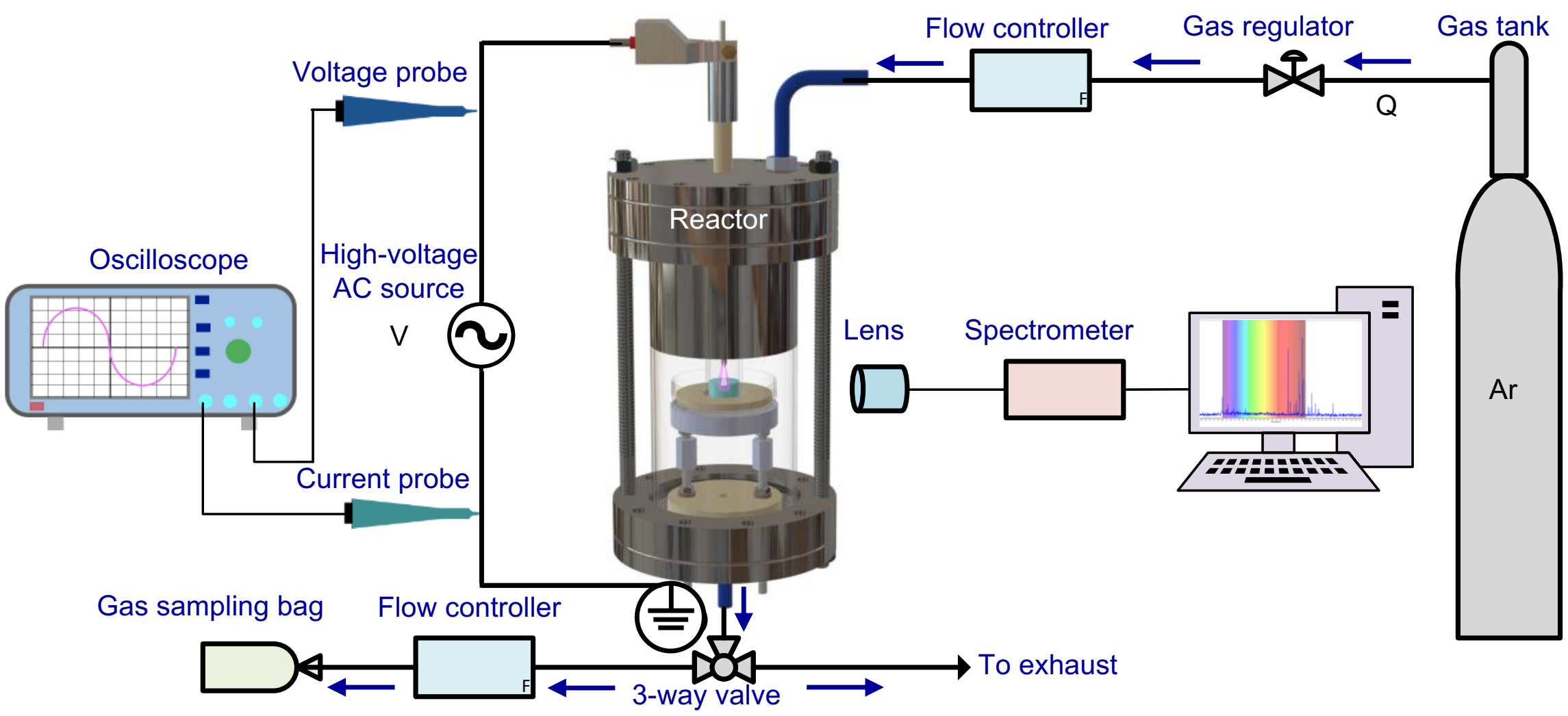

**Fig. 3. Experimental setup for hydrogen production from cellulose and LDPE via SDBD plasma.** Schematic of the experimental layout showing the SDBD reactor, the different diagnostics, and the gas and electrical lines.

## 3.3 Electrical model

An electrical model allows the determination of the power consumed by the plasma given the total power consumed measured by the oscilloscope. The model assumes that the plasma within the discharge gap (electrode-feedstock spacing $h_p$), feedstock, and dielectric, and can be electrically described as parallel-plate capacitors in series, as schematically depicted in Fig. 4a. Based on this assumption, the electrical capacitances of the components in the plasma circuit are defined by their geometrical configurations as

$$C_f = \frac{A_f}{h_f} \kappa_f \varepsilon_0 \tag{1}$$

and

$$C_d = \frac{A_d}{h_d} \kappa_d \varepsilon_0, \tag{2}$$

where $C_f$ and $C_d$ are the capacitances of the feedstock and dielectric, respectively; $A_f$ and $A_d$ are the cross-sectional areas of feedstock and dielectric, respectively; $\varepsilon_0$ is the permittivity of free space; and $\kappa_f$ and $\kappa_d$ are the dielectric constants of feedstock and dielectric, respectively. The dielectric constant for the dielectric crucible made of quartz is $\kappa_d = 3.8$ [27], whereas the dielectric constant of the feedstock is $\kappa_f = 2.5\text{-}2.6$ [28] for cellulose and $\kappa_f = 2.2\text{-}2.35$ [29] for LDPE.

Using Kirchhoff's law, the total input voltage $U_a(t)$ expressed in terms of plasma voltage $U_p(t)$, voltage across the dielectric $U_d(t)$, feedstock voltage $U_f(t)$, and effective dielectric voltage $U_D(t)$ is given by:

$$U_a(t) = U_p(t)+U_f(t)+U_d(t) = U_p(t)+U_D(t). \tag{3}$$

The total input current $I_t(t)$ is the sum of plasma current $I_p(t)$ and the displacement current through the gap $I_{p,g}(t)$, i.e.,

$$I_t(t) = I_p(t) + I_{p,g}(t). \tag{4}$$

The effective capacitance of the feedstock and dielectric $C_D$, given that these are assumed to operate in series, is given by:

$$\frac{1}{C_D} = \frac{1}{C_f} + \frac{1}{C_d}, \tag{5}$$

where $C_f$ and $C_d$ are the capacitances of the feedstock and dielectric, respectively. Liu and Neiger [30] derived the voltage across the feedstock and dielectric, as:

$$U_D(t) = \frac{1}{C_D} \int_0^t I_{t,a}(t')\, dt' + U_D(0), \tag{6}$$

where $U_D(0)$ is the memory voltage, which depends on an arbitrarily zero set time ($t = 0$) and is attributed to the memory charges deposited during the preceding AC voltage cycle. Considering that the negative voltage peak occurs at time zero, $U_D(0)$ becomes a constant and is defined in terms of the period $T$, i.e.,

$$U_D(0) = -\frac{1}{2C_D} \int_0^{\frac{T}{2}} I_{t,a}(t')\, dt'. \tag{7}$$

The plasma discharge current $I_p(t)$ can be determined from the input current by

$$I_p(t) = \left(1 + \frac{C_{p,g}}{C_D}\right) I_{t,a}(t) - C_{p,g} \frac{dU_a(t)}{dt}, \tag{8}$$

where the first and second terms on the right-hand side represent the total displacement current $I_{v,g}(t)$ and the gap displacement current $I_{p,g}(t)$, respectively. The total displacement current, sometimes referred as the external discharge current, is attributed to the effective capacitance of the plasma-gap, feedstock, and dielectric. Hence, it is generally erroneous to assume that the input current is the same as the plasma current, even when the gap displacement current is small and can be neglected.

The instantaneous input power can be calculated as

$$P_t(t) = U_a(t)I_t(t), \tag{9}$$

whereas the instantaneous plasma power is determined by

$$P_p(t) = U_p(t)I_p(t). \tag{10}$$

In the SDBD reactor, describing the plasma as mostly acting capacitively, with plasma-gap capacitance $C_{p,g}$, is a substantial approximation. Nevertheless, such an approximation is consistent with more conventional DBD electrical models and can be considered as reasonable as a first-order approximation to estimate the power consumed by the plasma. Adopting this approximation, $C_{p,g}$ is approximated as a constant determined from the experimentally-measured electrical response of the system by considering a fixed input power and calculating the range for which the plasma power is less or equal to the input power. The approach is schematically shown in Fig. 4b. The plasma power decreases to a minimum and subsequently increases monotonically with increasing $C_{p,g}$. This is attributed to the variation of gap displacement current $I_{p,g}(t)$ and total displacement current $I_{v,g}(t)$, hence changing the plasma current $I_p(t)$ and, subsequently, the plasma power $P_p(t)$. When $C_{p,g}$ is zero, the gap displacement current is zero, and hence the plasma power equals the input power. As $C_{p,g}$ increases, $I_{p,g}(t)$ increases more rapidly than $I_{v,g}(t)$, leading to a reduction in $I_p(t)$, and consequently to a decrease in plasma power $P_p(t)$. Further increasing $C_{p,g}$ leads to greater $I_{v,g}(t)$ than $I_{p,g}(t)$, and as a result, $P_p(t)$ also increases. The minimum point suggests that the increase in $I_{v,g}(t)$ and gap displacement current are equal.

Using experimental $U_a(t)$, $I_t(t)$, and $P_t(t)$ data during the treatment of cellulose and given that the plasma power cannot exceed the total power, acceptable values of $C_{p,g}$ are found within the range $0.1 \leq C_{p,g} \leq 2.9$ pF. This range of capacitance across the gap leads to corresponding plasma power between in the range $90\% P_{t,rms} < P_{p,rms} < 100\% P_{t,rms}$, where $P_{p,rms}$ is the root-mean-square (rms) of plasma power. The estimated range is comparable with the results by Ozkan *et al*. [31], which obtained 92% of absorbed power using the Lissajous method. Valdivia-Barrientos *et al*. [32] reported that the plasma voltage is 98% of the applied voltage, consistent with our model estimates. Therefore, during the operation of the SDBD reactor, it is expected that between 90 and 100% of the total power is consumed by the plasma.

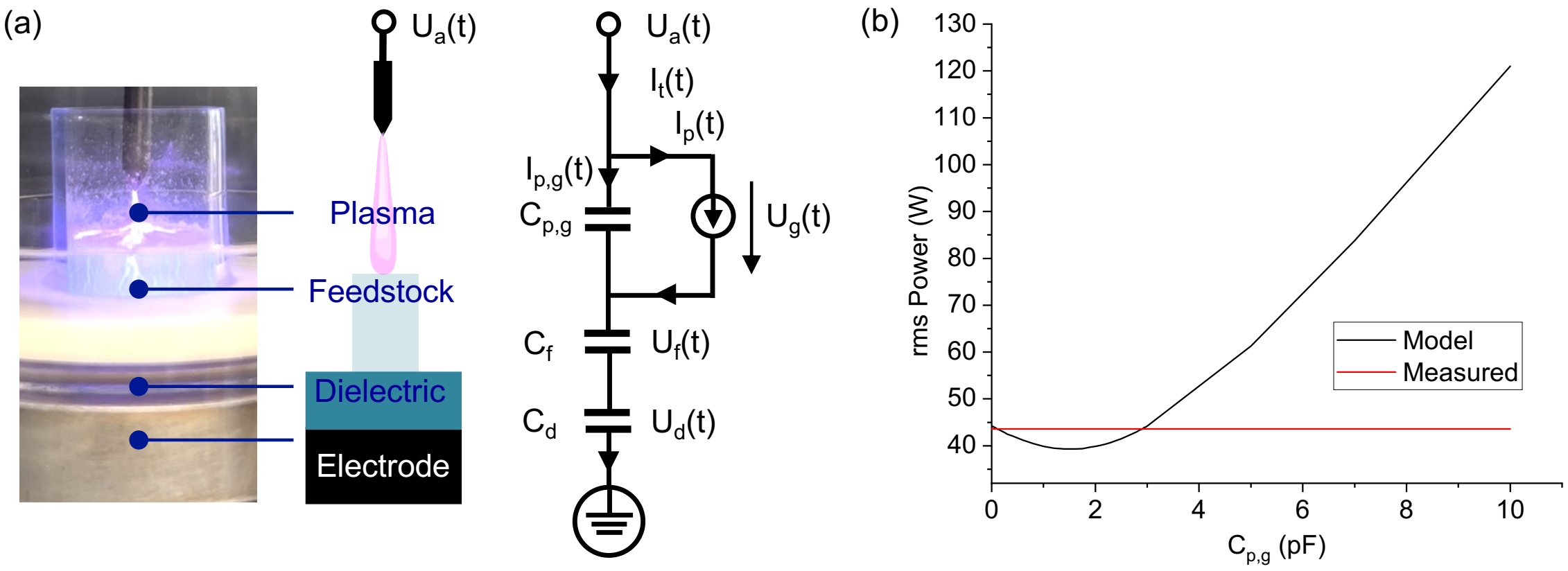

**Fig. 4. SDBD electrical model.** (a) Operation of the reactor during cellulose treatment and equivalent circuit diagram, with the feedstock and main reactor components represented as parallel plate capacitors. (b) Measured input power and the model as functions of capacitance plasma-gap capacitance $C_{p,g}$.

## 3.4 Spectroscopic diagnostics

Optical Emission Spectroscopy (OES) is used to determine the primary characteristics of the SDBD plasma, namely, representative values of excitation temperature $T_{exc}$, electron temperature $T_e$, and electron number density $n_e$. A spectrometer (Avantes ULS2048-USB2) with a wavelength range of 200-1100 nm and a grating of 300 lines/mm is used to measure the spectral emission from the plasma. The measurements are performed near the beginning ($t$ = 15 s) of the feedstock treatment. A representative spectrum is shown in Fig. 5a obtained during the treatment of cellulose feedstock under operational conditions of voltage level $V$ = 60% ($P_{t,rms}$ = 50 W), flow rate $Q$ = 0.01 slpm, and electrode-feedstock spacing $h_p$ = 5 mm. The peaks represent the relative intensity of radiative transitions, from some upper energy level $i$ to a lower energy level $j$, with the wavelengths of primary transitions used for the analysis indicated within the figure.

The estimation of plasma properties from OES data in the present work is based on the use of the Boltzmann plot method, which has been extensively used for determining $T_e$ and $T_{exc}$ in a wide range of plasmas [33–36]. The method is based on comparing the relative intensity of representative thermometric species. The Boltzmann plot method assumes local thermal equilibrium in which the excitation and de-excitation mechanism is controlled by the electronic collisions, and both $T_e$ and $T_{exc}$ are the same.

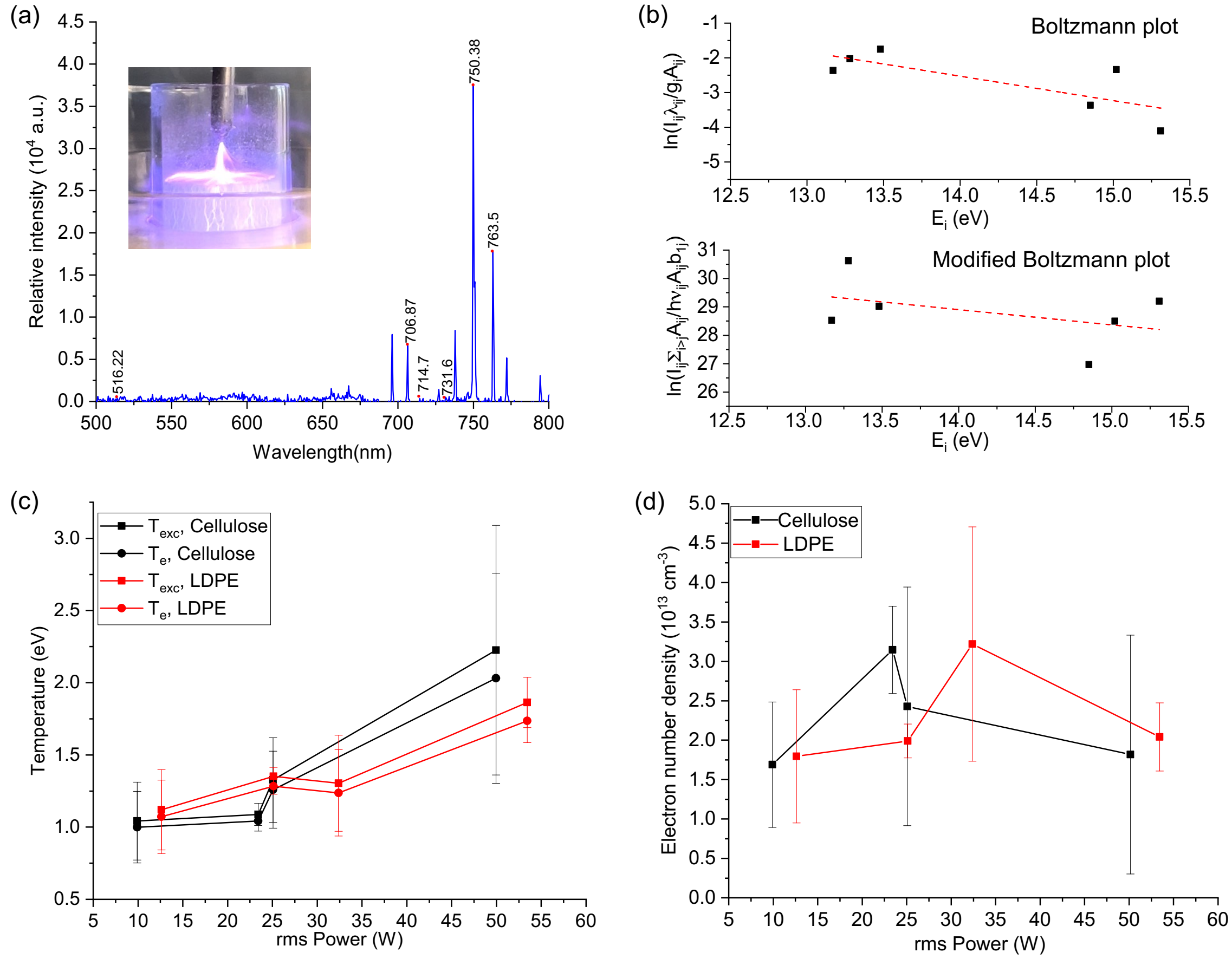

**Fig. 5. Spectroscopic characterization of SDBD plasma.** (a) Plasma spectra at the beginning of treatment of cellulose (t = 0.5 min) and representative Ar I spectral lines and corresponding (b) Boltzmann plot for the determination of the excitation temperature ($T_{exc}$ = 1.42 eV) and modified Boltzmann plot for calculation of electron temperature ($T_e$ = 1.35 eV). (c) Excitation and electron temperatures and (d) electron number density $n_e$ as a function of power during the treatment of cellulose and LDPE.

The excitation temperature $T_{exc}$ is calculated from the slope of the best fit of a Boltzmann distribution of excited states with the upper energy level *i* and lower energy level *j*, leading to the expression [33]:

$$ln\left(\frac{I_{ij}\lambda_{ij}}{g_i A_{ij}}\right) = -\frac{E_i}{k_B T_{exc}} + D, \tag{11}$$

where $g_i$ is the statistical weights of the upper level $i$ of the transition considered, $A_{ij}$ is the transition probability of the emitted spectra, $I_{ij}$ is the relative intensity of the spectral emission for upper to lower states, $\lambda_{ij}$ is the wavelength of the emitted spectra, $E_i$ is the excitation energy, $k_B$ is the Boltzmann constant, $T_{exc}$ is excitation temperature in electron volt (eV), and $D$ is a data-fitting constant.

Since the SDBD plasma is in a state of thermal nonequilibrium, i.e., different electron temperature from the heavy-species temperature – an intrinsic characteristic of nonthermal plasma, electronic collisions might not be the only processes controlling the excitation and de-excitation mechanism [33]. To estimate the electron temperature $T_e$, the modified Boltzmann plot method developed by Gordillo *et al.* [33] and used by several authors [34-36] is adopted. This approach assumes that the plasma is in the state of corona balance in which the populating and depopulating mechanisms are attributed to electron-impact collisional excitation from the ground state and spontaneous radiative emission, respectively, and that the two mechanisms are balanced. Gordillo *et al.* [33] modified Boltzmann plot method leads to the equation:

$$ln\left(\frac{I_{ij}\sum_{i>j}A_{ij}}{h_P v_{ij} A_{ij} b_{1i}}\right) = -\frac{E_i}{k_B T_e} + B, \qquad (12)$$

where $h_P v_{ij}$ is the energy gap between levels $i$ and $j$ (with $h_P$ as the Planck constant and $v_{ij}$ the collision frequency between species), $b_{1i}$ is a constant function of the electron-impact excitation rate coefficient, $\sum_{i>j}A_{ij}$ is the summation of the transition probabilities starting from the upper energy level $i$, and $B$ is a data-fitting constant. The quantity $\sum_{i>j}A_{ij}$ is determined by considering all the possible spontaneous radiative transitions from the upper energy levels associated with the measured lines and then summing up their respective transition probabilities [33]. The parameters $I_{ij}$, $A_{ij}$, and $E_i$ are obtained from NIST [37] atomic spectra database.

The excitation temperature $T_{exc}$ and the electron temperature $T_e$ are determined from the slope of the least-squares fit of equation (11) and equation (12), respectively, for a given number of spectral lines (radiative transitions). In this study, six Ar-I lines corresponding to wavelengths: 516.2 nm (electronic transition 6d→4p), 706.9 nm (6s→4p), 714.7 nm (4p→4s), 731.6 nm (6p→4s), 750.4 nm (4p→4s), and 763.5 nm (4p→4s) are chosen, as indicated in Fig. 5a. The lines are chosen such that they have the greatest gap between the upper energy levels (levels *i*) at the expense of considering higher relative intensities. This reduces the error in the estimation of electron excitation and electron temperature resulting from smaller differences (of 1 eV or less) between the upper energy levels of the transitions [33]. The Boltzmann plot and modified Boltzmann plot for the determination of $T_{exc}$ and $T_e$, respectively, are presented in Fig 5b for operating conditions: voltage level $V$ = 60% ($P_{t,rms}$ = 50 W), $Q$ = 0.01 slpm, and $h_p$ = 5 mm. The least-squares linear fitting is applied to obtain the slopes for each plot, leading to $T_{exc}$ = 1.42 eV and $T_e$ = 1.35 eV. Three different spectral measurements are performed, and the average $T_{exc}$ and $T_{e,}$ as well as the error, are determined and presented in Fig. 5c.

The dependence of $T_{exc}$ and $T_e$ with $P_{t,rms}$ during the treatment of cellulose and LDPE are shown in Fig. 5c. Both $T_{exc}$ and $T_e$ increase monotonically with $P_{t,rms}$ as the result of increased electron energy [36]. As $P_{t,rms}$ increases, the electrons gain more energy leading to the generation of a greater number of active species through inelastic electron-molecular species collisions [38]. The small difference in $T_{exc}$ and $T_e$ observed in Fig. 5c is consistent with reports by other authors [33-36]. The results in Fig. 5c show that $T_{exc}$ and $T_e$ in the treatment of both cellulose and LDPE are similar, implying comparable plasma conditions and that the type of feedstock has a relatively minor role in the electrical characteristics of the system.

The electron number density $n_e$ of the SDBD plasma is estimated using the approach by Kais *et al*. [34]. Their approach provides an expression for $n_e$ relating the sheath potential $V_{sh}$, the ionization energy $E_{ion}$ of the gas (15.7 eV in the case of argon), and the electron temperature according to:

$$n_e = \frac{P_{t,rms}}{A_s}\left( \begin{array}{c} \left(\frac{k_B T_e}{2\pi m_e}\right)^{\frac{1}{2}} \exp\left(\frac{eV_{sh}}{k_B T_e}\right)(2k_B T_e + E_{ion}) \\ +0.3 k_B T_e \left(\frac{k_B T_e}{m_i}\right)^{\frac{1}{2}} \left(\frac{k_B T_e}{2}\right) \left| \ln\left(\frac{2\pi m_e}{m_i}\right) + 1 \right| \end{array} \right)^{-1} \tag{13}$$

where $A_s$ is the substrate cross-sectional area, $e$ the elementary charge, $m_e$ electron mass, and $m_i$ the ion mass. Consistent with derivation leading to equation (13), the sheath potential $V_{sh}$ can be determined using the expression [34]:

$$V_{sh} = \left(\frac{k_B T_e}{2e}\right) \ln\left(\frac{m_i}{2\pi m_e}\right). \tag{14}$$

The electron number density for an electron temperature of 1.35 eV and $P_{t,rms}$ = 50 W of the SDBD plasma used in the production of hydrogen from cellulose is $1.82\times10^{13}$ cm$^{-3}$. The dependence of the average electron number density $n_e$ as a function of power $P_{t,rms}$ during the treatment of cellulose and LDPE is shown in Fig. 5d. The results show that $n_e$ remains approximately constant with varying power, with the lowest value of $1.69\times10^{13}$ cm$^{-3}$ corresponding to $P_{t,rms}$ = 10 W during the treatment of cellulose. The maximum $n_e$ of $3.33\times10^{13}$ cm$^{-3}$ is attained during the treatment of LDPE with $P_{t,rms}$ = 32 W. The estimated $n_e$ range is comparable with that reported for other nonthermal plasma, typically in the order of $10^9$ to $10^{13}$ $cm^{-3}$ [33–35, 39].

## 3.5 Hydrogen production from Cellulose

### 3.5.1 SDBD plasma-feedstock interaction

The interaction between SDBD plasma and cellulose at the end of each treatment is depicted in Fig. 6. The treatment process initially leads to the emission a mainly purple glow characteristic of argon plasma (as shown in Fig. 5a). As the plasma treatment progresses, the purple glow transitions to yellow for $t = 0.5$ min. Moon *et al*. [21, 40] noted that devolatilization and char reaction are the main regimes during biomass gasification. Therefore, the glow transition observed for $t = 0.5$ min probably indicates the beginning of carbonization. The rapid devolatilization at the beginning is probably attributed to the weak hydrogen bond (17-30 kJ/mol)[41] of the cellulose feedstock.

The images in Fig. 6 also show the production of fine particles depicted by the clouding of the reactor chamber (smoke), particularly noticeable at $t = 1$ min. The intensity of the smoke and yellow emissions increases with time for $t = 1$ to 4 min, then it decreases. The increase in smoke and yellow emissions is probably attributed to the dominance of devolatilization of the feedstock and the presence of oxygen atoms, respectively. However, the shift of devolatilization to char reaction could be responsible for the decreased smoke intensity. The red glow observed after $t = 5$ min probably depicts the emission from carbon particles.

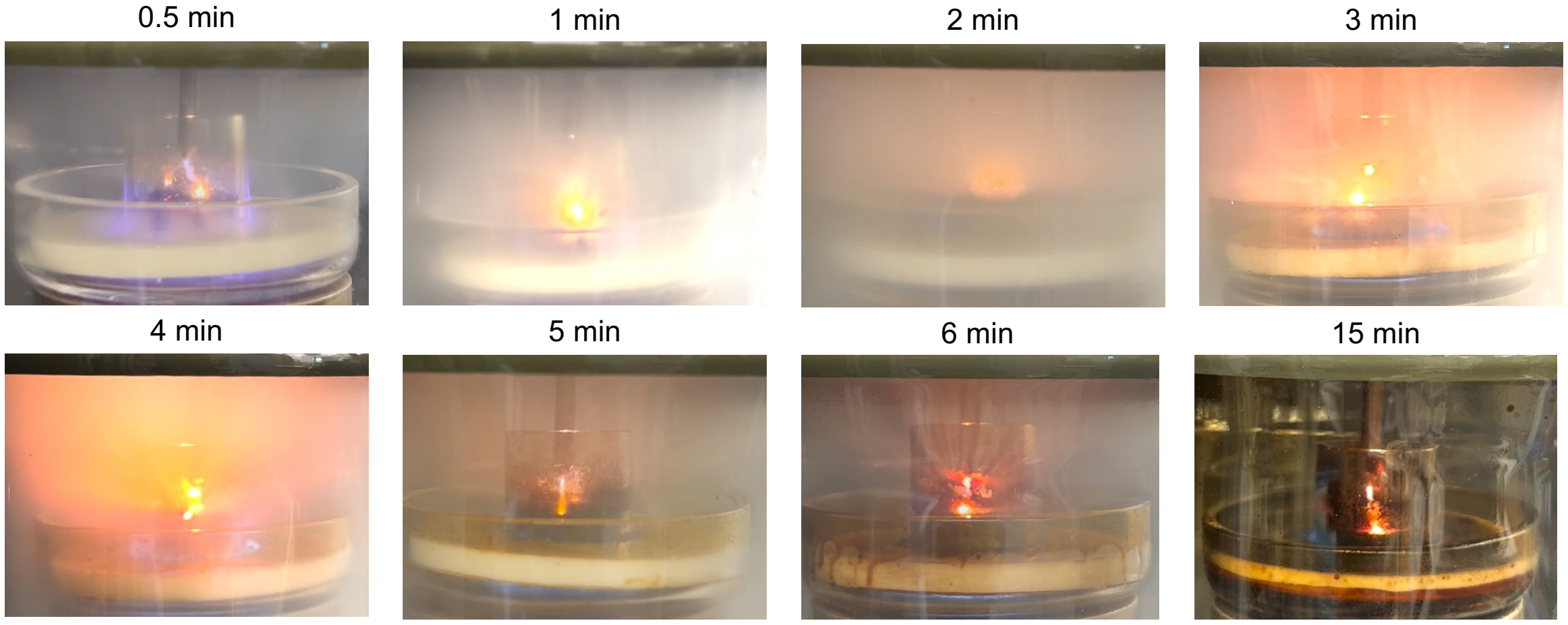

**Fig. 6. Optical imaging of plasma treatment of cellulose**. SDBD plasma interaction with cellulose at different times under operation conditions of $V$ = 60% ($P_{t,rms}$ = 50 W) and $Q$ = 0.01 slpm.

### 3.5.2 Hydrogen production and production efficiency

The performance of the SDBD reactor during the production of hydrogen from cellulose (cellulose dehydrogenation or carbonization) is next assessed. The set-up (Fig. 3) is operated at near atmospheric pressure. The temperature of the reactor chamber is measured using an infrared thermometer, ranging from 21 to 66 °C, depending on the time with a treatment (between0.5 and 15 min), confirming the relatively low temperature of the process.

The reactor's performance parameters considered are *Cumulative* $H_2$ *production*, $H_2$ *Production rate*, $H_2$ *Production efficiency*, and *Energy cost of* $H_2$ *production*. The *Cumulative* $H_2$ *production* quantifies the total amount of hydrogen collected in the sampling bag and analyzed via GC, and $H_2$ *Production rate* = *Cumulative* $H_2$ *production*/$t_{reatment}$. $H_2$ *production efficiency* (in units of mol/kWh) is defined as

$$H_2 \ Production \ efficiency = \frac{H_2 \ Production \ rate}{Input \ rms \ power} \tag{15}$$

and the *Energy cost of $H_2$ production* (kWh/kg $H_2$) is determined from

$$Energy \ cost \ of \ H_2 \ production = \frac{1000}{M_w(H_2 \ Production \ efficiency)}, \tag{16}$$

where $M_w$ is the molecular weight of hydrogen and 1000 is the conversion factor for grams to kilograms.

The obtained performance of the SDBD reactor as a function of treatment time is depicted in Fig. 7. The *Cumulative $H_2$ production*, namely the total amount of hydrogen produced for each $t_{treatment}$, is presented in Fig 7a. The *Cumulative $H_2$ production* rapidly increases for $t_{treatment} < 3$ min of and then depicts a significantly slower increase. As observed by Sun *et al*. [40] and indicated in section 3.1, the gas release process from biomass gasification occurs mainly in two regimes: devolatilization and char reaction. Devolatilization is the main process of gas release in cellulose gasification since cellulose is composed of largely volatile components. This can explain the sharp rise in cumulative hydrogen production in the first 3 minutes of treatment, which is comparable to the behavior observed by Moon *et al*. [21], who obtained a peak devolatilization time of 4 minutes for biomass steam gasification with an operating temperature of 900 $^{o}$C. The char reaction is a slow process but lasts for a longer time [40], which can explain why the cumulative hydrogen production remains almost constant during the rest of the treatment time. Overall, the trend of cumulative hydrogen production is comparable to hydrogen production from biomass pyrolysis at 900 $^{o}$C obtained by Moon *et al*. [21].

As hypothesized in section 2, the hydrogen production rate increases with treatment time to a maximum, reaching 40 mmol/h in 3 minutes of treatment before decreasing and eventually remaining almost constant, as illustrated in Fig. 7b. The rapid increase in hydrogen production rate is probably attributed to the weak hydrogen bond of cellulose. Moon *et al*. [21] observed similar behavior during the transient production of hydrogen from biomass via pyrolysis at 600 °C with peak hydrogen production rate of about 800 mmol/h. The occurrence of the peak production suggests that the plasma treatment of cellulose leading to syngas production is a two-stage process of devolatilization and char reaction, as noted by other authors [21], [40], [42]. The devolatilization, which accounts for the primary gas release process, dominates during the first 3 minutes of treatment. However, the char reaction becomes more significant as the cellulose treatment progresses, leading to decreasing hydrogen production rate. The peak hydrogen production rate at 3 minutes indicates the maximized synergistic effect of both devolatilization and char reaction, as noted in [21].

Hydrogen production efficiency quantifies the amount of energy required to produce a unit quantity of hydrogen. As shown in Fig. 3c, the *$H_2$ Production efficiency* increases rapidly to a maximum of 0.8 mol/kWh in 3 minutes and then starts decreasing. This efficiency is an order of magnitude smaller than the 2.1 mol/kWh obtained by Wu *et al*. [43] in the plasma reforming of n-pentane via DBD. The occurrence of maximum efficiency is also a manifestation of the existence of the two regimes of devolatilization and char reaction of biomass pyrolysis, as reported in [42], in which devolatilization dominates the initial stage of the process and char reaction becomes more pronounced after the peak efficiency is reached. The peak production efficiency, similar to the peak hydrogen production, is attributed to the synergistic effect of devolatilization and char reaction.

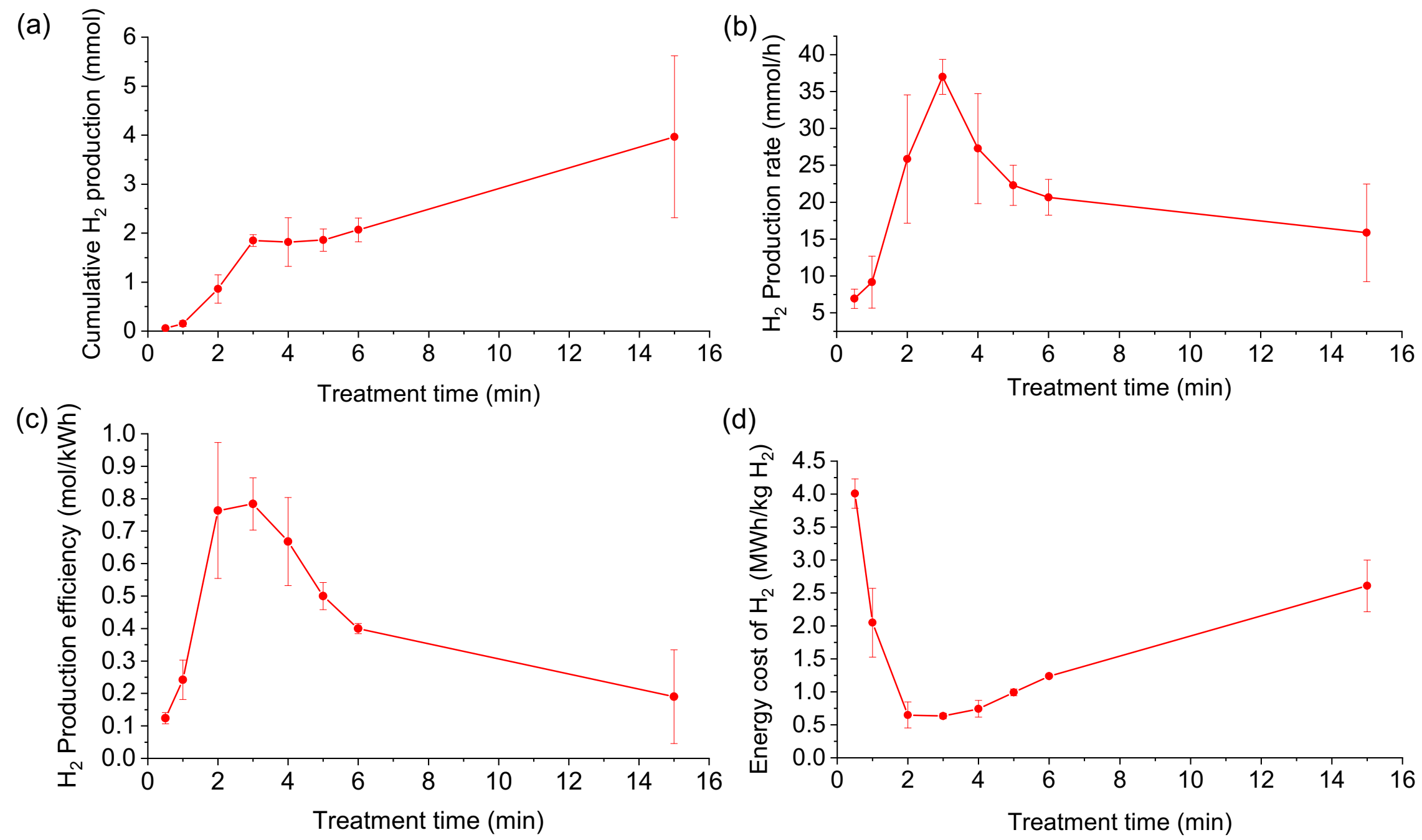

**Fig. 7. Hydrogen production performance of the SDBD plasma treatment of cellulose.** (a) Cumulative hydrogen production, (b) hydrogen production rate, (c) hydrogen production efficiency, and (d) energy cost of hydrogen production versus treatment time.

The *Energy cost of $H_2$ production*, defined as the amount of electrical energy required to produce 1 kg of hydrogen, is presented in Fig. 7d. The *Energy cost of $H_2$ production* decreases rapidly to a minimum of 630 kWh/kg $H_2$ for $t_{treatment}$ = 3, consistent with the maximum *$H_2$ Production efficiency*, as well as the maximum *$H_2$ Production rate*. Comparably, the energy cost of hydrogen production is about two orders of magnitude greater than for water electrolysis (41.6 kWh/kg) and methane steam reforming (21.9 kWh/kg). The higher energy cost of producing hydrogen from cellulose is expected, given the greater embedding of hydrogen within a polymeric solid. Nevertheless, the plasma valorization of solid feedstock such as cellulose, representative of biomass waste, could lead to environmental and economic benefits.

### 3.5.3 Gas product yield and hydrogen selectivity

To assess the selectivity of the production of hydrogen from cellulose via SDBD plasma, the area of the main gas products detected by gas chromatography is analyzed for the representative treatment times of $t_{treatment}$ = 3 min and $t_{treatment}$ = 15 min. The main gas products obtained during the plasma treatment of cellulose are $H_2$, $CO$, $CO_2$, $CH_4$, $C_2H_4$, and $C_2H_6$. The 3-minute treatment depicts the greatest hydrogen production efficiency and minimum energy cost (Fig. 7), whereas the 15-minute treatment depicts the greater charring (carbonization) of the feedstock.

*Gas product yield* is defined as the amount of gas product produced per gram of cellulose during the plasma treatment, and it is quantified by,

$$\text{Gas product yield} = \frac{\text{Moles of gas product}}{\text{Total mass of feedstock}}. \qquad (17)$$

Similarly, *Selectivity* is derived from *Gas product yield* as

$$\text{Selectivity} = \frac{\text{Gas product yield}}{\text{Total gas product yield}}. \qquad (18)$$

The *Gas product yield* for the plasma treatment of cellulose under the two treatment times of 3 and 15 min is presented in Fig. 8a. The hydrogen yield for $t_{treatment}$ = 3 min and 15 min is 1.8 and 4.0 mmol/g of cellulose, respectively, and it is significantly higher than the yield of other gas products. These results are comparable to the electrolytic pyrolysis of biomass. For instance, Zeng *et al*. [20] obtained a maximum hydrogen yield of 3.1 mmol/g at 650 $^oC$. Wei *et al*. [19] experimentally pyrolyzed cellulose in molten carbonate obtaining a maximum hydrogen yield of 8.3 mmol/g of cellulose at 600 $^oC$. Although thermochemical pyrolysis generally depicts significantly higher hydrogen yield, typically of 22 to 128 mmol/g cellulose, as reported by [24, 44–46], the SDBD plasma process operates under significantly lower temperatures (< 200 $^{o}$C)

The yields of CO and $CO_2$ are the second and third largest, respectively, for both treatment times. This is attributed to the large amount of oxygen atoms present in cellulose. The higher yield of CO, slightly greater for $t_{treatment}$ = 15 min than for $t_{treatment}$ = 3 min, is caused by the partial oxidation reaction of cellulose [47]. Light hydrocarbons, i.e., $CH_4$, $C_2H_4$, and $C_2H_6$ are also produced by the process, but with significantly smaller yield. The overall yield trend of $H_2 > CO > CO_2$ observed in the SDBD plasma process has also been observed by Du *et al*. [47] during the gasification of corn cob via nonthermal plasma.

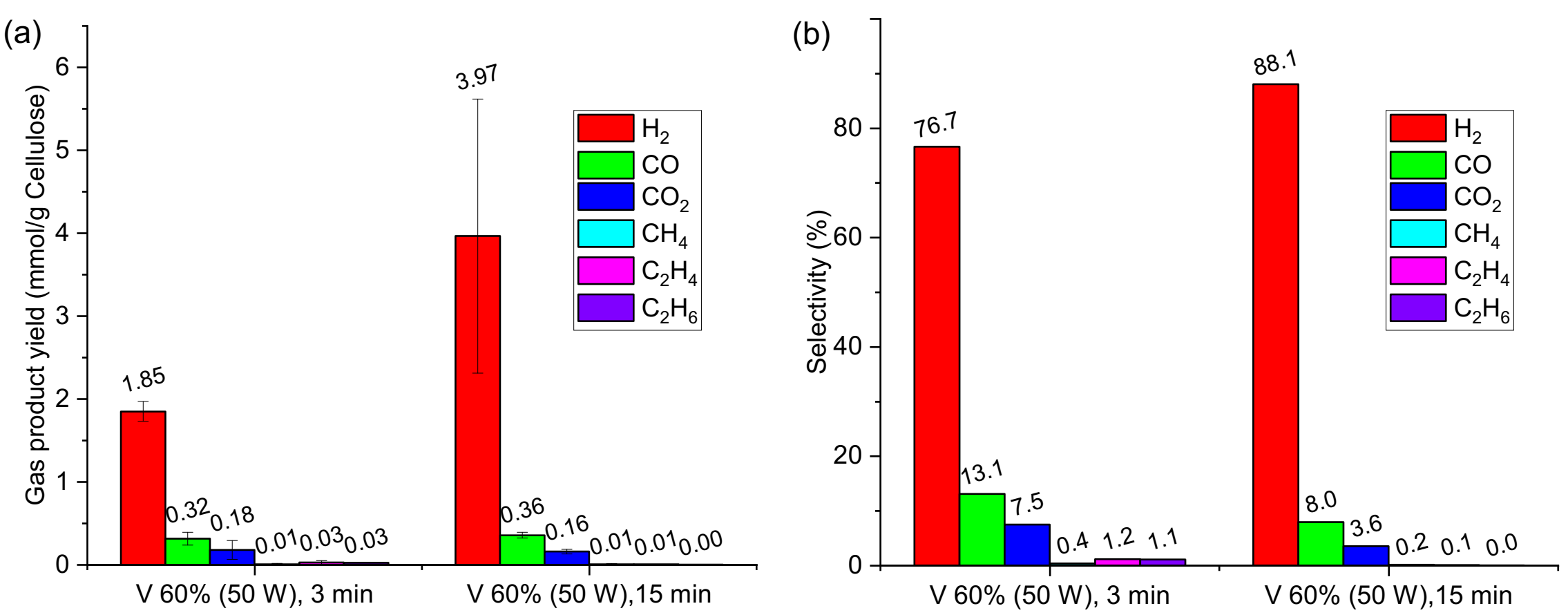

**Fig. 8. Gas product yield and selectivity.** (a) Yield and (b) selectivity of different gas products generated during the SDBD plasma treatment of cellulose.

*Selectivity* quantifies the relative yield of a gas product compared to the total gas products and is presented in Fig. 8b. Hydrogen has the greatest selectivity of 76.7% and 88.1% for $t_{treatment}$ = 3 and 15 min, respectively. The greater hydrogen selectivity for $t_{treatment}$ = 15 min suggests that some of the lower molecular weight hydrocarbons are further decomposed into hydrogen and carbon. Several studies on the pyrolysis of biomass reported hydrogen selectivity. Du *et al.*[47] gasified corn cob using nonthermal plasma and reported hydrogen selectivity of about 60%. Wu *et al*. [44] catalytically pyrolyzed cellulose for hydrogen production using nickel-based catalysts (Ni-Zn-Al, 1:1) and obtained a selectivity of about 55 vol.%. Zsinka *et al*. [48] reported the highest hydrogen selectivity of 19% during the pyrolysis of biomass waste using modified nickel catalysts at 800 $^oC$. Turn *et al*. [45] and Zeng e*t al*. [20] reported hydrogen selectivity of 57 and 26 vol.%, respectively. The results obtained in the present work indicate that the treatment of cellulose by SDBD plasma can lead to greater hydrogen selectivity than thermochemical and thermo-catalytic approaches.

### 3.5.4 Solid products characterization

Images of the cellulose samples before and after plasma treatment as a function of time are presented in Fig. 9a. The pristine white sample ($t_{treatment}$ = 0 min) starts charring almost right from the start of the treatment and becomes more pronounced as the treatment time increases, transforming almost completely into char for a 15-minute treatment. The rapid charring is probably owing to the weak hydrogen bond. The non-uniform treatment of the sample is clearly observed at $t_{treatment}$ = 0.5 and 1 min, probably due to the porous nature of the cellulose feedstock (i.e., compacted powder). The porosity of cellulose creates less-resistive electrical paths leading to a localized concentration of discharge filaments. However, as the treatment time increases, this effect becomes less significant.

Results of CHN-elemental analysis of the samples (by Midwest Microlab) are presented in Fig. 9b. The carbon content and carbon-to-hydrogen ratio (C/H) of the solid sample increases with treatment time to a maximum of 85.6% (for $t_{treatment}$ = 6 min) and 26.6% (for $t_{treatment}$ = 15min), respectively. This indicates a significant extent of carbonization, as noted by Nanda *et al*. [49], who obtained 76.4% of carbon content in the catalytic gasification of wheat straw in hot compressed water for hydrogen production. The sudden variation of both carbon content and C/H ratio indicates the non-uniformity of the plasma treatment of cellulose, which is also evidenced by the images in Fig. 9a.

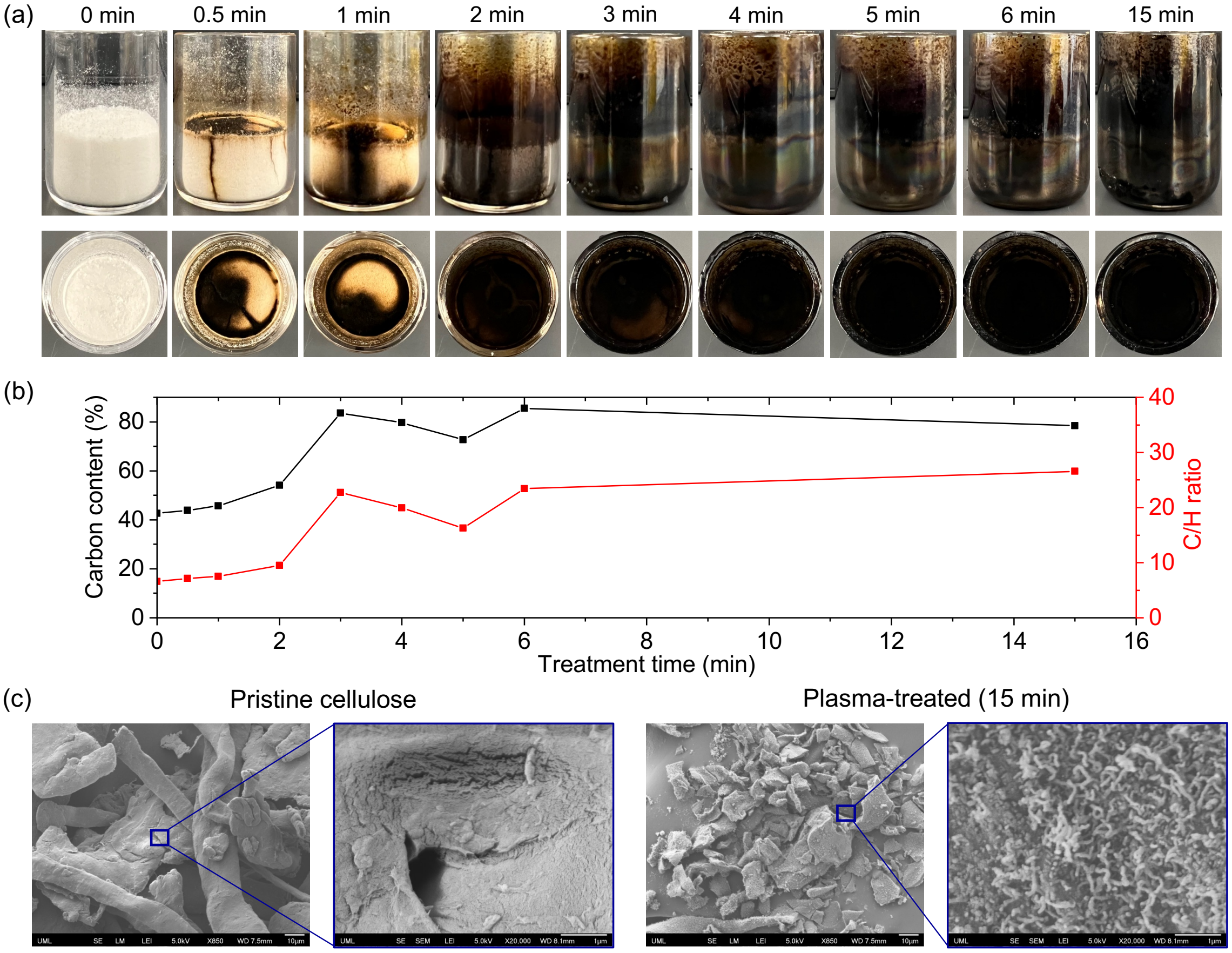

**Fig. 9. Characterization of treated cellulose feedstock**. (a) Optical images of cellulose samples before treatment ($t_{treatment}$ = 0 min) and at different treatment times. (b) Elemental characterization of cellulose as function of treatment time. (c) Surface morphological characteristics of pristine and plasma-treated cellulose samples obtained using FESEM under low- (850×)) and high-(20000×)) magnification.

Based on the optical observation of the solid residue with the greatest charring, the solid residue for $t_{treatment}$ = 15 min is selected for FESEM imaging together with the pristine sample. Representative imagining results are presented in Fig. 9c. The pristine sample consists of entanglements of cellulose fibers with empty spaces between fibers, as shown in the low-magnification images in Fig. 9c. The average width of the fibers is 100 $\mu$m. It is the crisscrossed network of fibers with empty spaces in-between fibers that lead to the high porosity of cellulose. The higher magnification image reveals that each fiber is made of bundled and indistinguishable fibrils, as observed by Du *et al.* [47]. The existence of pores creates the least resistive path for electric current, which probably accounts for the non-uniform treatment of the feedstock. Furthermore, the porous nature of cellulose is, in part, responsible for cellulose's weaker dielectric strength, which leads to lower power consumption and higher hydrogen production efficiency compared to that attained from LDPE treatment (discussed in section 4.4).

The plasma-treated cellulose has fragmented fibers of small pieces, shown in Fig. 9c in the image under low magnification. The higher magnification image shows fragmented fibers that consist of protruded fibrils with an average diameter of 50 nm. These fibrils are loosely tangled and clearly visible and increase the surface area of the remnant solids. Du *et al*. [47] observed a similar structure in gasified corn cob using nonthermal plasma. Also, Zhang *et al*. [50] obtained carbon nanotubes of an average diameter of 50 nm using microwave-assisted chemical vapor deposition of carbon nanotubes on pine nutshell char.

## 3.6 Hydrogen production from low-density polyethylene

### 3.6.1 SDBD plasma-feedstock interaction

The interaction between SDBD plasma and LDPE at the end of each treatment is depicted in Fig. 10. The sequence of images shows that initial violet emission from the plasma changes to yellow and eventually to red towards the end of 15 minutes of treatment. The yellow emission depicts the presence of oxygen attributed to the residual air in the reaction chamber and oxygen admixture during the pre-melting of the LDPE sample preparation. The presence of carbon particles is likely responsible for the red emission.

The emission of particulate matter (smoke) observed after 1 minute of operation can probably be attributed to the beginning of charring/carbonization of the feedstock. Unlike cellulose, the intensity of the smoke during the treatment of LDPE remains relatively low in the first 6 minutes, depicting lower devolatilization. The presence of smoke is more pronounced as the treatment progresses. Gunasee *et al* [51] noted that the devolatilization of cellulose occurs faster and at lower temperatures than that of LDPE. Such behavior is also observed in the present study. The slow devolatilization is probably attributed to the stronger carbon-hydrogen bond of 416.7 kJ/mol [52] exhibited by polyethylene. The results for $t$ = 10 and 15 min reveal the presence of waxy deposits inside the reactor chamber, suggesting the formation of hydrocarbons as often observed in the pyrolysis of plastics.

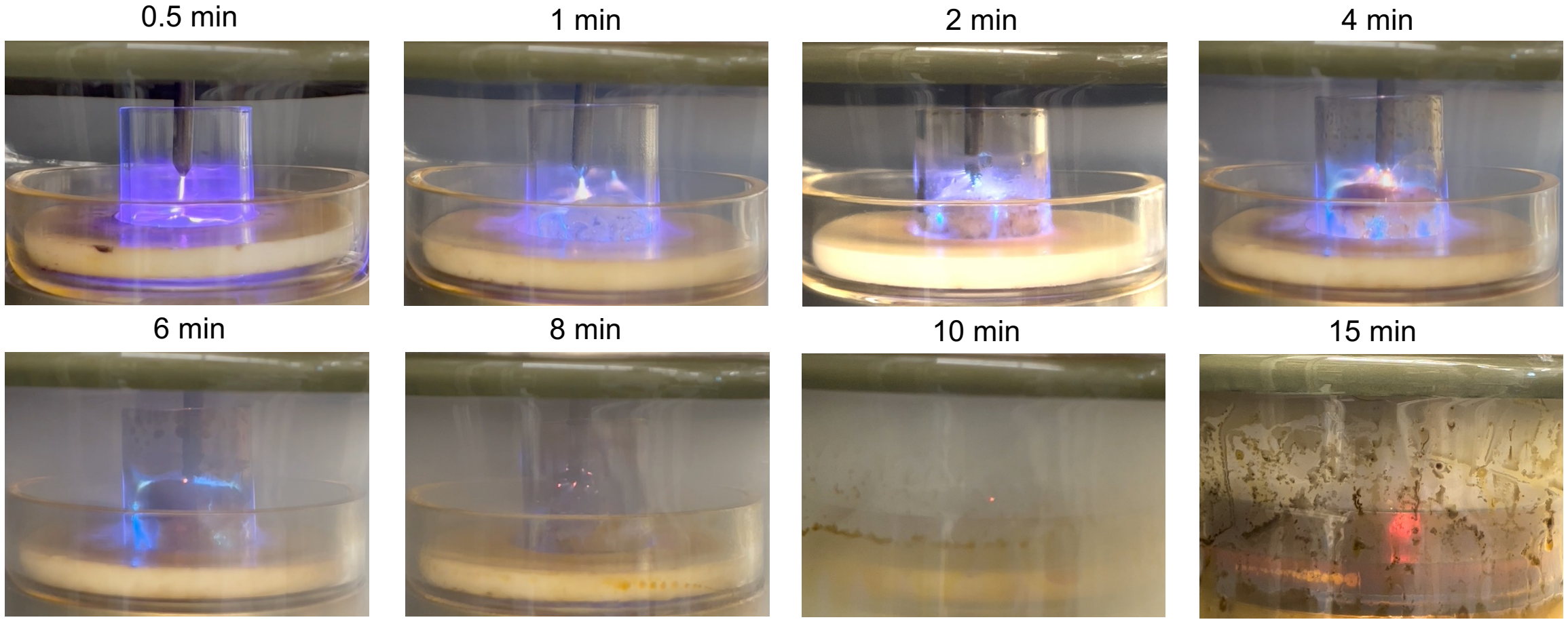

Fig. 10. Optical imaging of plasma treatment of LDPE. SDBD plasma interaction with LDPE for different times under operation conditions of $V$ = 60% ($P_{t,rms}$ = 53 W) and $Q$ = 0.01 slpm.

### 3.6.2 Hydrogen production and production efficiency

Similarly, as done for the treatment of cellulose, the performance of the SDBD reactor to produce hydrogen from LDPE is assessed in terms of *Cumulative $H_2$ production*, *$H_2$ Production rate*, *$H_2$ Production efficiency*, and *Energy cost of $H_2$*. The definition of these metrics is similar to those presented in section 3.2. The *Cumulative $H_2$ production* of LDPE in Fig. 11a increases gradually with treatment time to a maximum of 3.1 mmol for $t_{treatment}$ = 15. Unlike cellulose, no distinct regime between devolatilization and char reaction is observed. The negligible hydrogen production at $t_{treatment}$ = 0.5 min suggests that the plasma power is consumed in melting LDPE. The lower amount of hydrogen produced from LDPE compared to that obtained from cellulose, given the same amount of feedstock used, can be expected given the stronger hydrogen bond within LDPE.

The *$H_2$ Production rate*, shown in Fig. 11b, indicates a rapid increase during the first 2 minutes of treatment to a peak rate of 20 mmol/h and after which it slightly declines. The slight decline in the production rate suggests a small depletion of the amount of hydrogen in the feedstock. The maximum hydrogen production rate is three times greater than what Tabu *et al*. [26] obtained in the experimental extraction of hydrogen LDPE via two different nonthermal plasma processes.

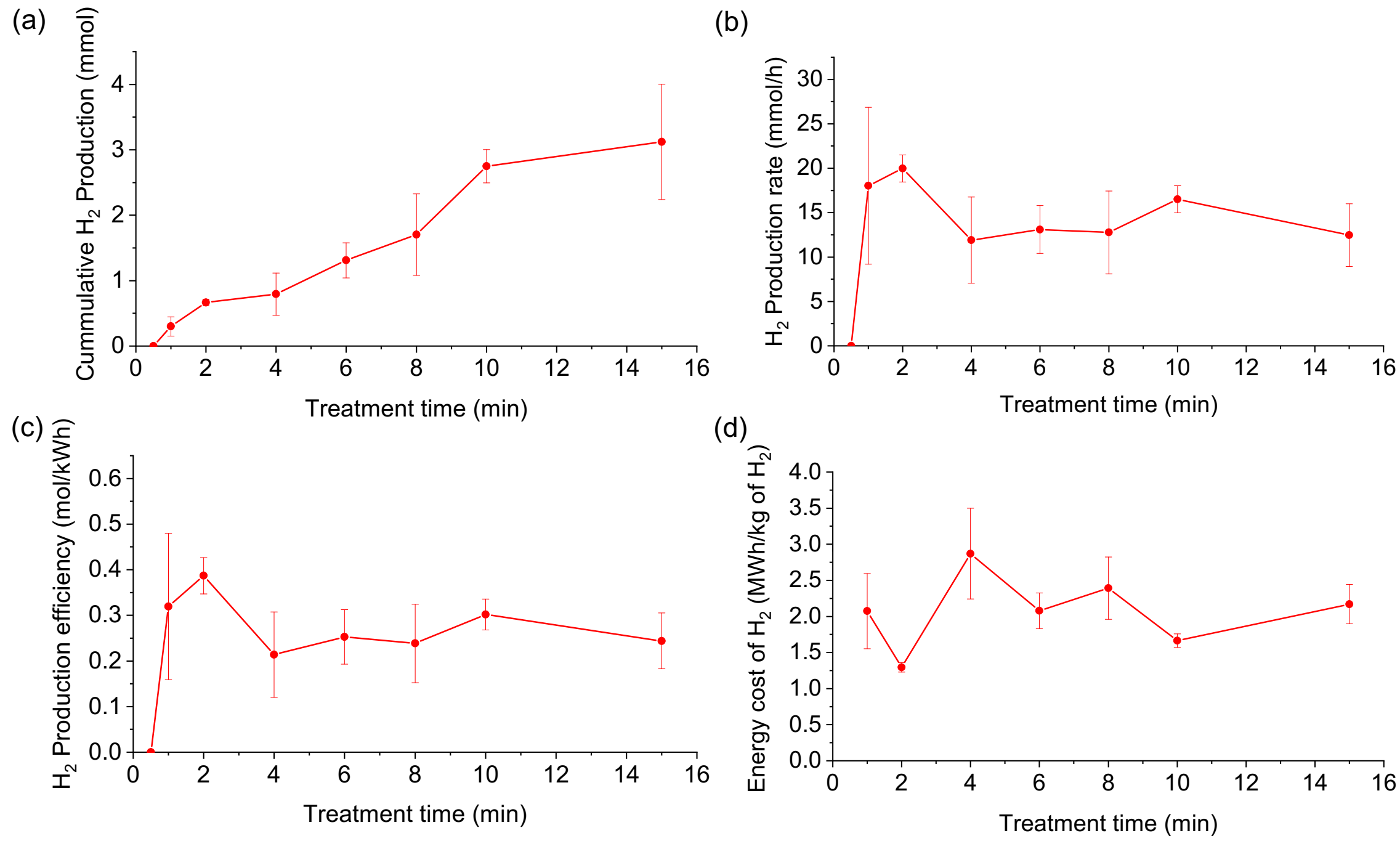

**Fig. 11. Hydrogen production performance of the SDBD plasma treatment of LDPE.** (a) Cumulative hydrogen production, (b) hydrogen production rate, (c) hydrogen production efficiency, and (d) energy cost of hydrogen production as a function of treatment time.

The *$H_2$ Production efficiency* is often used as a metric to evaluate the viability of the process. Following the same trend as the *$H_2$ Production rate* (Fig. 11b), the hydrogen production efficiency, depicted in Fig. 11c, increases rapidly in the first 2 minutes of treatment before decreasing. The peak *$H_2$ Production efficiency* is 0.4 mol/kWh, half that obtained for cellulose. This is probably attributed to the stronger carbon-hydrogen bond (416.7 kJ/mol) in LDPE compared to that of cellulose with weaker hydrogen bonds of 17-30 kJ/mol. The decrease in production efficiency at the later time of treatment implies that the same amount of energy is consumed in extracting hydrogen from the hydrogen-deprived feedstock. The results also suggest that treating solidified pre-melted LDPE is undesirable as it leads to more compact nonporous feedstock with limited surface area. The smaller surface area limits the interaction between reactive species produced from the plasma and the hydrogen and carbon within LDPE, lowering hydrogen production.

The *Energy cost of* $H_2$, depicted in Fig. 11d, represents the amount of energy required to produce one kilogram of hydrogen. As for cellulose, the *Energy cost of* $H_2$ has an inverse relation with $H_2$ *Production efficiency*. It should be noted that the energy cost of hydrogen production for $t_{treatment}$ = 0.5 min is undefined since no quantifiable amount of hydrogen is detected by gas chromatography. The minimum energy cost obtained is 1300 kWh/ kg $H_2$, approximately two times greater than that of cellulose, and about 30 times greater than that for water electrolysis. Also, the energy cost is a 2-factor less expensive than the 3300 kWh/kg $H_2$ obtained in a previous study by the authors [26]. The high energy cost of hydrogen production is attributed to the stronger carbon-hydrogen bonds of LDPE.

### 3.6.3 Gas product yield and selectivity

The *Gas product yield* and *Selectivity* are determined using the same definitions used for cellulose, namely equation (17) and equation (18), respectively. Yield and selectivity are determined for the hydrogen production experiments for $t_{treatment}$ = 4 and 15 min, and they are summarized in Fig 12. The gas species identified and quantified are $H_2$, $CO$, $CO_2$, $CH_4$, $C_2H_4$, and $C_2H_6$. Given that LPDE does not contain oxygen, the presence of CO and $CO_2$ in the gas products is ascribed to residual air in the reactor and oxygen admixture during the pre-melting, a process of LDPE sample preparation (section 2.2). The maximum hydrogen yield of 3.12 mmol/g LDPE obtained in the 15-minute treatment of LDPE is comparable to what has been reported by other authors. Aminu *et al*. [23] studied the plasma catalytic steam reforming of high-density polyethylene (HDPE) and obtained hydrogen yield of about 4.5 mmol/g of HDPE. Nguyen and Carreon [25] reported hydrogen yield of 8 mmol/g of HDPE from a catalytic deconstruction of HDPE via nonthermal plasma. Alvarez *et al*. [14], in the catalytic pyrolysis of biomass and plastic, reported a hydrogen yield of 25.5 mmol/g HDPE.

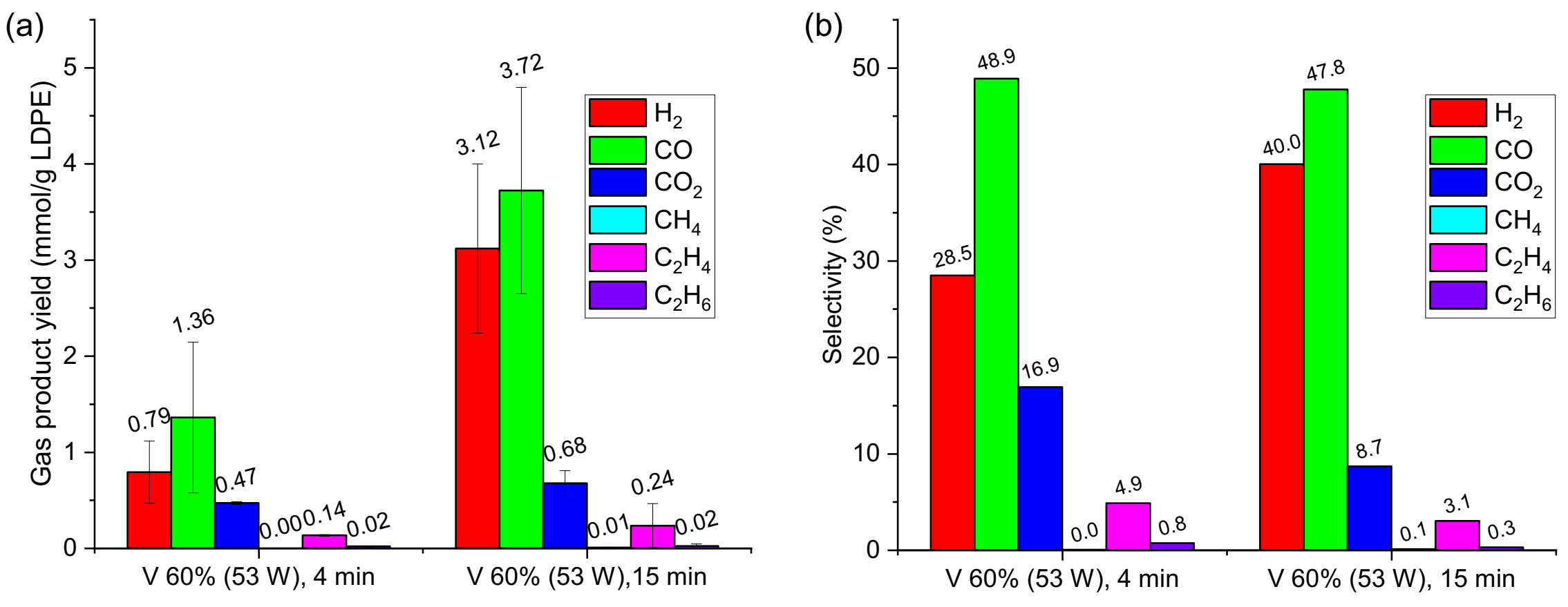

**Fig. 12. Gas product yield and selectivity**. (a) Yield and (b) selectivity of different gas products generated during the SDBD plasma treatment of LDPE.

The *Selectivity* of the different gas species produced is presented in Fig. 12b. The selectivity of hydrogen is 28.5% and 40.0% for $t_{treatment}$ = 4 and 15 min, respectively. The low hydrogen selectivity is attributed to the presence of residual air in the reactor chamber and oxygen admixture during the pre-melting of the LDPE sample leading to the formation of CO and $CO_2$. The highest yield of CO compared to $CO_2$ is probably attributed to the partial oxidation of LDPE in the presence of oxygen. The obtained selectivity is significantly lower than in other plasma-based processes. Nguyen and Carreon [25] obtained hydrogen selectivity of 50% in the catalytic deconstruction of HDPE via nonthermal plasma, whereas Farooq *et al*. [53] reported a hydrogen selectivity of 76 vol% in the catalytic pyrolysis of LDPE.

### 3.6.4 Solid products characterization

Results of the characterization of the LDPE samples before and after SDBD plasma treatment is summarized in Fig 13. As depicted in Fig. 13a, the white color of the pristine compact LDPE sample is transformed into brown as the treatment progresses and eventually to black for $t_{treatment}$ = 10 and 15 min. Contrary to cellulose, the plasma treatment of LDPE is more uniform, as depicted by an evenly-distributed browning of the sample for $t_{treatment}$ = 1 and 2 min. The solidified molten LDPE is denser and nonporous hence providing stronger dielectric resistance, which leads to the spreading of the plasma over the surface instead of leading to a localized electrical discharge. The light brown sample at $t_{treatment}$ = 1 min probably depicts the beginning of carbonization, whereas the blackening is attributed to charring.

The elemental characterization of the solids shows that the carbon content and C/H ratio vary minimally with treatment time, as shown in Fig. 13b. The unnoticeable increase in C/H depicts a very low degree of carbonization, and this suggests the treatment of solidified pre-melted LDPE is undesirable. The solidified pre-melted LDPE is associated with higher dielectric strength and limited reaction surface area, which require more energy and longer for a greater degree of carbonization.

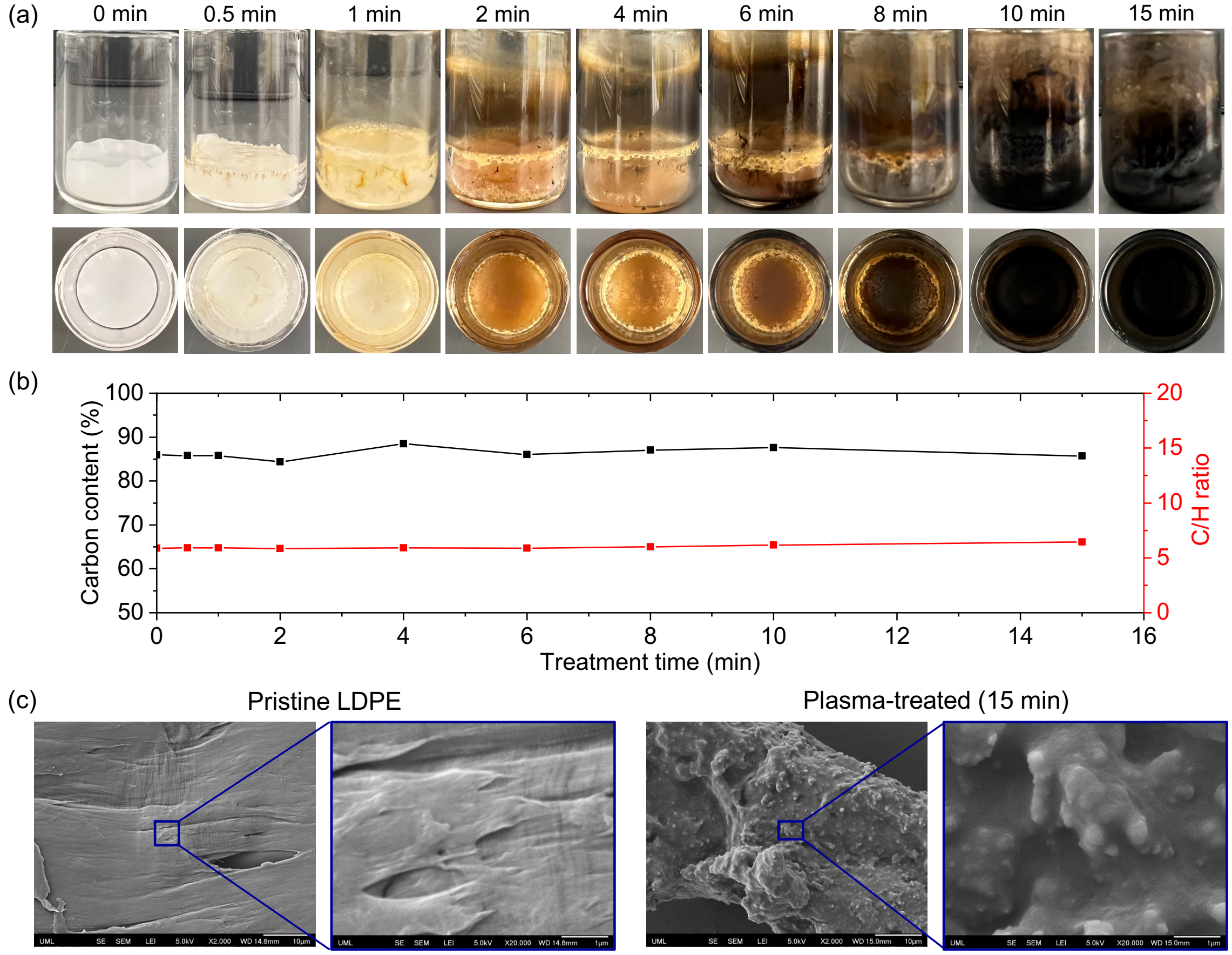

**Fig. 13. Characterization of treated LDPE feedstock.** (a) Optical images of LDPE samples before treatment (t = 0 min) and at different times. (b) Elemental characterization of LDPE as a function of treatment time. (c) Surface morphological characteristics of pristine and plasma-treated LDPE samples obtained using FESEM under low- (2000×) and high- (20000×) magnification

The pristine LDPE morphology depicted by the FESEM result, presented in Fig. 13c, is a smooth and dense surface under low magnification with no observable pores. However, under high magnification, the surface is less smooth and non-uniform, with some fine peelings, which can be attributed to the mechanical damage during the sample extraction for the FESEM imaging. The denser and nonporous nature of LDPE shown by the microscope imaging appears consistent with its high dielectric strength, leading to a more uniform treatment compared to that for cellulose. Moreover, LDPE's dense and nonporous nature also suggests greater power consumption during the plasma treatment (Fig. 11).

The microscopy images of the plasma-treated LDPE revealed substantive changes in the surface morphology shown in Fig. 13c. The surface is rough and embedded with micro-grains faintly visible under low magnification. It is interesting to note that, under higher magnification, the surface exhibited shallow dimples with micro-grains well embedded within the sample. This is probably attributed to the bombardment of the LDPE sample by the highly energetic electrons and ions during SDBD plasma treatment. The micro-grains are comparable to the carbon nanospheres observed by Kibria and Rashid [54] for the low-temperature synthesis of carbon nanomaterials and those by Panickar *et al*. [55] in the chemical vapor deposition synthesis of carbon spheres, respectively.

## 3.7 Conclusions

The valorization of polymeric solid waste via low-temperature atmospheric pressure plasma could lead to economic and environmental benefits, particularly when powered by renewable electricity. In this study, a Streamer Dielectric-Barrier Discharge (SDBD) reactor is designed and built to extract hydrogen and carbon co-products from cellulose and low-density polyethylene (LDPE) as model feedstocks of biomass and plastic waste, respectively.

Experimental characterization and modeling indicate that the plasma consumes between 90 to 100% of the input power; the electron and excitation temperatures depict approximately the same value, independent of feedstock, and increase with input power up to ~ 1.6 to 2.4 eV; and the electron number density is ~ 2.5 $10^{13}$ $cm^{-3}$ irrespective of the feedstock and input power.

The hydrogen production rate for plasma-treated cellulose and LDPE increases to a maximum, after which it declines. The occurrence of peak production suggests that the plasma treatment of cellulose depicts two regimes, namely devolatilization and char reaction, similarly as observed in other biomass pyrolysis studies. The peak hydrogen production rate for cellulose is 40 mmol/h, which is twice that of LDPE (20 mmol/h). Moreover, the energy costs of hydrogen production for cellulose and LDPE are 600 and 1300 kWh/kg of $H_2$, respectively. The lower energy cost for cellulose is probably owed to its high porosity, leading to weaker dielectric strength that promotes increased hydrogen production at lower input power. Additionally, the hydrogen selectivity of cellulose is about two times more than that of LDPE due to the presence of residual gas in the reaction chamber and oxygen admixture during LDPE sample preparation, resulting in the production of CO and $CO_2$ in addition to $H_2$ and hydrocarbons.

The characterization of solid products via field emission scanning electron microscopy reveals distinct morphological structures of the two feedstocks. Whereas pristine cellulose comprises fibrils bundled into fibers with porous and entangled structures, pristine LDPE is a nonporous, uniform, and dense-structured compound. The plasma-treated cellulose consists of protrusions of an average diameter of 50 nm, while residual LDPE has embedded micro-grains, and both present promising valuable solid residues which need further investigation. The results indicate that the use of SDBD plasma is an effective approach for the production of hydrogen from cellulose and from LDPE at near atmospheric pressure and relatively low-temperature conditions in rapid-response and compact processes.

## Acknowledgments

This work has been supported by the US Army Combat Capabilities Development Command (DEVCOM) Soldier Center Contracting Division through Contract W911QY-20-2-0005. The authors also acknowledge Danya Carla Maree for analyzing gas products from cellulose treatment.

# CHAPTER 4: SUMMARY, CONCLUSIONS, AND RECOMMENDATIONS

## 4.1 Summary and conclusions

The potential of using hydrogen as a sustainable energy carrier is attributed to its high energy density and its utilization without $CO_2$ emissions. Hydrogen is mainly produced by steam-methane reforming and water splitting by electrolysis leaving hydrogen-rich solids such as organic polymeric solids largely unexplored. Approaches based on nonthermal atmospheric pressure plasma powered by renewable electricity could lead to the production of green hydrogen more viably than current approaches, providing a sustainable alternative for upcycling the increasing amount of plastic and biomass waste.

This doctoral research dissertation focuses on the production of hydrogen from solids via atmospheric nonthermal plasma. It's envisioned as the initial step towards upcycling solid waste. Two low-temperature atmospheric pressure plasma reactors based on transferred arc (transarc) and gliding arc (glidarc) discharges are designed, built, and characterized to produce hydrogen from low-density polyethylene (LDPE) as a model plastic waste. The two reactors have complementary characteristics i.e transarc based on pin-to-plate configuration is electrically coupled, which allows direct control of the power delivered on the feedstock, and glidarc is electrically decoupled and can be used for the treatment of a stream of feedstocks. The maximum hydrogen production and minimum energy cost are 0.16 mol/kWh and 3100 kWh/kg $H_2$, respectively, for the transarc reactor and 0.15 mol/kWh and 3300 kWh/kg $H_2$, respectively, for the glidarc reactor. The performance of the two reactors is comparable despite significantly different modes of operation.

Subsequently, a Streamer Dielectric-Barrier Discharge (SDBD) reactor is devised to produce hydrogen and carbon co-products from LDPE and cellulose, the latter as a model of biomass waste feedstock. Experimental characterization and modeling indicate that the plasma consumes between 90 to 100% of the input power; the electron and excitation temperatures depict approximately the same value, independent of feedstock, and increase with input power up to ~ 1.6 to 2.4 eV; and the electron number density is ~ $2.5\ 10^{13}$ $cm^{-3}$ irrespective of the feedstock and input power.

The hydrogen production rate for plasma-treated cellulose and LDPE increases to a maximum, after which it declines. The occurrence of peak production suggests that the plasma treatment of cellulose depicts two regimes, namely devolatilization and char reaction, similarly as observed in other biomass pyrolysis studies. The peak hydrogen production rate for cellulose is 40 mmol/h, which is twice that of LDPE (20 mmol/h). Moreover, the energy costs of hydrogen production for cellulose and LDPE are 600 and 1300 kWh/kg of $H_2$, respectively. The lower energy cost for cellulose is probably owed to its high porosity, leading to weaker dielectric strength that promotes increased hydrogen production at lower input power. Additionally, the hydrogen selectivity of cellulose is about two times more than that of LDPE due to the presence of residual gas in the reaction chamber and oxygen admixture during LDPE sample preparation, resulting in the production of $CO$ and $CO_2$ in addition to $H_2$ and hydrocarbons.

The characterization of solid products via field emission scanning electron microscopy reveals distinct morphological structures of the two feedstocks. Whereas pristine cellulose comprises fibrils bundled into fibers with porous and entangled structures, pristine LDPE is a nonporous, uniform, and dense-structured compound. The plasma-treated cellulose consists of protrusions of an average diameter of 50 nm, while residual LDPE has embedded micro-grains, and both present promising valuable solid residues which need further investigation. The results indicate that the use of SDBD plasma is effective approach for the production of hydrogen from cellulose and from LDPE at near atmospheric pressure and relatively low-temperature conditions in rapid-response and compact processes.

## 4.2 Contributions of the research

The primary tasks pursued in this study are:

1. The design and characterization of two nonthermal plasma reactors with complementary characteristics, namely transarc and glidarc, depicting complementary characteristics, to produce hydrogen from LDPE. The maximum hydrogen production efficiency and minimum energy cost are 0.16 mol/kWh and 3100 kWh/kg $H_2$, respectively, for the transarc reactor and 0.15 mol/kWh and 3300 kWh/kg $H_2$, respectively, for the glidarc reactor. Overall, the reactors' performance in hydrogen production is comparable despite markedly different modes of operation.
2. The devising of a Streamer Dielectric Barrier Discharge (SDBD) reactor to produce hydrogen and carbon co-products from cellulose and LDPE. Cellulose and LDPE are plasma-treated for different treatment times to characterize the evolution of the hydrogen production process. The results show that the maximum hydrogen production efficiency and minimum energy cost for cellulose treated by the SDBD reactor are 0.8 mol/kWh and 600 kWh/kg of $H_2$, respectively, representing approximately twice the efficiency and half the energy cost attained during the SDBD treatment of LDPE.

3. Spectroscopic and electrical diagnostics and modeling are used to estimate representative properties of the plasma, including electron and excitation temperatures, number density, and power consumption. Experimental characterization and modeling indicate that the plasma consumes between 90 to 100% of the input power; the electron and excitation temperatures depict approximately the same value, independent of feedstock, and increase with input power up to ~ 1.6 to 2.4 eV; and the electron number density is ~ 2.5 $10^{13}$ $cm^{-3}$ irrespective of the feedstock and input power.
4. Characterization of solid products via scanning electron microscopy and CHN-elemental analysis. The results revealed the distinct morphological structure of the two feedstocks treated.

## 4.3 Recommendations for further work

Further research is suggested in the following:

1. The influence of catalysts on the performance of the reactors for hydrogen and carbon co-products should be explored. This may affect the hydrogen yield and selectivity of the reactors.

2. Assessing the performance of different feedstock morphology, such as powder and films, could have a significant effect on the production of hydrogen and carbon co-products.
3. Other spectroscopic analysis such as continuum spectrum, double line ratio, and line broadening should be studied. The different spectral peaks should also be tagged.
4. Investigation of the geometrical effects of the dielectric and the feedstock on the production of hydrogen and carbon co-products.
5. The energy balance and mass balance of the process should be evaluated.

# APPENDICES

## Appendix A: Plasma diagnostics

Plasma is fundamentally characterized by excitation temperature, electron temperature, vibrational and rotational temperatures, and electron number density. Some of the popular techniques for plasma diagnostics include optical emission spectroscopy (OES), Langmuir probe measurements, interferometry, and mass spectroscopy, depending on the nature of the plasma. According to some of the latter studies, there is a good agreement between the OES and Langmuir probe measurements [1]. However, the advantages of OES are least perturbative and simple, the method establishes relationships between plasma parameters and the radiation features such as the emission or absorption intensity, and it allows for temporal and spatial monitoring of plasma. On the other hand, using OES to determine the average energy of free electrons requires analysis of kinetic processes leading to population and depopulation of the excited states of species in the plasma based on the collisional-radiative model. For two energy levels of spontaneous emission spectral lines, $E_i$ and $E_j$, with atomic density $N_i$ and $N_j$ for upper and lower energy levels, respectively, in thermal equilibrium, Boltzmann's distribution is :

$$\frac{N_i}{N_j} = \frac{g_i}{g_j} \exp^{-\frac{1}{k_B T_{exc}}\left(E_i - E_j\right)}, \qquad (1)$$

where $g_i$ and $g_i$ are the statistical weights of upper and lower energy levels respectively, $k_B$ is the Boltzmann's constant ($1.38\times10^{-23}$ J/K) and $T_{exc}$ is excitation temperature in degrees Kelvin.

The total population distribution over atomic state is defined as:

$$\frac{N_i}{N}=\frac{g_i}{U(T)}\exp^{-\frac{E_i}{k_B T_{exc}}}, \tag{2}$$

where $N$ is the total population density and a partition function, $U(T)$ is a summation of the population of all the possible energy levels of atoms, ions, or molecule, which is expressed as:

$$U(T)=\sum_m g_m \exp^{-\frac{E_m}{k_B T_{exc}}}. \tag{3}$$

The intensity of the spectral emission line for upper to lower state is defined as:

$$I_{ij}=\frac{hc}{4\pi\lambda_{ij}}A_{ij}N_i, \tag{4}$$

where $\lambda_{ij}$ is the wavelength of the emitted light, $h$ is Plank's constant, $c$ is the speed of light in vacuum, and $A_{ij}$ is a transition probability of the emitted spectra. The equations (4) and (2) can be deduced to

$$\ln\left(\frac{I_{ij}\lambda_{ij}}{g_i A_{ij}}\right)=-\frac{E_i}{k_B T_{exc}}+D \tag{5}$$

where constant $D = \ln(\frac{hcN}{4\pi U(T)})$. $T_{exc}$ is calculated from the slope of the straight and in the case of thermal plasma in LTE, $T_{exc}$ is approximately the electron temperature, $T_e$.

In nonthermal plasma, the LTE approximation is not appropriate because the excitation and de-excitation of the ionic species is usually not controlled by collision with electrons since these species are not in thermal equilibrium. Hence the excitation temperature in the Boltzmann plot differs from the electron temperature. Gordillo *et al.* [1] applied the concept of corona balance to modify the Boltzmann plot for estimation of $T_e$ in nonthermal plasma. The corona balance equation [11] is expressed as:

$$N_e N_1 k_{1i} = N_i \sum_{i>j} A_{ij} \tag{6}$$

where $N_e$ is the electron population density, $N_1$ is the ground level population density, $N_i$ is the excited state population density, $k_{1i}$, electron-impact excitation rate coefficient from the ground state 1 to level *i.*

**Table 1. Extract from NIST Spectral Databases for Ar I species**

| $\lambda_{ij}$ (nm) | $E_i$ (eV) | $A_{ij}$ ($10^8$ $s^{-1}$) | $\sum_{i>j} A_{ij}$($10^8$ $s^{-1}$) | $g_i$ | $b_{1i}$ | Number radiative transitions |
|---|---|---|---|---|---|---|
| 763.5 | 13.17 | 0.245x10$^{-1}$ | 4.14x10$^{-1}$ | 5 | 6.31x10$^{-9}$[3] | 3 |
| 714.7 | 13.28 | 6.30x10$^{-3}$ | 5.81x10$^{-2}$ | 3 | 5.75x10$^{-11}$[2] | 4 |
| 750.4 | 13.48 | 4.50x10$^{-1}$ | 8.50x10$^{-1}$ | 1 | 2.9x10$^{-9}$[2] | 2 |
| 731.6 | 15.02 | 9.60x10$^{-3}$ | 4.82x10$^{-2}$ | 3 | 4.19x10$^{-10}$[2] | 4 |
| 706.9 | 14.85 | 2.00x10$^{-2}$ | 8.92x10$^{-2}$ | 3 | 7.9x10$^{-10}$[2] | 4 |
| 516.2 | 15.31 | 1.90x10$^{-2}$ | 4.29x10$^{-2}$ | 2 | 5.86x10$^{-11}$[1] | 2 |

The population density $N_i$ can be found from:

$$I_{ij} = \frac{h\nu_{ij} A_{ij} N_i}{4\pi} L, \tag{7}$$

where, $h\nu_{ij}$ is the energy gap between levels *i* and *j*, *L* the plasma length that the emitted light goes through. The electron-impact excitation rate coefficient $k_{1i}$ can be expressed as function $T_e$ as

$$k_{1i} = b_{1i} e^{-\frac{E_{1i}}{k_B T_e}} \tag{8}$$

where the constant $b_{1i}=E_{1i}^{a}p_{i}^{b}$. The quantum number, $p_i$, of the excited states is defined as:

$$p_i=\sqrt{\frac{E_H}{E_\infty\text{-}E_i}}, \quad (9)$$

where $E_\infty$ and $E_i$ are the ionization energy and the energy of the excited state $i$, and $E_H$ is the Rydberg constant (13.6 eV).

The modified Boltzmann's plot technique can be obtained from Equations (6) and (7) as:

$$\ln\left(\frac{I_{ij}\sum_{i>j}A_{ij}}{hv_{ij}A_{ij}b_{1i}}\right)=-\frac{E_{1i}}{k_BT_e}+B \quad (10)$$

and $T_e$ can be determined from the slope of the straight line and the intercept $B=\ln(\frac{N_eN_1L}{4\pi})$.

The parameters $I_{ij}$, $A_{ij}$,and $E_{1i}$ will be obtained from NIST [4] atomic spectra database. $k_{1i}$ is obtained from different authors, shown in Table 1.

## Appendix B: Hydrogen production from Cellulose and LDPE composites

### *Optical imaging*

Representative optical imaging results of the transarc reactor during the treatment of samples with different LPDE-CE compositions for operating conditions: $Q$ = 0.1 slpm, $H$ = 5 mm, and $V$ = 30% is shown in Fig. 1. The transarc plasma glows from yellow (100% CE) to purple (100% LPDE) as the proportion LDPE increases in the sample. The yellow appearance depicts more significant amount of oxygen species derived from cellulose, and the purplish glow is the characteristic of nitrogen plasma.

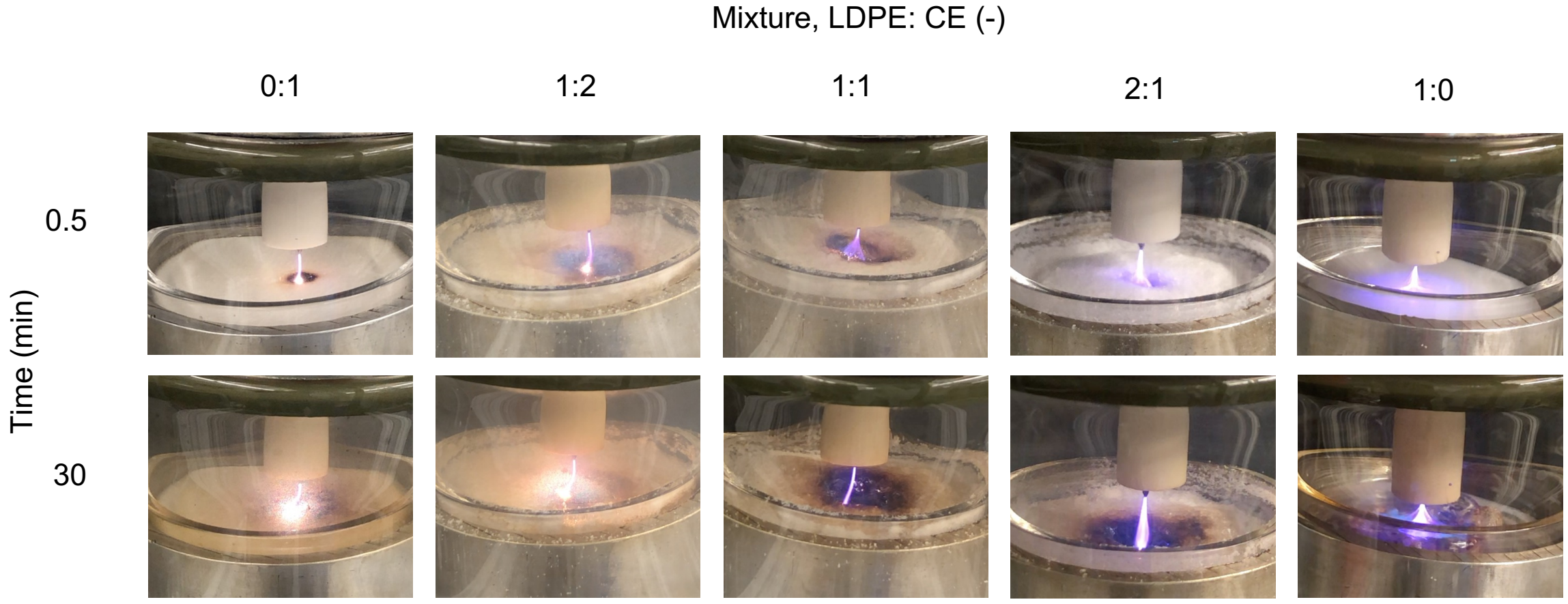

**Fig. 1. Characterization of the transarc reactor during the treatment of LDPE-CE samples**. Optical imaging under operational conditions of $Q$ = 0.1 slpm, $V$ = 30%, and $H$ = 5 mm for different compositions LPDE: CE by mass at the beginning (0.5 min) and end (30 min) of the experiment.

The interaction of plasma with CE generates a scintillating yellow glow that is more pronounced by the end of the experiment for samples consisting of 50%, 67% and 100% CE. The intensity of the glow is highest for 67% and 100% CE, as shown in Fig. 1. As the proportion of CE in the samples decreases, it is observed that plasma operates in a spark regime characterized by highly unstable filamentary electrical discharges. With 100% LPDE, the spark discharge covers a larger sample surface area, potentially leading to greater hydrogen production. The apparent increase in plasma length at the end of the experiment, particularly for samples containing CE, is attributed to the larger depth of the crater formed at the center of the sample. However, the small thickness of the 100% LPDE sample leads to the crater's formation at the bottom of the crucible with a smaller depth.

***Sample characterization***

Optical images of LDPE-CE samples of different compositions before and after 30 min treatment with the transarc reactor are shown in Fig. 2. The white pristine, untreated samples in Fig. 2 (first row) show the differences in volume with varying the amount of CE, which is significantly less dense than LDPE. The non-uniformity of the top surface of the samples with LDPE: CE = 1:2, 1:1, and 2:1 before the experiment depict the dissimilar characteristics of LDPE and CE particles after sedimentation.

The 30-minute sample treatment with nitrogen plasma results in the formation of a crater at the center of the samples, as shown in Fig. 2 (second row). For the 100% LDPE, the crater hole extends up to the bottom of the sample due to the smaller thickness of the sample. Since in the transarc reactor, the sample is electrically coupled to the plasma, the plasma interacts through the sample reaching the bottom of the sample, as shown in Fig. 2 (third row). The intense plasma-feedstock interaction at the center is an important attribute for redesigning the crucible to a smaller diameter. The dark appearance of the sample after 30 minutes of the experiment probably depicts the formation of carbon compounds (e.g., char, carbon black).

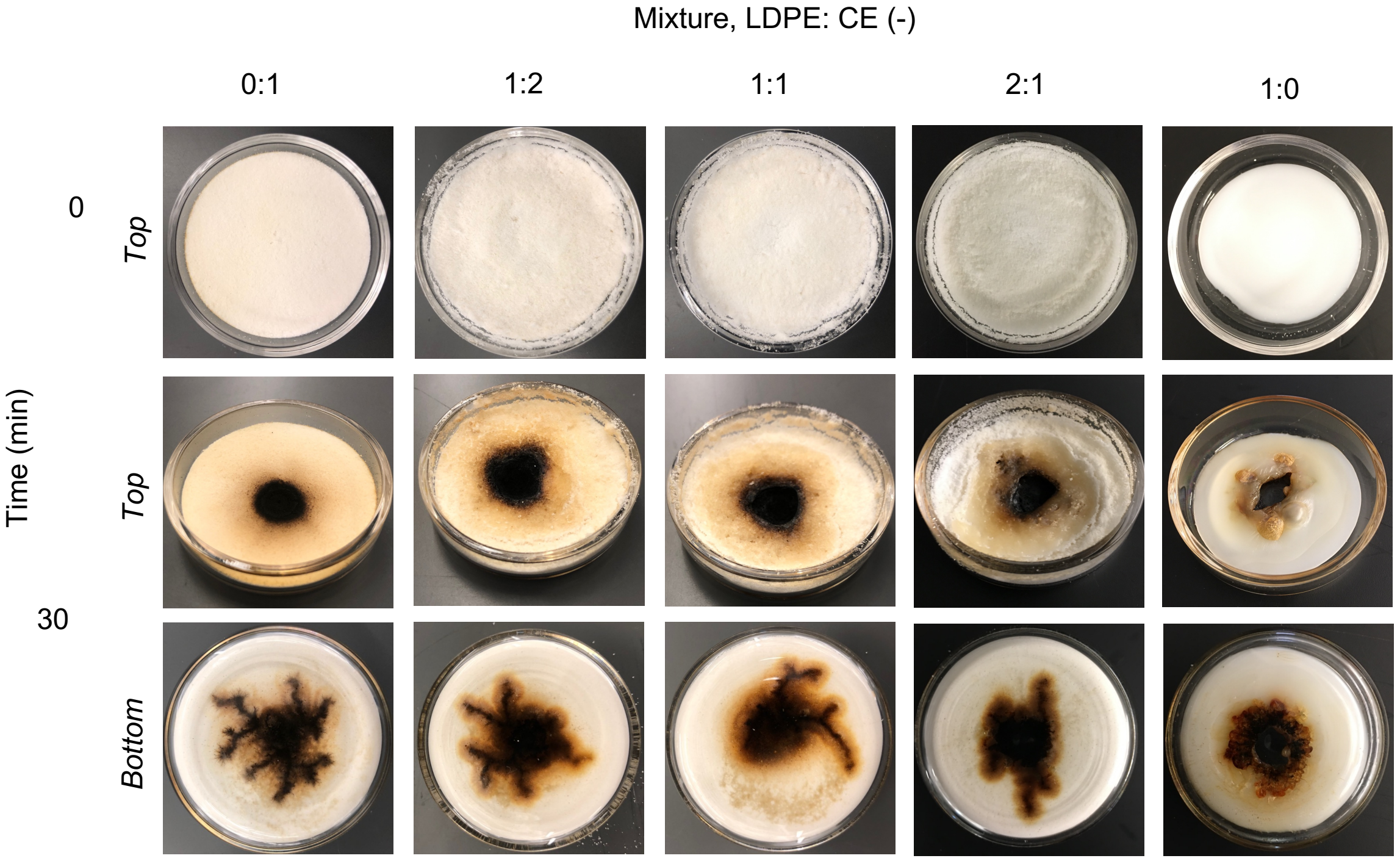

**Fig. 2. Characterization of LDPE: CE samples**. Optical imaging of samples before treatment (first row) and after 30 min treatment (second and third rows) under operational conditions of $Q$ = 0.1 slpm, $V$ = 30%, and $H$ = 5 mm for different compositions of LPDE-CE.

### *Electrical Characterization*

The electrical characterization of the reactor allows an indirect assessment of the plasma dynamics and their influence on hydrogen production. The overall electrical characterization showing the rms current and rms power for the different LPDE-CE compositions is shown in Fig. 3. The rms current fluctuates minimally throughout the experiments and it is generally highest and lowest for 100% LPDE and 100% CE, respectively, as shown in Fig. 3b. However, the LPDE-CE composites have comparable rms current irrespective of the proportion of each polymer in the sample.

The rms power delivered to the substrate also follows the same trend as the rms current. It is highest and lowest for 100% LPDE and 100% CE, respectively. This is attributed to the differences in the thermophysical and thermochemical properties of the two polymers. Additionally, the hardened surface of the sample of 100% LDPE results in an intense plasma interaction that covers the whole surface leading to more electrical power deposition.

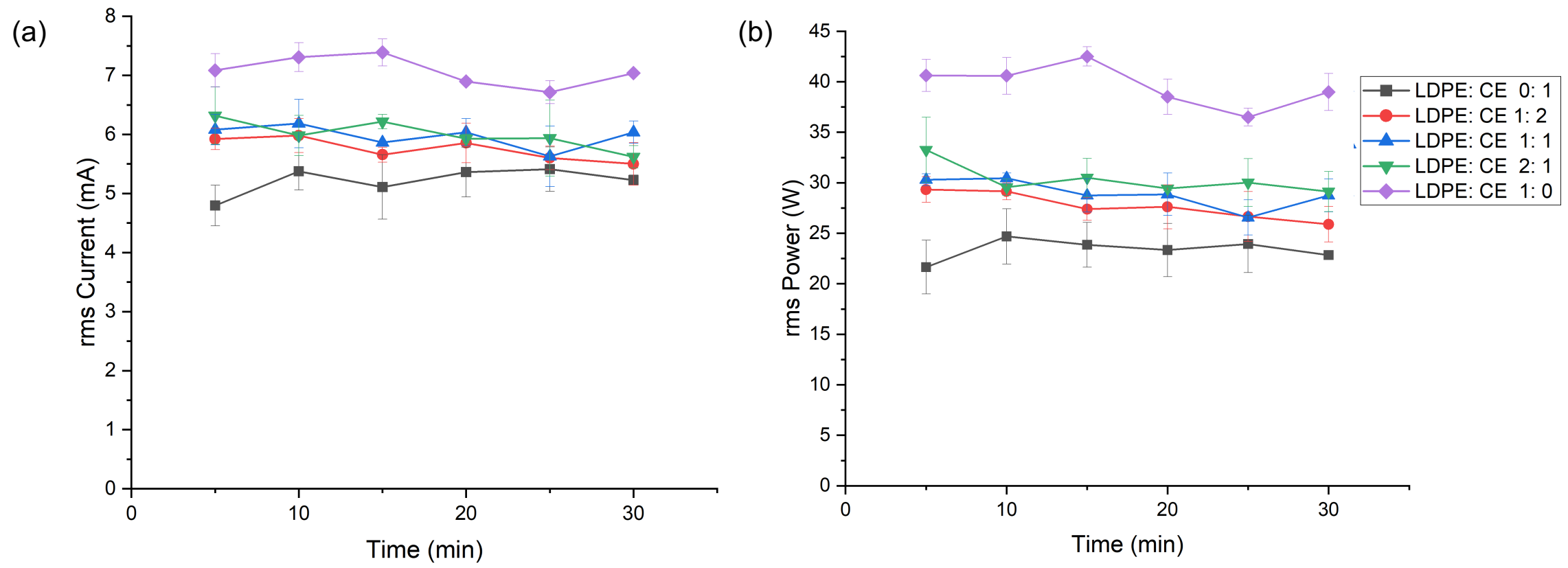

**Fig. 3. The overall electrical performance of the reactors during operation**. (a) current and (b) power as a function of time for samples with different LPDE: CE compositions.

### *Process performance*

The performance of the transarc reactor through the treatment of the LDPE-CE composite samples is evaluated in terms of hydrogen production rate and hydrogen production efficiency, as shown in Fig. 4. The hydrogen production rate for the mixtures of LDPE and CE is higher than that of 100% LPDE and 100% CE (Fig. 4a). This is probably attributed to the synergistic effect of the LDPE-CE composites that lowers the activation energy of the feedstock leading to greater hydrogen production as compared to individual materials (100% LDPE and 100% CE), as observed by Ma *et al.* [2]. Furthermore, the interaction between volatiles and fixed carbon of cellulose degradation results in radical donation leading to initiating and enhancing the polyethylene chain scission [3]. The lower hydrogen production in 100% CE as compared to 100% LDPE is owing to fewer hydrogen atoms and weaker hydroxyl bonds, favoring the production of OH instead of hydrogen.

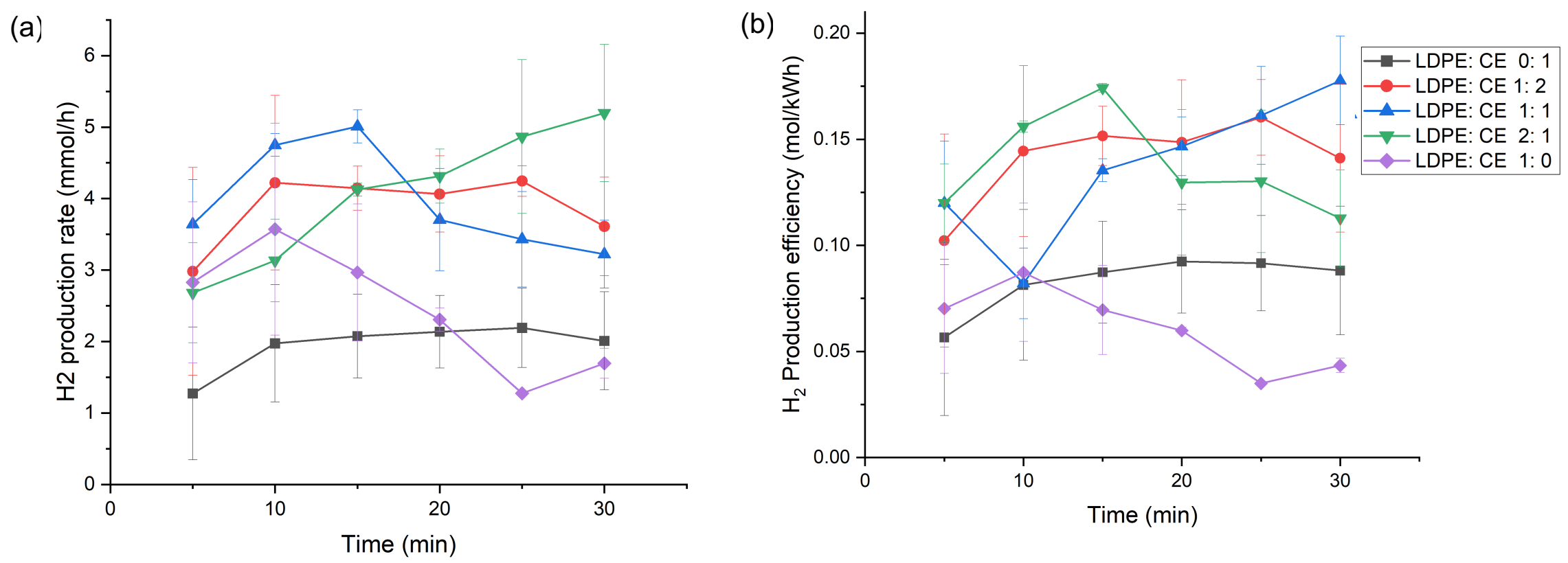

**Fig. 4. Hydrogen production rate and production efficiency**. (a) hydrogen production rate (b) hydrogen production efficiency for the treatment of LDPE-CE samples with the transarc reactor.

Hydrogen production efficiency, assessed as the amount of hydrogen produced per unit energy for different LDPE-CE composites, is shown in Fig. 4b. The high hydrogen production efficiency of LDPE-CE composite depicts the synergistic effect of lowering the activation energy of the mixtures, leading to greater hydrogen production and higher production efficiency. The lower hydrogen production efficiency of 100% LDPE signifies that larger energy wastage at the bottom of the crucible due to the complete etching of the sample at the center, as shown in Fig. 4.

# Appendix C: Lab-related photography

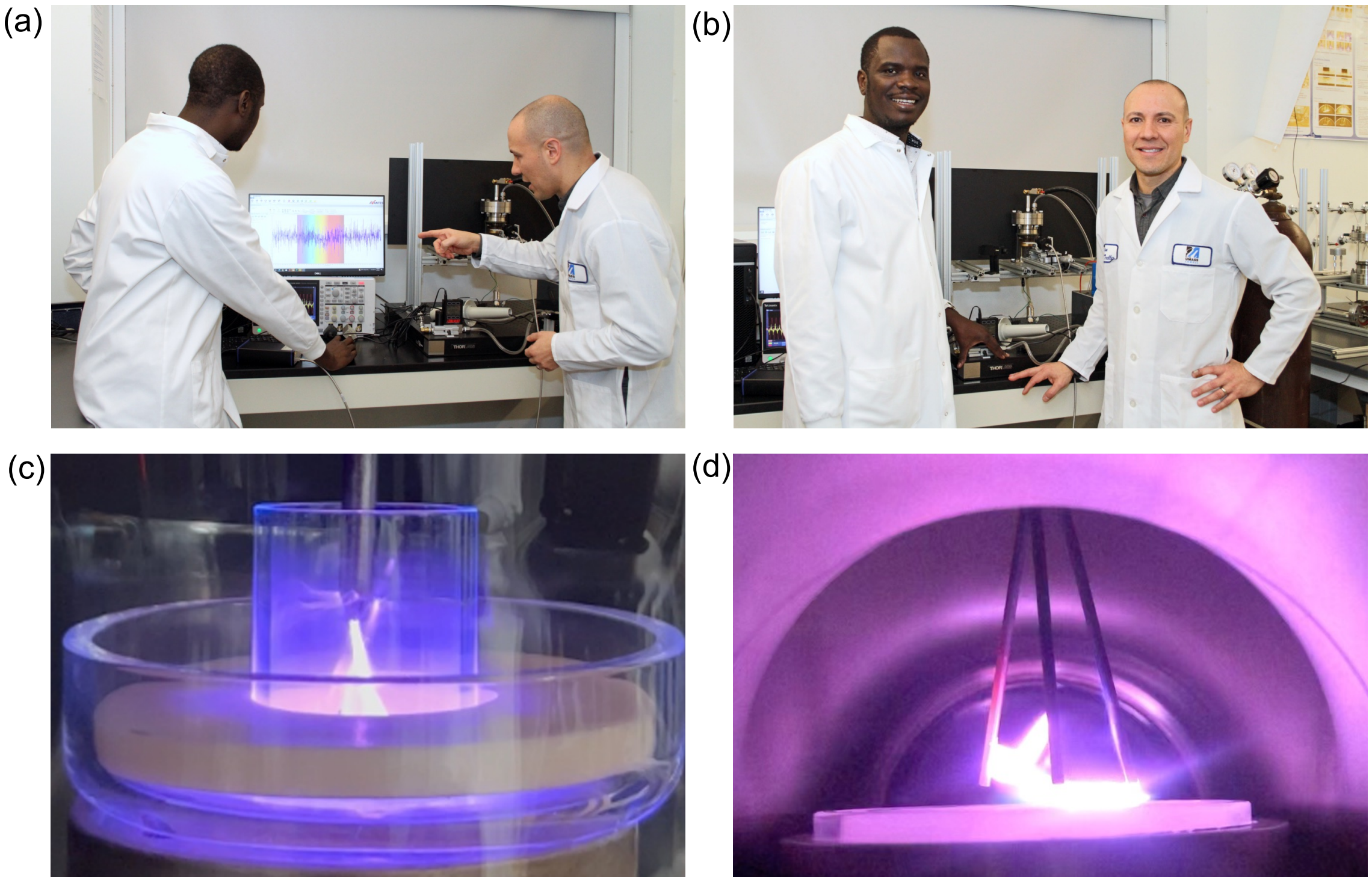

**Fig. 1. Lab-related photography**. (a) Benard and his advisor, Prof. Juan Pablo Trelles, analyzing in the spectroscopic signal. (b) Benard and his advisor post for a photo after the experiment. (c) Streamer DBD without feedstock (c) Gliding arc plasma interacting with a dummy feedstock.

# BIOGRAPHICAL SKETCH

## Research Interest

Plasma engineering, energy sustainability, solar thermal storage, hydrogen production, waste valorization, plasma diagnostics, and plasma-chemical synthesis.

## Education

- **MS: Physics**, (2017), Makerere University, Kampala Uganda
  Topic: Thermal performance of selected oils in Uganda for indirect solar domestic cooking application

- **BS: Science Education**, (Physical), (2010), Gulu University, Gulu, Uganda
  Topic: Design and construction of a fire detector

## Employment

- **Assistant Lecturer** (2018-2019), Department of Physics, Gulu University, Gulu, Uganda

- **Teaching Assistant** (2013-2018), Department of Physics, Gulu University, Gulu, Uganda

- **Education Officer** (2010-2013), Department of Physics, Gulu High School, Gulu, Uganda

**Selected Publications:**